\documentclass[JHEP,12pt]{article}
\pdfoutput=1
\usepackage{graphics, color,soul}
\usepackage{graphicx}
\usepackage{amssymb}
\usepackage{jheppub}
\usepackage{lmodern,mathtools}
\usepackage{caption}
\usepackage {subcaption}
\usepackage{booktabs}
\usepackage[english]{babel}
\usepackage{amsmath,amssymb,amsbsy,amstext, amsthm, simplewick}
\usepackage{hyperref}
\usepackage{tikz}
\usetikzlibrary{decorations.markings}
\usepackage{xcolor}
\usepackage{ulem}

\usepackage{caption}

\usepackage{amsfonts}
\usepackage{amssymb}
\usepackage{upgreek}
\usepackage{simplewick}
 \usepackage{exscale,relsize}
\usepackage{mathtools}
\usepackage{comment}

\usepackage[margin=1cm,labelfont={sf,bf,scriptsize},textfont={sf,scriptsize}]{caption}

\def\cO{\mathcal{O}}

\def \la {\langle}
\def \ra {\rangle}
\def \p {\partial}

\def\nn{{\nonumber}}

\DeclareRobustCommand{\Fig}[1]{Fig.~\ref{#1}}

\def\be#1\ee{\begin{align}#1\end{align}}

\begin{document}
\title{ 
A thermal product formula
}

\author{Matthew Dodelson$^{a}$,}
\author{Cristoforo Iossa$^{b,c}$,}
\author{Robin Karlsson$^{a}$, and}
\author{Alexander Zhiboedov$^a$}

\affiliation{$^a$CERN, Theoretical Physics Department, CH-1211 Geneva 23, Switzerland}
\affiliation{$^b$SISSA, via Bonomea 265, 34136 Trieste, Italy}
\affiliation{$^c  $INFN, sezione di Trieste, via Valerio 2, 34127 Trieste, Italy}

\abstract{
 We show that holographic thermal two-sided two-point correlators take the form of a product over quasi-normal modes (QNMs). Due to this fact, the two-point function admits a natural dispersive representation with a positive discontinuity at the location of QNMs. We explore the general constraints on the structure of QNMs that follow from the operator product expansion, the presence of the singularity inside the black hole, and the hydrodynamic expansion of the correlator. We illustrate these constraints through concrete examples. We suggest that the product formula for thermal correlators may hold for more general large $N$ chaotic systems, and we check this hypothesis in several models.
    }
    
    \begin{flushleft}
 \hfill \parbox[c]{40mm}{CERN-TH-2023-062}
\end{flushleft}
\maketitle

\section{Introduction}

A thermal two-point function describes the response of a system at finite temperature to a small perturbation \cite{kadanoff1963hydrodynamic}. In theories with simple holographic duals, it is captured by the wave equation in a black hole background \cite{Maldacena:1997re,Gubser:1998bc,Witten:1998qj,Witten:1998zw,Son:2002sd}.\\
\indent In this paper we mostly  (but not only!) concern ourselves with holographic CFTs, for which the relevant geometry is a black hole in AdS. For this case, the problem of computing the thermal response is reduced to a quantum-mechanical scattering problem in a potential $V(z)$ which encodes the black hole geometry. Very generally for non-extremal black holes, this potential has the following features,\footnote{Here $z$ is the tortoise coordinate, i.e. $d z = -{d r \over f(r)}$, where $f(r)$ is the black hole redshift factor.}
\be
V(z)\sim \begin{cases}
			\frac{1}{z^2} & z\to0 ~~~~\text{(AdS boundary)} ,\\
            \sum_{n=1}^\infty  a_n e^{-\frac{4\pi n}{\beta}z} &  z\to\infty ~~~\text{(BH horizon)} .
		 \end{cases}
\ee
The exponential decay close to the horizon is consistent with the periodicity of the potential $V(z) = V(z+i {\beta \over 2})$ under complex shifts of $z$.

Let us denote the two-sided thermal function that can be obtained by solving the wave equation as $G_{12}(\omega)$, where we keep the angular/spatial momentum dependence implicit. There are two basic facts about $G_{12}(\omega)$ that will be important for us in the present paper:

\begin{itemize}
\item It is a meromorphic function of $\omega$ \cite{Hartnoll:2005ju,festucciathesis,Festuccia:2005pi}. Its singularities, known as \emph{quasi-normal modes} (QNMs), encode the characteristic decay of perturbations in time \cite{Horowitz:1999jd,Berti:2009kk}.
\item It has \emph{no zeros}, or in other words, ${1 \over G_{12}(\omega)}$ is an entire function of $\omega$. This property is intimately related to the behavior of the potential $V(z)$ close to the black hole horizon, see \cite{festucciathesis} and Appendix \ref{nozeroesappendix}.
\end{itemize}
Even though these properties have been extracted by studying classical wave propagation on a black hole background, at the level of the thermal two-point function they can be considered more generally in situations where there is no well-defined notion of a classical geometry or black hole horizon. In fact, extrapolating beyond holography, we will find that both properties hold at finite coupling in several low-dimensional examples of large $N$ systems. 

The two properties above imply, modulo simple technical details deferred to the main text, that we can write a product formula representation for the two-sided correlator,\footnote{The condition $G_{12}(0)>0$ is a consequence of unitarity, see Section \ref{Sec:AnalyticStructure}.}
\be
\label{eq:ansatzprod}
G_{12}(\omega) =   {G_{12}(0) \over \prod_{n=1}^\infty(1 - {\omega^2 \over \omega_n^2})(1 - {\omega^2 \over (\omega_n^*)^2})} , ~~~~ G_{12}(0) > 0. 
\ee
In other words, the two-point function is \emph{fixed} by the QNMs up to an overall rescaling. No further data (such as the residues at the poles) is required. Note that the same statement translates to the retarded two-point function $G_R(t)$ for $t>0$, see Appendices \ref{app:2pointconventions} and \ref{app:probingomega}. In  frequency space we have the following relation
\be
G_{12}(\omega) =\frac{G_R(\omega)- G_R(-\omega)}{2 i \sinh(\beta\omega/2)}, ~~~~~~ \omega \in \mathbb{C} \ .\label{g12grIntro}
\ee

Another characteristic feature of the black hole geometry in $AdS_{d+1>3}/CFT_{d>2}$ is the presence of the curvature singularity. In \cite{Festuccia:2005pi} it was argued that the singularity leads to exponential decay of the two-sided two-point function at large imaginary $\omega$. An illuminating way to encode this property is to say that the two-sided correlator obeys the following dispersive sum rules
\be
\label{eq:BHSS}
\text{BH singularity}:~~~ \oint_{C_\infty} d\omega' \,(\omega')^m G_{12}(\omega') = 0 ,
\ee
see Figure \ref{singcontour}. Again, even though \eqref{eq:BHSS} was extracted by studying the classical wave equation, it is interesting to contemplate the possibility that it holds more generally and consider its violation as some sort of ``resolution'' of the singularity.

The purpose of this paper is to analyze the general properties of the product formula \eqref{eq:ansatzprod}, as well as to explore its workings in various examples. In particular, we study the constraints imposed on the structure of QNMs by the OPE; we discuss how the  singularity sum rules \eqref{eq:BHSS} are satisfied and violated; and finally, we explore the relation between the structure of QNMs and the low-energy (hydrodynamic) expansion of the correlator. In the latter case, we find that the hydrodynamic expansion encodes certain moments of the QNM density. 

After examining the general constraints, we specialize to several analytically solvable examples, including BTZ and Rindler space. We then proceed to discuss several examples in pure GR and beyond for which no analytic expression is available. To compute the QNMs for these cases, we use the package $\mathtt{QNMSpectral}$ \cite{Jansen:2017oag}.

In the final section of the paper, we show that the two-point functions in several SYK-type models satisfy the holographic properties mentioned above. We conclude by presenting a hypothesis regarding the two-point function in planar chaotic theories, and with a discussion of various open directions.

\section{Analytic structure of thermal two-point functions\label{Sec:AnalyticStructure}}

In this paper we are mainly interested in the two-sided thermal two-point function, defined as follows,
\be
G_{12}(t, \vec x) \equiv \left\langle {\cal O}\left(t-\frac{i\beta}{2}, \vec x\right) {\cal O}(0,0) \right\rangle_{\beta} = {1 \over Z(\beta)} {\rm tr}\Big[ e^{- \beta H} {\cal O}\left(t-\frac{i\beta}{2}, \vec x\right) {\cal O}(0,0) \Big] ,
\ee
where $\beta$ is the inverse temperature and the spatial coordinate $\vec x$ labels a point on the spatial manifold on which the CFT lives. We will mostly consider the case where $\vec x \in \mathbb{R}^{d-1}$ 
in the present paper.

We will be considering $G_{12}(t, \vec x)$ in Fourier space 
\be
G_{12}(\omega, k) &= \int d t\, d^{d-1} \vec x\, e^{i \omega t - i \vec k \cdot \vec x} G_{12}(t, \vec x), ~~~ \vec x \in \mathbb{R}^{d-1} ,
\ee
where $k \equiv |\vec k|$. We will be mostly concerned with the dependence of the two-sided correlator on $\omega$, while keeping $ k$ fixed. We will therefore sometimes write $G_{12}(\omega)$ with the spatial dependence kept implicit. We have chosen to work with $G_{12}(\omega,k)$ because of its nice properties, but all other two-point functions can be easily obtained once $G_{12}(\omega,k)$ is known, see Appendix \ref{app:2pointconventions}.

\subsection{Definitions and general properties}

Let us review some basic properties of the two-sided thermal two-point function $G_{12}(\omega)$ as a function of $\omega$:

\begin{itemize}
\item {\bf Unitarity}: on the real axis the two-sided correlator is real and non-negative
\be
\label{eq:unitarity}
G_{12}(\omega) \geq 0 , ~~~ \omega \in \mathbb{R}.
\ee
This property simply follows from the insertion of a complete set of states between the two operators in the definition of the two-sided correlator and unitarity of the underlying theory.\footnote{In the most general setting, $G_{12}(\omega)$ is a distribution, and therefore \eqref{eq:unitarity} should be understood in the distributional sense. In the present paper we will treat $G_{12}(\omega)$ as a function. In particular, we will study its analytic properties in the complex $\omega$ plane. We believe that this is justified for interacting CFTs on $S^1 \times \mathbb{R}^{d-1}$ and large $N$ interacting CFTs on $S^1 \times S^{d-1}$ above the Hawking-Page phase transition \cite{Hawking:1982dh}. It is true by inspection for holographic CFTs.} Assuming that $G_{12}(\omega)$ is real-analytic for real $\omega$, we get in the complex plane
\be
\label{eq:herman}
G_{12}^*(\omega) = G_{12}(\omega^*).
\ee
When $G_{12}(\omega)=0$ on the real axis something special happens, namely the local operator with a given energy $\omega$ annihilates the thermal state. We expect that this is only possible in free (or integrable) theories, and that in interacting theories $G_{12}(\omega)>0$ on the real axis.
\item {\bf KMS symmetry}: the two-sided correlator is even under $\omega \to - \omega$,
\be
\label{eq:kms}
G_{12}(\omega) = G_{12}(-\omega) . 
\ee
\item {\bf OPE limit}: the large frequency limit of the correlator is universal and is controlled by the unit operator,
\be\label{opebehavior}
\lim_{\omega \to + \infty} G_{12}(\omega) \sim e^{-{\beta \omega \over 2}} \omega^{2 \Delta - d} ,
\ee
where $\Delta$ is the conformal dimension of $\mathcal{O}$ and the proportionality coefficient is known exactly in terms of $d$ and $\Delta$. As we will see later, corrections to this formula at large $\omega$ can be systematically computed in the $1/\omega$ expansion, and are determined by the OPE coefficients and thermal one-point functions of operators that appear in the $\mathcal{O}\times \mathcal{O}$ OPE expansion.
\end{itemize}

The properties above hold in any CFT and do not rely on holography.

\subsection{Properties of holographic thermal correlators}

Next we list the properties of $G_{12}(\omega)$ which are specific for theories with a classical gravity dual:
\begin{itemize}
\item {\bf Meromorphy}: the only singularities of $G_{12}(\omega)$ are isolated simple poles in the complex plane \cite{Hartnoll:2005ju,festucciathesis,Festuccia:2005pi}.\footnote{Double poles are allowed on the imaginary axis (although imaginary poles are still generically simple). For instance in $AdS_3/CFT_2$ there are double poles on the imaginary axis when $k=0$.} These singularities correspond to \emph{quasi-normal modes} (QNMs), and they characterize the real-time decay of a perturbation to the thermal plasma \cite{Horowitz:1999jd,Berti:2009kk}. By virtue of \eqref{eq:herman} and \eqref{eq:kms}, QNMs come in families $(\omega_n, - \omega_n, \omega_n^*, - \omega_n^*)$. Similarly, the residues of $G_{12}(\omega)$ for various frequencies within one family are all related to each other, 
\begin{align}\label{residuerelation}
\text{Res}_{\omega_n}G_{12}(\omega)=-\text{Res}_{-\omega_n}G_{12}(\omega)=(\text{Res}_{\omega_n^*}G_{12}(\omega))^*=-(\text{Res}_{-\omega_n^*}G_{12}(\omega))^*.
\end{align}
\item {\bf No zeros:} the two-sided correlator does not have zeros in the complex $\omega$-plane \cite{festucciathesis}. This property follows from the fact that the two-sided correlator is obtained by solving the wave equation in the black hole background and from the universal properties of the scattering potential close to the black hole horizon, see Appendix \ref{nozeroesappendix}. Put differently, $1/G_{12}(\omega)$ is an entire function. 
\item {\bf Asymptotic behavior:} along any ray in the complex plane that asymptotically avoids poles, the two-sided correlator satisfies 
\begin{align}\label{asymptoticg12}
\left|\frac{1}{G_{12}(\omega)}\right|\sim e^{\lambda(\theta) r},\hspace{5 mm}\omega=re^{i\theta}\to \infty,\hspace{5 mm}\lambda(\theta)\ge 0,
\end{align}
up to a polynomial prefactor. In other words, $1/G_{12}(\omega)$ is an entire function of order one. The asymptotic behavior (\ref{asymptoticg12}) follows from the large $\omega$ expansion of the wave equation, see \cite{festucciathesis}.
\end{itemize}

The three properties above imply the following product formula for holographic thermal correlators\footnote{An analogous product formula appeared in the context of thermal functional determinants \cite{Denef:2009kn}, but for thermal correlators we are not aware of such results in the literature.}, 
\be
\label{eq:ansatzprod}
G_{12}(\omega) =  {G_{12}(0)  \over \prod_{n=1}^\infty(1 - {\omega^2 \over \omega_n^2})(1 - {\omega^2 \over (\omega_n^*)^2})} . 
\ee
To see this, we use Hadamard's factorization theorem, see e.g. \cite{conway2012functions}, which states that an entire function $f(z)$ of order $m$ and zeroes $a_n$ can be written as 
\begin{align}
f(z)= z^\ell e^{P(z)}\prod_{n=1}^\infty E_{\left \lfloor m\right\rfloor}(z/a_n),
\end{align}
where $P(z)$ is a polynomial of degree $q\leq m$ and 
\begin{align}
E_{\left \lfloor m\right\rfloor}(z)=(1-z)\prod_{k=1}^{\left \lfloor m\right\rfloor}e^{z^k/k}
\end{align}
are elementary factors. By assumption, $1/G_{12}$ is of order $m=1$. Imposing evenness, $\ell=0$, and noting that $E_1(z)E_1(-z)=E_0(z)E_0(-z)=1-z^2$, the product formula follows.

In writing \eqref{eq:ansatzprod} we assumed that $\omega^*_n \neq - \omega_n$, which is not true for purely imaginary QNMs. It is trivial to extend the ansatz by including the product of purely imaginary simple modes $\omega_n = i \tilde \omega_n$ as follows, $\prod_{n=1}^\infty \left(1+{\omega^2 \over \tilde \omega_n^2}\right)^{-1}$.

\subsection{Black hole singularity sum rules}
\indent In \cite{Festuccia:2005pi,festucciathesis} it was shown that for the Schwarzschild black hole, $G_{12}(\omega)$ at large imaginary $\omega$ is controlled by a real geodesic that bounces off the black hole singularity \cite{Kraus:2002iv,Fidkowski:2003nf}. As a result, the correlator decays exponentially as $\omega\to \pm i\infty$ \cite{Festuccia:2005pi,festucciathesis},
\begin{align}
|G_{12}(\omega)|\sim e^{-\tilde{\beta}|\omega|/2},\hspace{10 mm}\omega\to \pm i\infty,\hspace{10 mm}\tilde{\beta}=\beta\cot\left(\frac{\pi}{d}\right).
\end{align}
The two-point function is also exponentially decaying in the region $\omega\to \pm \infty$, where it is controlled by the OPE behavior (\ref{opebehavior}). Therefore in the presence of the curvature singularity there exists a positive number $\kappa$ such that (avoiding the poles)
\begin{align}
\frac{|G_{12}(\omega)|}{e^{-\kappa |\omega|}}\to 0\text{ as }|\omega| \to \infty,\hspace{10 mm}\kappa>0.\label{expdecay}
\end{align}
This is a stronger condition than (\ref{asymptoticg12}), which allows for a vanishing decay rate $\lambda(\theta)=0$ along some complex ray. \\
\indent We can use the condition (\ref{expdecay}) to write down a simple set of sum rules,\footnote{By singularity here we mean the curvature singularity. For example, \eqref{eq:presumrules} does not hold for the BTZ black hole.}
\begin{figure}
\centering
\begin{tikzpicture}
\draw[gray] (-2.5,0) -- (2.5,0);
\draw[gray] (0,-2.5) -- (0,2.5);
\draw[red, very thick, decoration={markings, mark=at position 0.55 with {\arrow{>}}},
        postaction={decorate}] (0,0) circle (90 pt);
\foreach \Point in {(-.5,.5), (-1,1), (-1.5,1.5), (-2,2),(.5,.5), (1,1), (1.5,1.5), (2,2),(.5,-.5), (1,-1), (1.5,-1.5), (2,-2),(-.5,-.5), (-1,-1), (-1.5,-1.5), (-2,-2)}{
    \node at \Point {\textbullet};
       }
       \node[red, very thick] at (.6,2.5)  {$C_\infty$};
\end{tikzpicture}
 \caption{The contour integral along $C_\infty$ in (\ref{eq:presumrules}) vanishes due to the asymptotic decay of the correlator. By deforming the integration contour inside and picking up the contribution of QNMs we get the sum rules \eqref{sumrules}.  \label{singcontour}}
\end{figure}
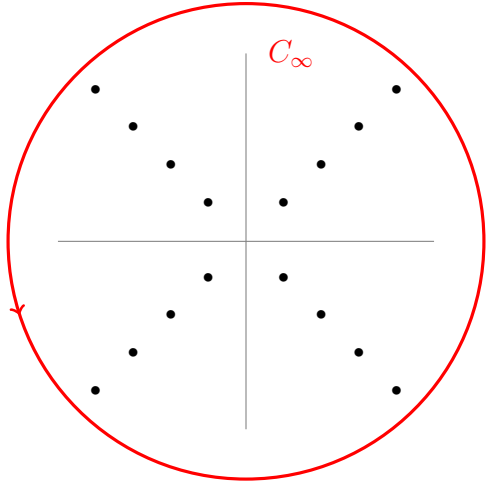
\be
\text{BH singularity}:~~~ \oint_{C_\infty} d\omega' \,(\omega')^m G_{12}(\omega') = 0 ,
\label{eq:presumrules}
\ee
see Figure \ref{singcontour}. The asymptotic behavior (\ref{expdecay}) guarantees that the contour integrals (\ref{eq:presumrules}) vanish. We can then deform the contour and rewrite these sum rules in terms of the residues of $G_{12}(\omega)$, $\lambda_n$, at the QNMs $\omega = \omega_n$. For $m$ even the sum vanishes identically, but for $m$ odd we find a nontrivial constraint:
\begin{align}
\text{Re}\sum_{n}\omega_n^m \lambda_n=0,\hspace{10 mm}\text{ for odd }m>0,\label{sumrules}
\end{align}
where we used the relations (\ref{residuerelation}) between residues of poles in the same family. Note that the sum rules \eqref{sumrules} immediately imply that the residues $\lambda_n$ decay exponentially fast in $n$. 
As we will see in Section \ref{complexbehavior}, the singularity sum rules follow directly from the product formula (\ref{eq:ansatzprod}) with some mild assumptions on the asymptotic structure of QNMs. In Section \ref{complexbehavior} we will use the exponential decay at $\omega\to i\infty$ along with the OPE to uniquely fix the leading high-energy asymptotics of QNMs.

Let us also notice that we can consider finite energy singularity sum rules by placing the integration contour $C_{\Lambda}$ in \eqref{eq:presumrules} at some finite radius $|\omega|=\Lambda$.\footnote{See e.g. \cite{Dolen:1967jr,Mukhametzhanov:2018zja,Noumi:2022wwf} for discussions of finite energy sum rules in the context of scattering amplitudes. In particular, in \cite{Mukhametzhanov:2018zja} they were rigorously derived using Tauberian theorems.}  In this case the sum rules are satisfied up to exponentially small corrections $e^{- c \Lambda}$. The advantage of considering sum rules at finite $\omega$ is that we expect them to hold also at finite 't Hooft coupling (or string length) as long as $\Lambda \lesssim {1 \over l_{s}}$.

\section{QNMs and black hole geometry}\label{qnmbh}
In the next section we will analyze various constraints placed on the QNMs by the product formula (\ref{eq:ansatzprod}). To set the stage for these constraints, let us first give a brief overview of QNMs and their connection to the geometry of the black hole background, closely following \cite{Festuccia:2008zx,festucciathesis}. For a detailed review, see \cite{Berti:2009kk}.\\
\indent We consider a spherically symmetric asymptotically $AdS_{d+1}$ black hole, with metric 
\begin{align}
ds^2=-f(r)\, dt^2+\frac{dr^2}{f(r)}+r^2\, d\Omega_{d-1}^2.
\end{align}
In order to write the scalar field wave equation $(\Box-\Delta(\Delta-d))\phi=0$ in a convenient form, let us Fourier expand
\begin{align}
\phi(t,\vec{x},r)=e^{-i\omega t}Y_{J \vec{m}}(\Omega) r^{-(d-1)/2}\psi(\omega,J,r),
\end{align}
where $Y_{J\vec{m}}$ are spherical harmonics on $S^{d-1}$. In terms of $\psi$, the wave equation takes the form of the Schrodinger equation for a particle in a potential,
\begin{equation}
\label{waveequation}
    (-\partial_z^2+V(J, z)-\omega^2)\psi=0,
\end{equation}
where $dz=-\frac{dr}{f(r)}$ is the tortoise coordinate. The horizon is located at $z=\infty$ and the boundary at $z=0$. The potential is given by 
\begin{align}\label{genpotential}
V(J,z)=f(r)\left(\frac{J(J+d-2)}{r^2}+\nu^2-\frac{d^2}{4}+\frac{(d-1)f'(r)}{2r}+\frac{(d-1)(d-3)}{4r^2}f(r)\right),
\end{align}
where $\nu=\Delta-d/2$. The potential for a black brane takes the same form, with $J(J+d-2)$ replaced with $k^2$.\\
\indent Now let us discuss the behavior of the potential (\ref{genpotential}) near the horizon and infinity. For a general AdS black hole, the asymptotic behavior is universally given by
\begin{equation}\label{eq:VAsymp}
V(J,z)\sim \begin{cases}
			\frac{\nu^2-\frac{1}{4}}{z^2} & z\to0\\
            \sum_{n=1}^\infty  a_n(J) e^{-\frac{4\pi n}{\beta}z} &  z\to\infty.
		 \end{cases}
\end{equation}
The behavior at the boundary follows from the asymptotics $f(r)\sim r^2$ at large $r$. The structure of the potential near the horizon can be justified as follows. Since the redshift factor $f(r)$ has a simple zero at the horizon $r_S$, we can expand 
\begin{align}
f(r)=\frac{4\pi}{\beta}(r-r_S)+\sum_{n=2}^{\infty}\frac{1}{n!}\frac{d^nf}{dr^n}(r_S)(r-r_S)^{n}.
\end{align}
The tortoise coordinate is then 
\begin{align}
z&=z_0-\int^r dr\, \left(\frac{\beta}{4\pi (r-r_s)}-\frac{\beta^2 f''(r_S)}{32\pi^2}+\ldots\right)\\
&=z_0-\frac{\beta}{4\pi}\log(r-r_S)+\frac{\beta^2}{32\pi^2}f''(r_S)(r-r_S)+\ldots
\end{align}
Solving for $r(z)$, we find 
\begin{align}
r=r_S+\exp\left(-\frac{4\pi}{\beta}(z-z_0)\right)+\frac{\beta}{8\pi}f''(r_S)\exp\left(-\frac{8\pi}{\beta}(z-z_0)\right)+\ldots,
\end{align}
where the higher corrections are of the form $\exp\left(-4\pi n (z-z_0)/\beta\right)$. Since the potential is an analytic function of $r$, we therefore find that the potential takes the form shown in (\ref{eq:VAsymp}). \\
\indent Because perturbations can fall into the horizon, we cannot define normal modes of the scalar field. Instead, we consider quasi-normal modes, which are ingoing at the horizon and normalizable at the boundary, 
\begin{align}
\psi(z)&\sim e^{i\omega z},\hspace{10 mm}z\to \infty\\
\psi(z)&\sim z^{\frac{1}{2}+\nu},\hspace{10 mm}z\to 0.
\end{align}
Such resonances can only exist at a discrete set of complex energies, which satisfy $\text{Im }\omega<0$ for real $J$ \cite{Horowitz:1999jd}. Note that $\text{Im }\omega<0$ implies that perturbations decay in time, whereas poles in the upper half plane would lead to a growing mode, signifying an instability of the black hole. There are also constraints on QNMs from causality \cite{Brigante:2007nu,Brigante:2008gz,Heller:2022ejw}, but we do not explore them in this paper.\\
\indent In order to obtain a qualitative understanding of the structure of QNMs, it is useful to consider the WKB limit $\Delta \to \infty$ with $\omega/\Delta$ and $J/\Delta$ fixed. The potential (\ref{genpotential}) then takes the form
\begin{align}
V(z)= f(r)\left(\frac{J^2}{r^2}+\Delta^2\right).
\end{align}
For $z_0$ extremizing the potential, there exists a line of QNM poles emanating from $\sqrt{V(z_0)}$. However, not all extrema contribute. To find the relevant values of $z_0$, one must first solve the equation $V(r)=u^2$ for the turning point $r(u)$. The physical turning point is defined to be the solution $r(u)$ which behaves as $r(u)\sim u/\Delta $ for large real $u$. Then the extrema of the potential corresponding to QNMs are those at which the physical turning point merges with another turning point at some value of $u$ in the lower half plane. We refer the reader to \cite{Festuccia:2008zx} for further details. 

The first few modes of this line of poles are given by the expression 
\begin{align}\label{BSapprox}
\omega_n=\sqrt{V(z_0)}+\left(n+\frac{1}{2}\right)\sqrt{\frac{V''(z_0)}{2V(z_0)}},\hspace{10 mm}n=0,1,\ldots
\end{align}
The branches of the square roots should be chosen so that $\omega_n$ is in the lower half plane. \\
\indent Let us now discuss several important cases of the Bohr-Sommerfeld approximation to QNMs.

\begin{figure}
\centering
\begin{subfigure}{.45\textwidth}
  \centering
  \includegraphics[width=1\linewidth]{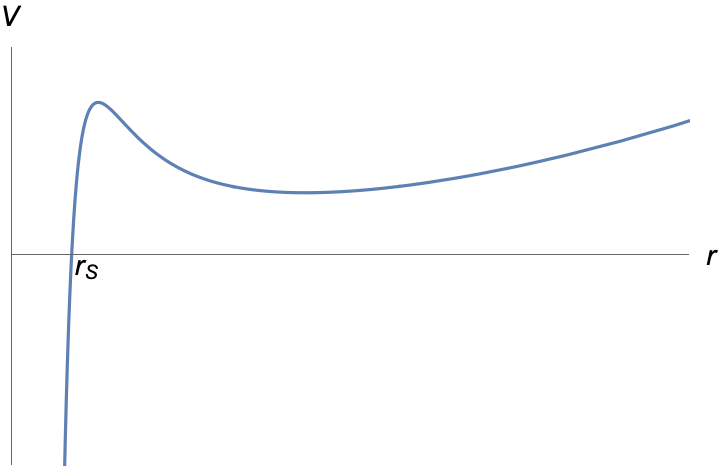}
  \caption{}
  \label{stableorbit}
\end{subfigure}\hfill
\begin{subfigure}{.45\textwidth}
  \centering
  \includegraphics[width=1\linewidth]{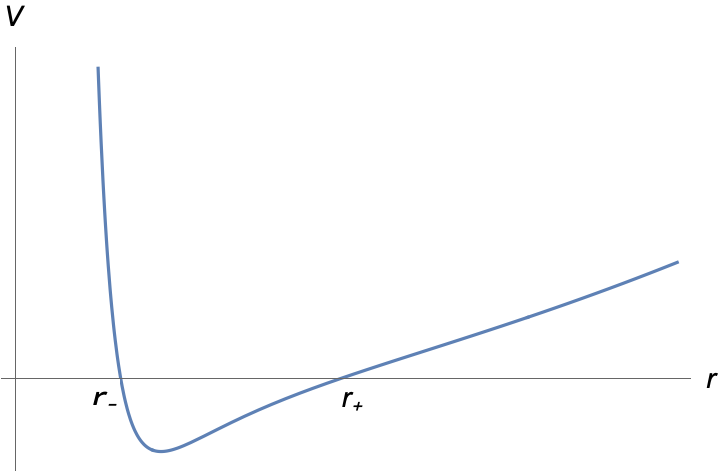}
  \caption{}
  \label{chargedmin}
\end{subfigure}
\caption{(a) A potential with a metastable minimum outside the horizon, leading to a line of weakly damped QNMs. The potential goes to $+\infty$ at the boundary, to $-\infty$ at the singularity, and is zero at the horizon $r=r_S$.  (b) A potential with a global stable minimum, corresponding to a line of virtual bound states. The minimum is in between the inner horizon $r_-$ and the outer horizon $r_+$.}
\end{figure}
\begin{itemize}
\item {\bf Weakly damped modes:} The first example we consider is where $V(z)$ has a metastable minimum outside the horizon, as depicted in Figure \ref{stableorbit}. Then in the approximation (\ref{BSapprox}), the associated QNMs are purely real. The imaginary part is exponentially small since it is related to tunneling over the potential barrier, and it is not captured by the leading WKB approximation. Since these modes decay very slowly, they represent the leading contribution to the two-point function at late times. One case where such a metastable minimum arises is at large $J$. The corresponding modes are related to stable orbits around the black hole \cite{Festuccia:2008zx,festucciathesis,Dodelson:2022eiz,Gannot:2012pb,Berenstein:2020vlp}.
\item {\bf Virtual bound states:} In the previous example, the minimum of the potential was outside the horizon. We can also consider the case where the minimum is behind the horizon in a region where $V(z_0)<0$, see Figure \ref{chargedmin}. We see from (\ref{BSapprox}) that the modes are then purely imaginary. Charged black holes provide an example of this phenomenon, as we will see later. In this case, there is a stable minimum of the potential between the inner and outer horizon. 
\item {\bf Complex extrema:} Generically, extrema of the potential are at some arbitrary point in the complex plane off the real $r$ axis. In this case there is no clean physical interpretation of the QNMs in terms of the real section of the geometry. Note that the extrema come in complex conjugate pairs, since $V^*(r)=V(r^*)$. Therefore a complex extremum leads to two lines of poles related by $\omega_n\to -\omega_n^*$, at an angle determined by (\ref{BSapprox}). For example, consider the QNMs of the black brane metric $f(r)=r^2-1/r^{d-2}$ at zero spatial momentum. There are $d$ complex extrema at $r^d=1-d/2$, but the only two that contribute according to the prescription described above have phase $e^{\pm i \pi/d}$. This leads to two lines of poles in the lower half plane at angles $e^{-\pi i/d}$ and $e^{-(d-1)\pi i/d}$ \cite{Festuccia:2008zx}.
\end{itemize}

\indent Let us now comment on the QNMs beyond the lowest lying modes (\ref{BSapprox}). In the case of a metastable minimum, the low-lying QNMs have exponentially small imaginary part, but at some $n_{\text{max}}$ the modes go off into the complex plane at a finite angle. Therefore there are only finitely many weakly damped modes. In contrast, the virtual bound states and modes associated to complex extrema go on forever. As we will see later, there must be infinitely many QNMs in order to reproduce the operator product expansion.\\
\indent In this section we have discussed the QNMs in the large $\Delta$ limit. The highly damped QNMs can also be computed for order one $\Delta$ by solving the wave equation near the singularity and near the boundary and then matching the two solutions. See \cite{Natario:2004jd,Cardoso:2004up} for details.

\section{Asymptotic Minkowski OPE}
\label{sec:asymminkope}

Let us consider a CFT two-point function of scalar primary operators on $S^1_\beta \times \mathbb{R}^{d-1}$. We can write the following OPE expansion,
\be
\label{eq:OPEfiniteT}
\langle {\cal O}(\tau, \vec{x}) {\cal O}(0,0) \rangle_{\beta} = \sum_{{\cal O}_{\Delta,J}} {a_{\Delta,J} \over \beta^{\Delta}} C_J^{({d-2 \over 2})}\Big( {\tau \over \sqrt{\tau^2 + \vec x^2}} \Big) (\tau^2 + |\vec x|^2)^{{\Delta - 2 \Delta_{{\cal O}} \over 2}} ,
\ee
where $C_J^{({d-2 \over 2})}(\cos \theta)$ are the usual Gegenbauer polynomials and $\tau$ is the Euclidean time $\beta > \tau \geq 0$. The sum goes over the primary operators that appear in the OPE of ${\cal O} \times {\cal O}$, and the expansion coefficients $a_{\Delta,J}$ are given by the product of the corresponding three-point function and thermal expectation value $\langle {\cal O}_{\Delta,J} \rangle_\beta$, see \cite{Iliesiu:2018fao} for details.

In this paper we are interested in the Lorentzian correlator. More precisely, we consider the two-sided correlator, which is related to the Euclidean correlator above as follows,
\be
G_{12}(\omega, k) &= \int_{-\infty}^{\infty} d t\,  e^{i \omega t} \left\langle {\cal O}\left({\beta \over 2} +i t, \vec{k}\right) {\cal O}(0,0) \right\rangle_{\beta}  \ . 
\ee
We would like to understand how the OPE expansion \eqref{eq:OPEfiniteT} constrains the form of $G_{12}(\omega, k)$. We will be interested in the limit $\omega \beta \gg 1$ with 
\be
\zeta \equiv {k \over \omega } \geq 0 
\ee
kept fixed. 

It has been suggested in \cite{Caron-Huot:2009ypo} that in this limit the two-sided correlator admits the following \emph{asymptotic Minkowski OPE} expansion
\be
\label{eq:OPEmink}
G_{12}(\omega, k) &= e^{- {\beta \omega \over 2}} \omega^{2 \Delta_{{\cal O}}-d} \theta(1-\zeta) \sum_{{\cal O}_{\Delta,J}} {a_{\Delta,J} \over (\beta \omega)^{\Delta} } G_{\Delta,J}(\zeta) + O(e^{- \beta \omega}) ,
\ee
where the conformal blocks $G_{\Delta,J}(\zeta)$ take the following form (see Appendix \ref{app:OPEderiv} for the derivation),
\be
G_{\Delta,J}(\zeta) &= \frac{\pi ^{\frac{d}{2}+1} 2^{d+\Delta -2 \Delta_{{\cal O}} +1} \Gamma (d+J-2)}{\Gamma
   (d-2) \Gamma (J+1) \Gamma \left(\frac{J}{2}-\frac{\Delta }{2}+\Delta_{{\cal O}} \right) \Gamma \left(-\frac{d}{2}-\frac{J}{2}-\frac{\Delta
   }{2}+\Delta_{{\cal O}} +1\right)} \nn \\
  &\times  \left(1-\zeta ^2\right)^{\frac{1}{2} (2\Delta_{{\cal O}} -d-\Delta -J )}  \ _2 F_1\left({1-J \over 2},-{J \over 2}, {d-1 \over 2}, \zeta^2\right) .
\ee
 The support of $G_{12}(\omega, k) $ in \eqref{eq:OPEmink}, namely the presence of $\theta(1-\zeta)$, was discussed in \cite{Manenti:2019wxs}. This expansion is supported by the known holographic and perturbative examples and we will assume it to be true in this paper. Note that $G_{12}(\omega, k) $ is nonzero for $k> \omega$, but it is expected to be exponentially suppressed in this region \cite{Son:2002sd,Papadodimas:2012aq,Banerjee:2019kjh}.

The basic idea behind \eqref{eq:OPEmink} is the following. The coordinate space OPE leads to the large frequency expansion of the Euclidean correlator. The behavior of the Euclidean correlator at large Matsubara frequencies controls the behavior of the retarded correlator via the relation $G_E(\omega_n) = G_R(i \omega_n)$. Consider next the subtracted dispersion relation for the retarded correlator. The large frequency behavior at imaginary frequencies controls the large behavior of the spectral density or, equivalently, the two-sided correlator via complex Tauberian theorems for the Stieltjes transform \cite{Mukhametzhanov:2018zja}. \emph{Assuming} that the $G_{12}(\omega, k) $ itself admits a power-like expansion leads to the formulas above. More generally, rigorous formulas can be derived as in \cite{Mukhametzhanov:2018zja}. This is particularly relevant for finite $c_T$ CFTs on a sphere, where $G_{12}(\omega)$ is given by a sum of $\delta$-functions.

We next discuss the universal contribution to the OPE as well as its structure in holographic theories.

\subsection{The unit operator and the stress tensor}

The leading contribution at large frequencies comes from the unit operator $\Delta = J = 0$. The first subleading correction comes from the stress tensor. The leading terms in the expansion of the 
two-point function thus take the following form
\be
\label{eq:OPEunitandT}
{G_{12}(\omega, k) \over e^{- {\beta \omega \over 2}}\omega^{2 \Delta_{{\cal O}}-d}} &\stackrel{1+T}{=} {4^{{d+1 \over 2}-\Delta_{{\cal O}}} \pi^{{d \over 2}+1} (1 - \zeta^2)^{\Delta_{{\cal O}}-d/2} \over \Gamma(\Delta_{\cal O}) \Gamma(\Delta_{\cal O}+1-{d \over 2})} \nn \\
&+ {a_{T} \over (\beta \omega)^{d}} {4^{d-\Delta_{{\cal O}}}(d-1)(d-2) \pi^{{d \over 2}+1} \over \Gamma(\Delta_{\cal O}-d) \Gamma(\Delta_{\cal O}+1-{d \over 2})} (1 - \zeta^2)^{\Delta_{{\cal O}}-d} \Big( 1 + {d \over d-1} {\zeta^2 \over 1-\zeta^2} \Big),
\ee
\noindent where the coefficient of the stress tensor is fixed by the Ward identities as follows,
\be
a_T = - f S_d {2 \Delta_{\cal O} \over d-2} {c_{\text{free}} \over c_T}.\label{atholo}
\ee
Here $S_d = \text{vol}(S^{d-1}) =  {2 \pi^{d/2} \over \Gamma(d/2)}$ and $c_{\text{free}} = {d \over d-1} {1 \over S_d^2}$ is the free boson two-point function of the stress tensor. Finally, $f$ is related to the free energy density as follows
\be
f \equiv {F \over T^{d}}.
\ee
\indent For holographic theories a convenient formula was found in \cite{Kovtun:2008kw}
\be
f_{\text{holo}} = - \frac{2^{2 d-3} (d-1) \left(\frac{1}{d}\right)^{d+1} \pi ^{d/2} \Gamma \left(d/2\right)^3}{(d+1) \Gamma (d)} S_d^2 c_T ,
\ee
where we used that $f = - {s\over d} $, and $s$ is the entropy density. Plugging this into (\ref{atholo}) we get
\be
a_T^{\text{holo}}= \frac{2^{d-2} \left(\frac{1}{d}\right)^d \pi ^{d+\frac{1}{2}} \Gamma \left(\frac{d}{2}-1\right)}{\Gamma \left(\frac{d+3}{2}\right)}  \Delta_{\cal O} .
\ee
We have tested this formula in the case of an AdS black brane by numerically solving the wave equation at large $\omega$. Let us also notice that for $\Delta_{\cal O} = d, d-1, ...$, the contribution of the stress tensor drops out.

\subsection{Multi-stress tensor operators}

Next we discuss the contribution of multi-stress tensor operators $T^n$ to the OPE. These have scaling dimension $\Delta_{T^n} = n d$ and spin $0 \leq J \leq 2 n$. In a generic CFT it is not known how to compute the contribution of these operators to the OPE.

For theories with gravity duals it can be done, see \cite{Fitzpatrick:2019zqz}. Indeed, their contribution is controlled by the physics close to the AdS boundary. To keep the discussion simple, let us present the result for the contribution of $T^2$ operators in $d=4$. Combining the results above with the ones in \cite{Fitzpatrick:2019zqz} we get\footnote{We set in \cite{Fitzpatrick:2019zqz} $f_0 = (\pi/\beta)^4$, which is valid for the black brane.}
\be
{G_{12}^{\text{holo}}(\omega, k) \over e^{- {\beta \omega \over 2}}\omega^{2 \Delta_{{\cal O}}-4}} &\stackrel{T^2}{=} {1 \over (\omega \beta)^8} \frac{\pi ^{11} 4^{3-\Delta_{{\cal O}} } \Delta_{{\cal O}} 
   \left(1-\zeta ^2\right)^{\Delta_{{\cal O}} -8}}{1575
   \Gamma (\Delta_{{\cal O}} -5) \Gamma (\Delta_{{\cal O}} -1)} (c_0 + c_2 \zeta^2 + c_4 \zeta^4)  \\
   c_0 &=7 \Big(9 \Delta_{{\cal O}} ^4-95 \Delta_{{\cal O}} ^3+240 \Delta_{{\cal O}} ^2+80
   \Delta_{{\cal O}} +96 \Big) \ , \nn \\
   c_2 &= 6 \Big( 7 \Delta_{{\cal O}} ^4-105 \Delta_{{\cal O}} ^3+280 \Delta_{{\cal O}} ^2+400 \Delta_{{\cal O}} +288 \Big) \ , \nn \\
   c_4 &= 7 \Delta_{{\cal O}} ^4 -65 \Delta_{{\cal O}} ^3+160 \Delta_{{\cal O}} ^2+240 \Delta_{{\cal O}} +288 \ . \nn
\ee
Notice that the effect that we observed at the level of a single stress tensor, namely vanishing of its contribution for $\Delta_{{\cal O}}=3$ and $\Delta_{{\cal O}}=4$, continues here as well. We see that for $\Delta_{{\cal O}} = 3,4,5$ the expression above vanishes.

We believe the same effect continues for $T^n$, and therefore we can write simple closed form expressions (up to nonperturbative corrections) for the OPE expansion. It follows that \eqref{eq:OPEunitandT} represents the full answer for $\Delta_{{\cal O}} = 3,4,5$ in $d=4$.

\subsection{Leading nonperturbative correction}

We now consider the leading nonperturbative correction to the large frequency expansion above. In coordinate space this correction has a regular expansion at small separations and therefore we expect that it could be related to the double trace operators. 
These are operators schematically of the form $\mathcal{O} \Box^n \partial^{\mu_1} \partial^{\mu_2}\dots \partial^{\mu_J} \mathcal{O}$. In the limit we are working their scaling dimension is $2 \Delta_\mathcal{O} + J + 2n$. Therefore, we see from \eqref{eq:OPEfiniteT} that they will contribute to the OPE with analytic terms. 

For theories with gravity duals these corrections can be computed by analyzing the wave equation in the large $\omega$ regime, see \cite{festucciathesis} and references therein. The solution to the wave equation close to both the singularity and the boundary can be approximated by Bessel functions in the tortoise coordinate $z$. For large $\omega$ the regions of validity of these approximations overlap and it is possible to obtain a solution valid for all $z$. Finally, one can use such a solution to extract the leading order contribution to \eqref{eq:OPEmink} with its nonperturbative corrections (but no power law corrections). The leading term with the first nonperturbative correction reads
\be\label{eq:nonpertfestuccia}
G_{12}^{\text{holo}}\left(\omega, 0\right) \simeq e^{-\frac{\beta \omega}{2}} \omega^{2\Delta_{\mathcal{O}}-d} {4^{{d+1 \over 2}-\Delta_{\mathcal{O}}} \pi^{{d \over 2}+1} \over \Gamma\left(\Delta_{\mathcal{O}}\right) \Gamma\left(\Delta_{\mathcal{O}}+1-{d \over 2}\right)} \left(1 -4 e^{- {\beta \omega \over 2}} \cos \left( \pi \left( \Delta_{\mathcal{O}} - {d \over 2} \right) - {\tilde \beta \omega \over 2} \right) \right) \,.
\ee
In Appendix \ref{exactexpression} we show that this correction can be reproduced from the exact expression for $G_{12}(\omega,0)$ found in \cite{Dodelson:2022yvn}.

\section{QNMs and the OPE}\label{qnmfromope}
We have argued that the quasi-normal modes determine the holographic Wightman function up to a constant. By providing additional input for $G_{12}$, we can place constraints on the spectrum of QNMs. 

In this section, we will input the constraint from the operator product expansion at large real frequency. As $\omega\to +\infty$, we should recover the zero temperature answer up to an overall factor of $e^{-\beta\omega/2}$, see \cite{Katz:2014rla,Caron-Huot:2009kyg,Manenti:2019wxs}, 
\begin{align}\label{identity}
 G_{12}(\omega)\sim \omega^{2\Delta-d}e^{-\beta \omega/2},\hspace{10 mm}\omega\to +\infty ,
\end{align}
where the proportionality constant and subleading in ${1 \over \omega}$ corrections can be found in the previous section.

\indent The goal is to use this constraint to derive conditions on the asymptotic behavior of QNMs. The first step is to convert the product over QNMs into a convenient sum. From (\ref{eq:ansatzprod}), we have 
\begin{align}\label{logsum}
\partial_\omega \log G_{12}(\omega)=-\sum_n \left(\frac{1}{\omega-\omega_n}+\frac{1}{\omega+\omega_n}+\frac{1}{\omega-\omega_n^*}+\frac{1}{\omega+\omega_n^*}\right).
\end{align}
On the other hand, from (\ref{identity}) we have 
\begin{align}\label{logope}
\partial_\omega \log G_{12}(\omega)\sim -\frac{\beta}{2}+\frac{2\Delta-d}{\omega}+\ldots
\end{align}
Let us now compare these two expressions. It is immediately clear that \eqref{logope} implies that the number of QNMs has to be infinite and that they must extend all the way to infinity. Indeed, imagine that there are a finite number of QNMs. Then $\partial_\omega \log G_{12}(\omega) \sim {1 \over \omega}$ at large $\omega$, which is not consistent with \eqref{logope}.

Next we consider several simple ways in which QNMs can approach infinity and see how this approach is constrained by the OPE.

\subsection{A single line of asymptotic QNMs}
\begin{figure}
\centering
\begin{tikzpicture}
\draw[gray] (-2.5,0) -- (2.5,0);
\draw[gray] (0,-2.5) -- (0,2.5);
\foreach \Point in {(-.462,.191), (-.924,.383), (-1.39,.574), (-1.85,.765),(-2.31,.957), (.462,.191), (.924,.383), (1.39,.574), (1.85,.765),(2.31,.957),(-.462,-.191), (-.924,-.383), (-1.39,-.574), (-1.85,-.765),(-2.31,-.957), (.462,-.191), (.924,-.383), (1.39,-.574), (1.85,-.765),(2.31,-.957)}{
    \node at \Point {\textbullet};
     \draw [red,very thick,domain=0:22.5] plot ({1.5*cos(\x)}, {1.5*sin(\x)});
       \node[red, very thick] at (1.7,.35)  {$\theta$};
       \draw[blue, very thick] (-.462,.191) -- (-.924,.383);
              \node[blue, very thick] at (-.6,.5)  {$r$};
       }
\end{tikzpicture}
 \caption{A single line of evenly spaced QNMs with angle $\theta$ and spacing $r$. \label{anglespacingfig}}
\end{figure}
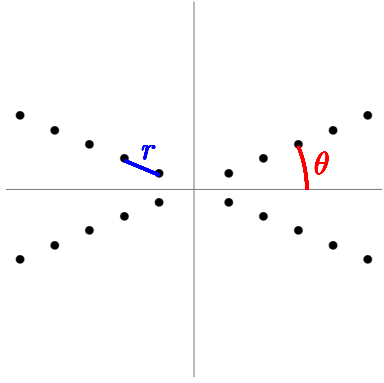
We start with the simplest case, where the QNMs are asymptotically organized into a single line at angle $\theta$, see Figure \ref{anglespacingfig}. We consider the following ansatz\footnote{Strictly speaking these are the images of QNMs under KMS symmetry, not QNMs themselves.}
\begin{align}\label{ansatzsingle}
\omega_n=r e^{i\theta} n^\alpha+\ldots,\hspace{10 mm}n\gg 1,~~~0<\theta<\pi/2.
\end{align}
Computing the large $\omega$ behavior of the sum we get
\begin{align}\label{integraln}
\partial_\omega \log G_{12}(\omega)&= -\sum_{n=1}^\infty \left(\frac{1}{\omega-re^{i\theta}n^\alpha}+\text{images}\right)\notag \\
&\approx-\frac{2\pi}{\alpha \omega}\left(\frac{\omega}{r}\right)^{1/\alpha}\frac{\cos\left(\frac{\pi-2\theta}{2\alpha}\right)}{\sin\left(\frac{\pi}{2\alpha}\right)},
\end{align}
where the leading asymptotic does not depend on the lower limit in the sum.

Matching to the leading term in (\ref{logope}), we find
\be
\alpha=1    
\ee
and
\begin{align}\label{spacingangle}
\beta=\frac{4\pi \sin\theta}{r}.
\end{align}
For a line of simple poles on the imaginary axis, one should divide the right hand side of \eqref{spacingangle} by two, since there is only one image. \\
\indent Let us now consider the first subleading term at large $n$. For this purpose, we take the improved ansatz 
\be
\label{eq:expansionlargenw}
\omega_n=re^{i\theta}n+se^{i\phi}+\ldots  \ . 
\ee
Using the Euler-Maclaurin formula, the sum (\ref{logsum}) to order $1/\omega$ is
\begin{align}
\partial_\omega \log G_{12}(\omega)&\approx -\frac{1}{2}\left(\frac{1}{\omega-\omega_1}+\text{images}\right)-\frac{2\pi\sin\theta}{r}+\frac{4}{\omega} \nn \\ 
&\hspace{ 5mm}-\int_{1}^{\infty}dn\,\left( \frac{se^{i\phi}}{(\omega-re^{i\theta}n)^2}+\text{images}\right)\notag\\
&\approx -\frac{2\pi \sin\theta}{r}+\frac{4s\cos(\theta-\phi)+2r}{r\omega}.
\label{eq:subleadingcomputation}
\end{align}
Comparing with (\ref{logope}), we find 
\begin{align}\label{deltas}
2\Delta-d=\frac{4s\cos(\theta-\phi)+2r}{r}.
\end{align}
To shed more light on this result, let us imagine that we add an extra QNM at $\omega_0$. A single QNM contributes at large $\omega$ as $\partial_\omega \log G_{12}(\omega) \sim -{4 \over \omega}$. For the computation above adding a single QNM corresponds to effectively relabeling $n \to n-1$, which shifts the subleading constant as $s e^{i \phi}\to s e^{i \phi}-r e^{i \theta}$. Plugging this expression into \eqref{eq:subleadingcomputation} indeed leads to the expected shift $-{4 \over \omega}$. In this sense the subleading term in \eqref{eq:expansionlargenw} is sensitive to the global structure of QNMs, but it is not sensitive to the details of their distributions at low energies.

\indent To summarize, we have learned that for a single line of QNMs, the spectrum must be asymptotically linearly spaced. Moreover, the spacing and angle must be related in terms of the temperature, and the first subleading term in the expansion is related to the conformal dimension of the operator under consideration. These conditions apply for any black hole solution in which there is only one line of asymptotic QNMs. For example, the QNMs of massless fields with $\Delta=d$ in large $AdS_{d+1}$/Schwarzschild black holes asymptotically approach the line\footnote{In flat space a similar formula was first derived in \cite{Motl:2003cd}.}  \cite{Cardoso:2004up,Natario:2004jd}
\begin{align}\label{eq:asympAdS}
\omega_n \simeq \frac{4\pi}{\beta}\sin\left(\frac{\pi}{d}\right)e^{i\pi/d}\left(n+\frac{d-2}{4}-i\frac{\log 2}{2\pi}\right) + ... \  ,
\end{align}
which is easily seen to satisfy (\ref{spacingangle}) and (\ref{deltas}). Notice that the OPE expansion effectively maps to the ${1 \over n}$ expansion of the QNMs. Moreover, we see that at each order in the expansion we have two real parameters to fix, and we have only one equation.\\
\indent To proceed to subleading orders we can write 
\be
\omega_n = re^{i\theta} n + se^{i\phi}  + \frac{te^{i\psi}}{n^{\gamma}}+\ldots\hspace{10 mm}\gamma>0.
\ee
If $\gamma$ is an integer, then it is straightforward to check that subleading terms in $\partial_\omega \log G_{12}(\omega)$ receive contributions from all $n$ in the sum, so we cannot fix $t$ purely in terms of the asymptotics of the QNMs. Let us now consider the case where $\gamma$ is fractional. At large $\omega$, the sum (\ref{logsum}) is dominated by $n\sim \omega$, so we can expand the summand in $n^{-\gamma}$.  Expanding $\partial_\omega \log G_{12}$ at large $\omega$ and plugging in (\ref{deltas}) and (\ref{spacingangle}) gives
\begin{align}
\partial_\omega \log G_{12}(\omega)&\approx -\frac{\beta}{2}+\frac{2\Delta-d}{\omega}-\int_{1}^{\infty}dn\,\left(\frac{te^{i\psi}n^{-\gamma}}{(\omega-re^{i\theta}n)^2}+\text{images}\right)\notag\\
&\approx-\frac{\beta}{2}+\frac{2\Delta-d}{\omega}+\frac{2\pi t \gamma r^{\gamma-1}\cos\left(\theta-\psi+\gamma\left(\frac{\pi }{2}-\theta\right)\right)}{\omega^{\gamma+1}\sin\left(\frac{\pi\gamma}{2}\right)}.
\end{align}
Since fractional powers of $\omega$ do not appear in the OPE expansion of $\partial_\omega \log G_{12}(\omega)$ for holographic correlators, it must be the case that 
\begin{align}\label{fractionalangle}
\psi=\pm\frac{\pi}{2}+\theta+\gamma\left(\frac{\pi}{2}-\theta\right).
\end{align}
In Section \ref{opesumrules}, we will check this relation for AdS black branes.\\
\indent We have seen that fractional powers can appear in the expansion of the QNMs at subleading orders. In fact, if there are two asymptotic lines of QNMs, fractional powers can appear in the leading behavior of one of the lines. We turn to this case next.
\subsection{Adding another asymptotic line}
\begin{figure}
\centering
\begin{tikzpicture}
\draw[gray] (-2.5,0) -- (2.5,0);
\draw[gray] (0,-2.5) -- (0,2.5);
\foreach \Point in {(-.462,.191), (-.924,.383), (-1.39,.574), (-1.85,.765),(-2.31,.957), (.462,.191), (.924,.383), (1.39,.574), (1.85,.765),(2.31,.957),(-.462,-.191), (-.924,-.383), (-1.39,-.574), (-1.85,-.765),(-2.31,-.957), (.462,-.191), (.924,-.383), (1.39,-.574), (1.85,-.765),(2.31,-.957)}{
    \node[red] at \Point {\textbullet};
       }
       \foreach \Point in {(0,.125),(0,.5),(0,1.125),(0,2),(0,-.125),(0,-.5),(0,-1.125),(0,-2)}{
    \node[blue] at \Point {\textbullet};
       }
\end{tikzpicture}
 \caption{The ansatz (\ref{omega1}) and (\ref{omega2}) for $\alpha=2$. The red points are evenly spaced, and the blue points scale quadratically with $n$. \label{twolinesfig}}
\end{figure}
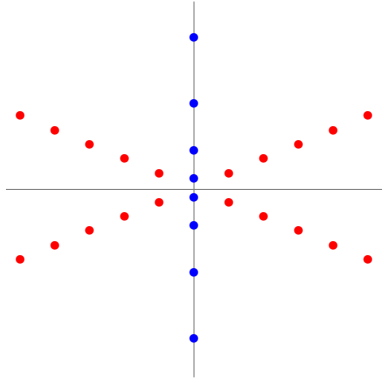
We now consider the case of several asymptotic lines of poles. If all the lines are asymptotically linearly spaced, then the previous calculation generalizes immediately, and we find 
\begin{align}\label{betamultilinlines}
\beta=4\pi\sum_{i}\frac{\sin\theta_i}{r_i}.
\end{align}
As mentioned above, a line of simple imaginary poles needs to be treated separately, and contributes $2\pi/r_i$ to this sum.  Similarly, we have the subleading constraint
\be
2\Delta-d=\sum_i \frac{4s_i\cos(\theta_i -\phi_i)+2r_i}{r_i} \ .
\ee
\indent The more interesting situation is when not all lines are asymptotically linearly spaced. Let us consider the case of two asymptotic lines. We take one of the lines along the imaginary axis for simplicity, and choose the ansatz 
\begin{align}
\omega_{1,n}&=ian^\alpha+\ldots,\hspace{10 mm}n\gg 1, ~a>0,~\alpha>1\label{omega1}\\
\omega_{2,n}&=re^{i\theta}n+se^{i\phi}n^\gamma+\ldots,\hspace{10 mm}n\gg 1,~0<\theta<\pi/2,~0<\gamma<1\label{omega2}.
\end{align}
Proceeding as above, we find
\begin{align}\label{twolinesintegral}
\partial_\omega \log G_{12}(\omega)&\approx -\int_{1}^{\infty}dn\,\left[\left(\frac{1}{\omega-ian^\alpha}+\text{image}\right)+\left(\frac{1}{\omega-re^{i\theta}n}+\text{images}\right)\right.\notag\\
&\hspace{30 mm}\left.-\left(\frac{e^{i\phi}n^\gamma s}{(\omega-e^{i\theta}nr)^2}+\text{images}\right)\right]\\
&\approx -\frac{2\pi \sin\theta}{r}-\frac{\pi}{\alpha \omega}\left(\frac{\omega}{a}\right)^{1/\alpha}\frac{1}{\sin\left(\frac{\pi}{2\alpha}\right)}+\frac{2\pi s\gamma}{\omega r}\left(\frac{\omega}{r}\right)^\gamma\frac{\cos\left(\phi-\theta+\gamma\left(\frac{\pi}{2}-\theta\right)\right)}{\sin\left(\frac{\pi \gamma}{2}\right)}.\notag
\end{align}
\indent We now match powers of $\omega$ to the OPE expansion (\ref{logope}). The first term in (\ref{twolinesintegral}) reproduces the result (\ref{spacingangle}) from before. The other terms must cancel, giving the conditions $\gamma=1/\alpha$ and 
\begin{align}\label{predictiontwolines}
s=r\left(\frac{r}{a}\right)^{1/\alpha}\frac{1}{2\cos\left(\phi-\theta+\frac{1}{\alpha}(\pi/2-\theta)\right)}.
\end{align}
We will confirm this prediction later by analyzing the QNMs in the presence of a higher derivative correction $W^2\phi^2$, where $W$ is the Weyl tensor. This higher derivative term introduces a new line of QNMs on the imaginary axis with asymptotic $n^3$ scaling, and we will see numerically that (\ref{predictiontwolines}) is satisfied to high accuracy in this case.
\subsection{Behavior in the complex $\omega$ plane}\label{complexbehavior}
So far, we have analyzed the behavior of the correlator at large real $\omega$. Let us now discuss what happens when $\omega$ is taken to infinity along some complex ray. In the case of a single line of QNMs, the lines of poles divide the complex plane into four asymptotic regions. For example, as $\omega\to i\infty$ with the ansatz (\ref{ansatzsingle}) with $\alpha=1$, we find
\begin{align}
\partial_\omega \log G_{12}(\omega)\approx \frac{2\pi i \cos\theta}{r},\hspace{10 mm}\omega\to i\infty.
\end{align}
It follows that the correlator exponentially decays at $\omega=i\infty$ with rate $2\pi \cos\theta/r$. As long as $\theta\not=\pi/2$, this decay rate is nonzero. Similar remarks apply for the other three regions of the complex plane, so the singularity sum rules (\ref{sumrules}) are satisfied when $\theta\not=\pi/2$. We conclude that meromorphy and no zeroes imply the singularity sum rules, given a line of asymptotic QNMs that is not parallel to the imaginary axis.\\
\indent In fact, the same conclusion holds if there are several lines of asymptotic QNMs, as long as \emph{at least one line is at angle $\theta\not=\pi/2$}. The lines divide the complex $\omega$ plane into different asymptotic regions, and in each region the correlator exponentially decays. Therefore the singularity sum rules will be satisfied.

It is instructive to compare the large frequency expansions along the real and imaginary axis. To this end we consider the ratio
\be
\log {G_{12}(\omega) \over G_{12}(i \omega)} = {\tilde \beta - \beta \over 2} \omega + C_0 + ...\hspace{10 mm}\beta \omega\gg 1   ,
\ee
where we wrote only the leading contribution to the ratio and we defined $\tilde \beta$ through the leading asymptotic of $\log G_{12}(i \omega) \simeq - {\tilde \beta \omega \over 2}$. Both $\tilde \beta$ and $C_0$ are sensitive only to the large $n$ tails of the large QNM expansion, and therefore can be safely computed using the same methods that we used above.

Using $\omega_n=re^{i\theta}n+se^{i\phi}$ we get 
\be
\log {G_{12}(\omega) \over G_{12}(i \omega)} ={\beta \over 2} (\cot \theta - 1)  \omega + \pi \left(\Delta -{ d+ 2 \over 2}\right) \tan (\theta - \phi)  \ . 
\ee
The holographic computation in the AdS-Schwarzschild background gives \cite{Festuccia:2005pi,festucciathesis}
\be
\tilde \beta &= \beta \cot {\pi \over d}, ~~~ \nn \\ 
C_0 &= \log 2 \ ,
\ee
which leads to the following expression for the asymptotic behavior of the QNMs,
\be
\omega_n \simeq \frac{4\pi}{\beta}\sin\left(\frac{\pi}{d}\right)e^{i\pi/d}\left(n+ {\Delta \over 2} - \frac{d+2}{4}-i\frac{\log 2}{2\pi}\right) .
\ee
For $\Delta =4$ this agrees with  \eqref{eq:asympAdS}.

\section{OPE sum rules}\label{opesumrules}

In the previous section we explored the basic relationship between the leading terms in the OPE expansion of the correlator and the structure of the QNMs. In this section we show how inputting more data about the asymptotic expansion for the quasi-normal modes leads to exact sum rules for QNMs. These sum rules encode the subleading terms in the OPE expansion, going beyond the previous section.

\subsection{OPE sum rules for QNMs }

Let us imagine that we have worked out the asymptotic expansion of the QNMs as above to a certain order in ${1 \over n}$, such that 
\be
\omega_n = \omega_n^{\text{asy}} + \delta \omega_n ,
\ee
such that
\be
\lim_{n \to \infty} n^{k_*} \delta \omega_n = 0.
\ee
We can then use the OPE to write down a set of sum rules for $\delta \omega_n$. More precisely, we write
\begin{align}\label{logsum2}
\partial_\omega \log G_{12}(\omega)-\partial_\omega \log G_{12}^{\text{asy}}(\omega) &=-\sum_n \left(\frac{1}{\omega-\omega_n^{\text{asy}} -  \delta \omega_n} - \frac{1}{\omega-\omega_n^{\text{asy}}} + \text{images}\right),
\end{align}
where $G_{12}^{\text{asy}}(\omega)$ is obtained by replacing $\omega_n$ with $\omega_n^{\text{asy}}$ in the product formula (\ref{eq:ansatzprod}) without the prefactor $G_{12}(0)$. Given the OPE and $\omega_n^{\text{asy}}$ we can compute the large $\omega$ expansion of the left hand side. 
Expanding the right hand side at large $\omega$ under the sum
we get
\be
\label{eq:sumrulesOPEfinite}
\partial_\omega \log G_{12}(\omega)-\partial_\omega \log G_{12}^{\text{asy}}(\omega) &= - \sum_{k=1}^\infty {4 \over \omega^{2k+1}} \sum_{n=1}^\infty {\rm Re} \Big(  (\delta \omega_n+\omega_n^{\text{asy}})^{2k} - (\omega_n^{\text{asy}})^{2k} \Big) \ . 
\ee
Such an expansion only makes sense if the sums over $n$ converge. The most dangerous contribution comes from $\delta \omega_n (\omega_n^{\text{asy}})^{2k-1} \sim \delta \omega_n n^{2k -1}$. For this to converge we require
\be
k_* \geq 2k . 
\ee
For example, to write down the sum rule corresponding to the contribution of the stress tensor in $d=4$ we need $k=2$, which requires $k_* = 4$.

Let us write more explicitly the first sum rule. We set
\be
\label{eq:asymqnms}
\omega_n^{\text{asy}} \simeq \frac{4\pi}{\beta}\sin\left(\frac{\pi}{d}\right)e^{i\pi/d}\left(n+ {\Delta \over 2} - \frac{d+2}{4}-i\frac{\log 2}{2\pi}\right) - {c \over n^{{d \over d-1}}} e^{i {(d-2) \pi \over 2 d (d-1)} },
\ee
where $c$ is computed in Appendix \ref{subleadingcorr}. Note that the phase of the last term in this expression agrees with the prediction (\ref{fractionalangle}). 

Using \eqref{eq:asymqnms}, we can write down the first nontrivial sum rule for $k=1$. The left hand side of \eqref{eq:sumrulesOPEfinite} takes the form
\be
\label{eq:firstsumruleOPE}
\partial_\omega \log G_{12}(\omega)-\partial_\omega \log G_{12}^{\text{asy}}(\omega) =- {\delta c_3 \over \omega^3}, 
\ee
where $\delta c_3$ is given explicitly in \eqref{eq:deltac3explicit}. Therefore we see that by combining the asymptotic large $n$ expansion of the QNMs with the asymptotic OPE expansion we get exact nonperturbative QNM sum rules, which are sensitive to all QNMs.\\

\noindent {\bf Example: $\Delta=4$, $d=4$} \\

\begin{figure}[t!]
\centering
  \includegraphics[]{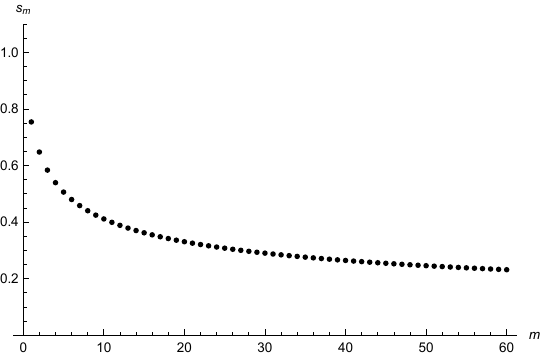}
\caption{We plot the $k=1$ OPE sum rule partial sums $s_m$ for $\Delta=d=4$ as a function of $m$.}
 \label{fig:opesumrule}
\end{figure}

Let us consider the $k=1$ sum rule explicitly for $\Delta=4$ and $d=4$. Setting $\beta=1$, we get the following equation
\be
\sum_{n=1}^\infty {\rm Re} \Big( \omega_n^{2} - (\omega_n^{\text{asy}})^{2} \Big) = {\delta c_3 \over 4} \simeq 15.281 .
\ee
To check this sum rule let us introduce partial sums $s_m \equiv 1-{\sum_{n=1}^m {\rm Re} \Big(  \omega_n^{2} - (\omega_n^{\text{asy}})^{2} \Big) \over 15.281}$. For $m \to \infty$, the partial sums $s_m$ approach 0. We can compute the first few QNMs numerically using $\mathtt{QNMSpectral}$ and check how well the sum rules are satisfied. The reader can find these modes in the accompanying file modes.txt. We plot $s_m$ in Figure \ref{fig:opesumrule}.

We also checked numerically that the leading large $n$ asymptotic of $\delta \omega_n$ in this case takes the form $\simeq 0.340 {e^{- i {\pi \over 48}} \over n^{7/3}}$ and this tail correctly accounts for the $\sim 23 \%$ of the sum rule left for the $k>60$ modes in Figure \ref{fig:opesumrule}.

\subsection{Overall constant}

There is one last piece of data that comes from the OPE that we have not used. It is the overall normalization of the correlator at large $\omega$,
as given by \eqref{eq:OPEunitandT}. Similarly, on the product formula side \eqref{eq:ansatzprod}, $G_{12}(0)$ has not entered in any of the equations above because it does not contribute to $\partial_\omega \log G_{12}(\omega)$. Likewise, it canceled in the ratio $G_{12}(\omega)/G_{12}(i \omega)$. 

As above, we can imagine that the asymptotic large $n$ expansion of QNMs is known. We then can write
\be
\log G_{12}(\omega)- \log G_{12}^{\text{asy}}(\omega) &=\log G_{12}(0) + \sum_{n=1}^{\infty} \log {(1 - {\omega^2 \over (\omega_n^{\text{asy}})^2})(1 - {\omega^2 \over (\omega_n^{\text{asy}*})^2}) \over (1 - {\omega^2 \over \omega_n^2})(1 - {\omega^2 \over (\omega_n^*)^2})} .
\ee
We can now take the large $\omega$ limit to get the following equation 
\be
\label{eq:sumruleconstant}
\log G_{12}(0) =\log G_{12}^{\text{OPE}} + 2 \sum_{n=1}^{\infty} \log { |\omega_n^{\text{asy}} |^2 \over  |\omega_n|^2 } ,
\ee
where $\log G_{12}^{\text{OPE}}$ can be computed given the normalization of the vacuum two-point function of the operator of interest and the explicit form of $\omega_n^{\text{asy}}$ via the following formula
\be
\log G_{12}^{\text{OPE}} =\lim_{\omega \to \infty} \Big( \log G_{12}(\omega)- \log G_{12}^{\text{asy}}(\omega)  \Big) .
\ee

\noindent {\bf Example: shear viscosity in ${\cal N}=4$ SYM at strong coupling}  \\

\begin{figure}[t!]
\centering
  \includegraphics[]{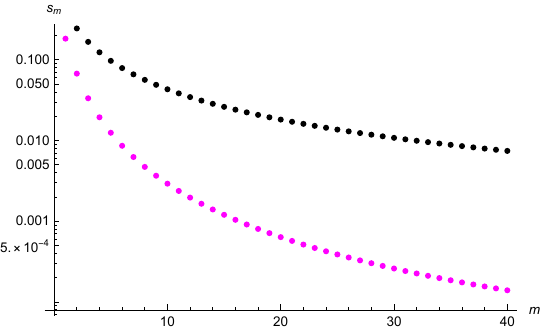}
\caption{We plot the partial sums $s_m$ for shear viscosity as a function of $m$. The black dots correspond to the choice \eqref{eq:shearstraight} for the asymptotic modes, whereas the magenta dots include one extra correction as in \eqref{eq:correctionshearqnm}. We see that for the corrected model (in \textcolor{magenta}{magenta}), knowing the first 10 QNMs allows us to satisfy the sum rule with $0.3 \%$ error.}
 \label{fig:shearsumrule}
\end{figure}

Let us consider the scalar two-point function for $\Delta=4$ in $d=4$. As reviewed in Appendix \ref{app:shearvis}, this is related to the shear viscosity in ${\cal N}=4$ SYM at strong coupling. Setting $d=4$ in (\ref{g120deltad}), the prediction from gravity is 
\be
\log G_{12}(0) = \log {\pi^5 \over 12} \simeq 3.23874  ,
\ee
where we set $\beta =1$. We would now like to understand how this result is reproduced from the sum rule \eqref{eq:sumruleconstant}. The asymptotic form of the QNMs takes the form
\be
\label{eq:shearstraight}
\omega_n^{\text{asy}} = 2 \pi (1+i)\left(n+{1 \over 2} - i {\log 2 \over 2 \pi} \right)  ,
\ee
see \eqref{eq:asymqnms} for $d=4$. Computing the large $\omega$ limit and using (\ref{eq:OPEunitandT}) we get
\be
\log G_{12}^{\text{OPE}} = \log {\pi^3 \over 96} - \log \frac{9}{4 \sqrt{2} \left(\pi ^2+\log ^2(2)\right)^2} \simeq 3.07947 .
\ee
Therefore we get the following sum rule
\be
\sum_{n=1}^{\infty} \log { |\omega_n^{\text{asy}} |^2 \over  |\omega_n|^2 } = \frac{1}{2} \log \left(\frac{9 \sqrt{2} \pi ^2}{\left(\pi ^2+\log ^2(2)\right)^2}\right) \simeq 0.079637.
\ee
To see how the sum rule is satisfied let us introduce partial sums $s_m \equiv 1 - {\sum_{n=1}^{m} \log { |\omega_n^{\text{asy}} |^2 \over  |\omega_n|^2 } \over 0.079637 }$, so that $\lim_{m \to \infty} s_m = 0$. We compute $\omega_n$ numerically using $\mathtt{QNMSpectral}$, and we plot $s_m$ in Figure \ref{fig:shearsumrule}.

In principle we can repeat the same computation as above by including the next term in the large $n$ expansion,
\be
\label{eq:correctionshearqnm}
\delta \omega_n^{\text{asy}}  =  6 \pi \frac{ 2^{1/6} 3^{5/6} e^{-\frac{11 i \pi }{12}} \Gamma \left(\frac{5}{6}\right)}{7 n^{4/3} \Gamma \left(\frac{1}{6}\right) \Gamma
   \left(\frac{1}{3}\right)} . 
\ee
With this correction the sum rule becomes
\be
\sum_{n=1}^{\infty} \log { |\omega_n^{\text{asy}} + \delta \omega_n^{\text{asy}} |^2 \over  |\omega_n|^2 } \simeq -0.03873.
\ee
We can again construct partial sums $s_m \equiv 1 - {\sum_{n=1}^{m} \log { |\omega_n^{\text{asy}} + \delta \omega_n^{\text{asy}} |^2 \over  |\omega_n|^2 } \over -0.03873 }$, and we plot $s_m$ in Figure \ref{fig:shearsumrule}.

\section{Positive moments for hydrodynamics}
\label{sec:qnmhydro}
 
It is interesting to see how the location of QNMs translates into the low-energy (hydrodynamic) expansion of the correlator.  In this section, we find that by restricting the location of QNMs to a subregion of the complex plane, we get nontrivial bounds on the hydrodynamic expansion coefficients.

The basic observation is that the \emph{no-zero} property of the thermal correlator can be thought as \emph{positivity} of the discontinuity of $\log G_{12}(\omega)$. The product formula \eqref{eq:ansatzprod} corresponds to the dispersive representation of $\log G_{12}(\omega)$, with $\log G_{12}(0)$ being the subtraction constant.
We then find that the coefficients of the low-energy expansion of the two-sided correlator are related to certain moments of the non-negative density of QNMs. 

The mathematical structure that emerges is similar to bounds that appear in the context of scattering amplitudes \cite{Adams:2006sv}. The role of Wilson coefficients is played by the hydrodynamic expansion, whereas the role of dispersion relations is played by the product formula \eqref{eq:ansatzprod}.

To follow this idea, it is convenient to introduce a positive-definite measure
\be
\rho(\omega) &= \sum_{n=1}^{\infty} \delta^{(2)}(\omega-\omega_n) \ , \nn \\
\tilde \rho(\omega) &=\sum_{n=1}^\infty \delta(\omega- \tilde \omega_n) \ , 
\ee
where $\omega_n$ are QNMs which satisfy $\text{Re }\omega_n > 0$, $\text{Im }\omega_n > 0$. As before it is also convenient to separately consider purely imaginary QNMs, for which $\omega=i\tilde{\omega}_n$ for real $\tilde{\omega}_n$. In terms of these densities we can write
\be
\log {G_{12}(\omega) \over G_{12}(0)} &= - \int d^2 \hat \omega\, \rho( \hat \omega) \log \left[\left(1 - {\omega^2 \over \hat \omega^2}\right)\left(1 - {\omega^2 \over (\hat \omega^*)^2}\right)\right]\notag \\
&\hspace{5 mm}- \int d \tilde \omega\,\tilde \rho(\tilde \omega) \log \left(1 + {\omega^2 \over \tilde \omega^2}\right) .
\ee

Let us consider next the low energy expansion of the two-sided correlator. It takes the following form,
\be
\log {G_{12}(\omega) \over G_{12}(0)} = \sum_{k=1}^\infty (-1)^k {2 \mu_k + \tilde \mu_k \over k} \omega^{2 k},
\ee
where 
\be
\mu_k &= \int d^2 \hat \omega\, \rho(\hat \omega) {\rm Re} {1 \over (-i \hat \omega)^{2k}} \ , \nn \\
\tilde \mu_k &= \int d \tilde \omega\, \tilde \rho(\tilde \omega)  {1 \over \tilde \omega^{2 k}} \ . 
\ee

We immediately see that the hydrodynamic expansion is naturally organized in powers of the smallest QNM. To this end, let us introduce $\omega_{\min}$ and $\tilde \omega_{\min}$ for the QNMs closest to the origin. It is convenient to define the moments in units of $|\omega_{\text{min}}|$. For this purpose we can write
\be
\log {G_{12}(\omega) \over G_{12}(0)} = \sum_{k=1}^\infty {(-1)^k \over k} \left(2 \mu_k {\omega^{2 k} \over | \omega_{\min}|^{2k}} + \tilde \mu_k {\omega^{2 k} \over \tilde \omega_{\min}^{2k}}\right),
\ee
where the moments have been defined as follows,
\be
\mu_k &= \int_{|\hat \omega|\geq 1} d^2 \hat \omega\, \rho(\hat \omega) {\rm Re} {1 \over (-i \hat \omega)^{2k}} \ , \nn \\
\tilde \mu_k &= \int_1^{\infty} d \tilde \omega\, \tilde \rho(\tilde \omega)  {1 \over \tilde \omega^{2 k}} \ . 
\ee

We therefore see that the low-energy expansion is related to moments of the QNM density. In fact, the moment problem for $\tilde \mu_k$ is nothing but the simplest Hausdorff moment problem, which can be seen by switching to $\tilde y = {1 \over \tilde \omega^2} \in [0,1]$. 

The moment problem for $\mu_k$ is more nontrivial since it involves the geometry of the 2d plane. Below 
we restrict our considerations to the cases where the QNMs are supported in the region $\text{Re }\omega_n \leq \text{Im } \omega_n$. For this configuration $\mu_1 \geq 0$ and it is natural to consider bounds on the ratios $({\mu_2 \over \mu_1}, {\mu_3 \over \mu_1})$. We have derived the bounds in two steps. First, we derived the bounds numerically by discretizing the region of support of QNMs and then solving the coupling maximization problem using $\mathtt{LinearProgramming}$ in Mathematica. We then have identified the boundary of the allowed regions analytically, finding that for the cases considered here they are very simple.

\begin{figure}[t!]
\centering
\begin{subfigure}{.5\textwidth}
  \centering
  \includegraphics[width=1\linewidth]{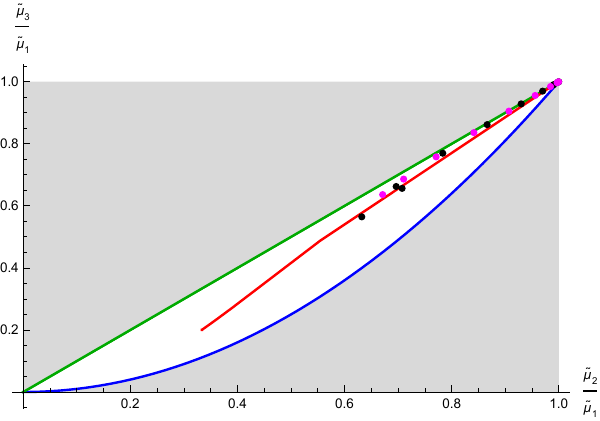}
  \caption{}
\end{subfigure}%
\begin{subfigure}{.5\textwidth}
  \centering
  \includegraphics[width=1\linewidth]{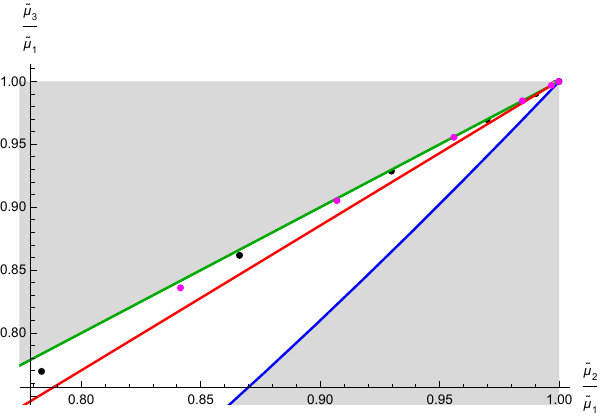}
  \caption{}
 \end{subfigure}
\caption{(a) We plot the allowed region for the ratios $({\tilde{\mu}_2 \over \tilde{\mu}_1}, {\tilde{\mu}_3 \over \tilde{\mu}_1})$ of the first three low-frequency expansion coefficients for the case when all QNMs are purely imaginary, see \eqref{eq:simplemomentsim}. The boundary of the allowed region can be realized with \textcolor{green!70!black}{two QNMs} and \textcolor{blue}{one QNM}. The \textcolor{red}{red curve} corresponds to the zero momentum BTZ correlator \eqref{eq:BTZreszeromom} as $\Delta$ is varied. For small $\Delta \to 0$ the moments reach the upper-right cusp. The black and \textcolor{magenta}{magenta} dots are the results for the SYK chain at different momenta for the couplings $v=0.8$ and $v=0.2$ correspondingly, with the momentum chosen in a range where all the poles are on the imaginary axis. As the momentum becomes small the points approach the the tip of the allowed region.  (b) The same plot but zoomed closer to the tip of the allowed region. We see that for small momenta the SYK chain results are close to saturating the upper bound.  
\label{fig:imagmoments}}
\end{figure}

\subsection{Purely imaginary QNMs}

We first consider the simplest case when QNMs are purely imaginary so that $\mu_i=0$. Restricting our attention to $(\tilde \mu_1,\tilde \mu_2,\tilde \mu_3)$, we can write down a set of optimal constraints \cite{Bellazzini:2020cot}:
\be
\label{eq:simplemomentsim}
\tilde \mu_1 \geq \tilde \mu_2 \geq \tilde \mu_3 \geq 0 , \nn \\
\tilde \mu_1 \tilde \mu_3 - (\tilde \mu_2)^2 \geq 0 . 
\ee
The allowed region is depicted in Figure \ref{fig:imagmoments}.

An example of this type is provided by the BTZ correlator at zero momentum, for which we have 
\be
\label{eq:BTZreszeromom}
\tilde \mu_1 &= {\Delta^2 \over 2} \psi^{(1)}\left({\Delta \over 2}\right) ,~~~\tilde \mu_2 = {\Delta^4 \over 48} \psi^{(3)}\left({\Delta \over 2}\right) , ~~~\tilde \mu_3 = {\Delta^6 \over 3840} \psi^{(5)}\left({\Delta \over 2}\right) \ . 
\ee
Doing a shift $\Delta \to \Delta - {d-2 \over 2}$ one recovers the Rindler space correlator, and by replacing $\Delta \to {2 \over q}$ we get the result in the SYK model. 

Let us quickly comment on the structure of the boundary of the allowed region in Figure \ref{fig:imagmoments}. Let us introduce $0 \leq x={1 \over \tilde \omega} \leq 1$. Then the green upper line corresponds to $\tilde \rho(x) = \alpha \delta(x) + (1-\alpha) \delta(1-x)$ with $0 \leq \alpha \leq 1$, with the upper cusp corresponding to $\alpha=0$ and the lower cusp corresponding to $\alpha=1$. The blue boundary corresponds to $\rho(x) = \delta(x-x_0)$.

\begin{figure}[t!]
\centering
\begin{subfigure}{.5\textwidth}
  \centering
  \includegraphics[width=1\linewidth]{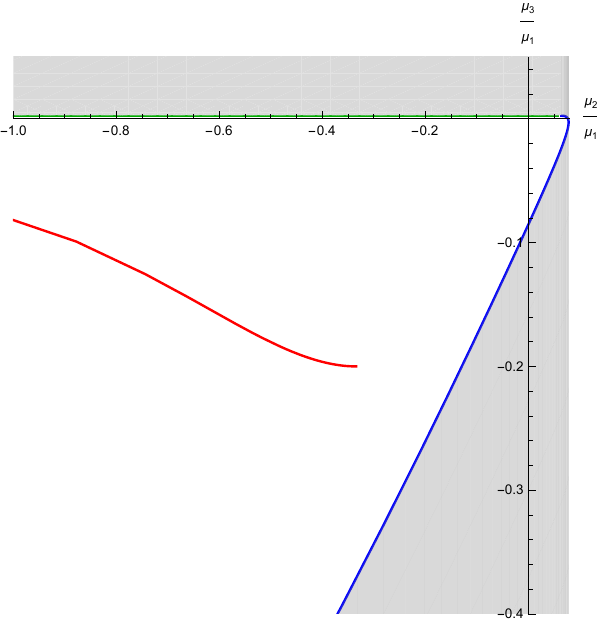}
  \caption{}
  \label{fig:linemomentsa}
\end{subfigure}%
\begin{subfigure}{.5\textwidth}
  \centering
  \includegraphics[width=1\linewidth]{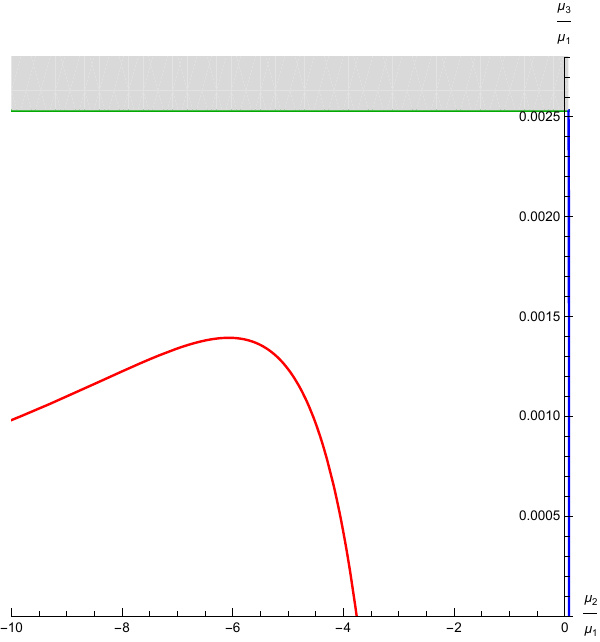}
  \caption{}
  \label{fig:linemomentsb}
\end{subfigure}
\caption{(a) We plot the allowed region for the ratios $({\mu_2 \over \mu_1}, {\mu_3 \over \mu_1})$ of the first three low-frequency expansion coefficients for the case when all QNMs lie on the line \eqref{eq:straightlinelocus} with $\phi = {\pi \over 4}$ and $\theta = {\pi \over 2}$. The boundary of the allowed region can be realized with \textcolor{green!70!black}{two QNMs} and \textcolor{blue}{one QNM}. The \textcolor{red}{red curve} corresponds to the $k=\Delta$ momentum BTZ correlator \eqref{eq:BTZreszeromom} as $\Delta$ is varied. For small $\Delta \to 0$ the moments approach ${\mu_2 \over \mu_1} \to - \infty$ and ${\mu_3 \over \mu_1} \to 0$. (b) The same plot but zoomed closer to the boundary of the allowed region. Notice that slightly positive values of ${\mu_2 \over \mu_1}$ are realized for certain $\Delta$s. }
\label{fig:linemoments}
\end{figure}

\subsection{QNMs on a line}

For the next model we consider the case where all QNMs are located on a single line in the complex plane. In other words, we consider
\be
\label{eq:straightlinelocus}
\omega_n = e^{i \phi}  + a_n e^{i \theta},
\ee
where $a_n \geq 0$ and real. 
This case again effectively reduces to a one-dimensional moment problem. This time, however, it is not easily treated analytically. 
It is nevertheless straightforward to work out the desired bounds numerically, by reformulating it as a linear programming problem. 

To be concrete let us consider the particular case $\phi = {\pi \over 4}$ and $\theta = {\pi \over 2}$. We first notice that in this case $\mu_1 \geq 0$, where $\mu_1 = 0$ corresponds to a degenerate case where we have a single QNM at $a=0$, where $a$ is defined in \eqref{eq:straightlinelocus}.
We can then derive the bounds for $({\mu_2 \over \mu_1}, {\mu_3 \over \mu_1})$ shown in Figure \ref{fig:linemoments} .

Let us compare the derived bounds with the explicit results for the BTZ black hole for momentum $k=\Delta$ (which corresponds to $\phi = {\pi \over 4}$)
\be
\mu_1 &= {\Delta^2 \over 4} \left[\psi^{(1)}\left({1+i \over 2} \Delta\right) + \psi^{(1)}\left({1-i \over 2} \Delta\right) \right] , \nn \\ 
\mu_2 &= {\Delta^4 \over 48} \left[ \psi^{(3)}\left({1+i \over 2} \Delta\right) + \psi^{(3)}\left({1-i \over 2} \Delta\right) \right] , \nn \\
\mu_3 &= {\Delta^6 \over 1920} \left[ \psi^{(5)}\left({1+i \over 2} \Delta\right) + \psi^{(5)}\left({1-i \over 2} \Delta\right) \right] \ . 
\ee
We see that we get a two-sided bound for ${\mu_3 \over \mu_1}$, but only a one-sided bound for ${\mu_2 \over \mu_1}$. Moreover, the simplest example of the BTZ black hole correlator realizes arbitrarily negative ${\mu_2 \over \mu_1}$.

\begin{figure}[t!]
\centering
  \includegraphics[scale=.7]{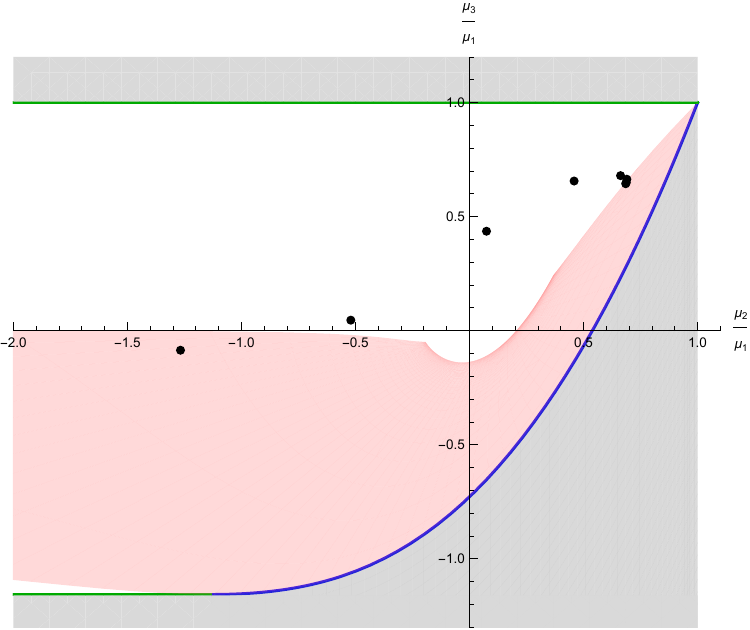}
\caption{The allowed region for the ratios $({\mu_2 \over \mu_1}, {\mu_3 \over \mu_1})$ for the case when all QNMs lie in the sectorial domain \eqref{eq:sectorialqnm}. The boundary of the allowed region can be realized with \textcolor{green!70!black}{two QNMs} and \textcolor{blue}{one QNM}. The \textcolor{pink!90!black}{pink} domain corresponds to the BTZ black hole for $0 \leq k \leq \Delta$ and arbitrary $\Delta$. The black dots represent the results for charged AdS black holes for $\Delta=4$ and different values of charge $Q$ ranging from $.42$ to $1.27$. In this range the QNMs are all contained within the sectorial region \eqref{eq:sectorialqnm}. }
 \label{fig:secQNM}
\end{figure}

Let us explain the structure of the boundary of the allowed region in Figure \ref{fig:linemoments}. The green upper line corresponds to a pair of QNMs $\rho(a) = \alpha \delta(a-{\sqrt{7}+1 \over \sqrt{2}}) + (1-\alpha) \delta(a)$, where $0 \leq \alpha \leq 1$. As $\alpha \to 0$ the point moves to minus infinity, whereas for $\alpha = 1$ it touches the blue boundary which is described by a single QNM $\rho(a) = \delta(a-a_0)$ with $0 \leq a_0 \leq {\sqrt{7}+1 \over \sqrt{2}}$.

\subsection{Sectorial QNMs }

Finally, we consider the case where the allowed region of support of QNMs is given by a sectorial region:
\be
\label{eq:sectorialqnm}
\text{Im } \omega_n \geq 1, ~~~ \text{Re } \omega_n \leq \text{Im } \omega_n .
\ee
It includes the two cases considered above but also allows for two-dimensional configurations of QNMs (not necessarily located on a line). We plot the results in Figure \ref{fig:secQNM}.

Let us comment on the structure of the boundary in Figure \ref{fig:secQNM}. The upper green line corresponds to a pair of QNMs $\rho(\omega) =\alpha \delta(\omega -i) + (1-\alpha) \delta(\omega - (1+i) )$ with $0 \leq \alpha \leq 1$. As $\alpha \to 0$ the point on the curve moves to the left, while $\alpha \to 1$ corresponds to the cusp at $(1,1)$, where we have only one QNM at $i$. The blue segment corresponds to a single QNM at $\omega_0 + i$, where $0 \leq \omega_0 \leq \sqrt{7}-2$. Finally, the lower green line is described by a pair of QNMs $\rho(\omega) =\alpha \delta(\omega - (\sqrt{7}-2 + i)) + (1-\alpha) \delta(\omega - (1+i) )$ for $0 \leq \alpha \leq 1$.

An obvious question is: what happens to the bounds as we relax the condition \eqref{eq:sectorialqnm}? For example, a natural minimal condition to consider is $\text{Im } \omega_n \geq 1$ which simply corresponds to the statement that there is a minimal characteristic timescale for perturbations to decay. We have not explored thoroughly what can be said about the structure of the hydrodynamic expansion in this case. However, we observed that for $({\mu_2 \over \mu_1}, {\mu_3 \over \mu_1})$ we do not get any nontrivial bounds in this case.

\section{BTZ and Rindler space}

Next we would like to discuss how the general structure discussed in the previous sections is realized in various examples. We start by considering the simplest examples in which the correlator is known analytically.

The two-sided correlator in the  BTZ background is given by \cite{Son:2002sd} (here $\beta=2\pi$)
\be
G_{12}(\omega,k)&=\frac{1}{\pi \Gamma(\Delta-1)^2}\Gamma\left(\frac{\Delta\pm i(\omega+k)}{2}\right)\Gamma\left(\frac{\Delta\pm i(\omega-k)}{2}\right),\label{eq:BTZ}
\ee
where $\Gamma(a\pm b)=\Gamma(a+b)\Gamma(a-b)$. The poles are located at $\omega_{n}=k+i(2n-2+\Delta)$ for $n=1,2,\ldots$ and the three images $(-\omega_n,\omega_n^*,-\omega_n^*)$, with residues 
\begin{align}
\lambda_n=\frac{2i(-1)^{n}\Gamma(ik-n+1)\Gamma(n-1+\Delta)\Gamma(-ik+n+\Delta-1)}{\pi (n-1)!\Gamma(\Delta-1)^2}.
\end{align}
The pole spacing is $r=2$ at angle $\theta=\pi/2$, and the subleading constant correction is $se^{i\phi}=k+(\Delta-2)i$, so the OPE constraints (\ref{spacingangle}) and (\ref{deltas}) indeed hold. Note that the singularity sum rules (\ref{sumrules}) are not satisfied, since there is no curvature singularity in the BTZ black hole. For instance, truncating the first sum rule at $n=n_{\text{max}}$ gives
\begin{align}
\text{Re}\sum_{n=1}^{n_{\text{max}}}\omega_n \lambda_n\sim -\frac{2k}{\sinh(\pi k)\Gamma(\Delta-1)^2}n_{\text{max}}^{2\Delta-1},\hspace{10 mm}n_{\text{max}}\to\infty
\end{align}
This diverges as $n_{\text{max}}\to \infty$, reflecting the fact that the contour integral along $C_{\infty}$ does not vanish.\\
\indent It is straightforward to verify that the BTZ two-sided correlator \eqref{eq:BTZ} can equivalently be written using the product formula \eqref{eq:ansatzprod} with $\omega_{n}=k+i(2n+\Delta)$ (note that $G_{12}(0,k)\neq 1$ in \eqref{eq:BTZ}, so that the product formula will differ from (\ref{eq:BTZ}) by a normalization). For $k=0$ the simple poles merge into double poles on the imaginary axis. 

An example of known higher-dimensional thermal correlators comes from CFTs on $S^1_\beta \times \mathbb{H}^{d-1}$ with $\beta=2\pi$ which can be conformally mapped to a Rindler wedge in $\mathbb{R}^d$, see e.g.\ \cite{Casini:2011kv}. The scalar two-point function on $S^1_\beta \times \mathbb{H}^{d-1}$ was studied in \cite{Ohya:2016gto} and is given by ($\beta=2\pi$)
\be
G_{12}(\omega,k)&=\frac{\pi^{\frac{d-2}{2}}}{\Gamma(\Delta)\Gamma(\Delta-\frac{d-2}{2})}\cr 
\times&\Gamma\left(\frac{\Delta-(d-2)/2\pm i(\omega+k)}{2}\right)\Gamma\left(\frac{\Delta-(d-2)/2\pm i(\omega-k)}{2}\right),\label{eq:scalarRindler}
\ee
which reduces to the scalar correlator in BTZ \eqref{eq:BTZ} for $d=2$. Here $k$ labels the principal series representation of $SO(1,d-1)$, see \cite{Ohya:2016gto} for details. This has no zeroes, and has poles at $\omega=k+i \left(-\frac{d}{2}+\Delta +2 n-1\right)$ for $n=1,2,\ldots$ and the corresponding images. Note also that the OPE  expansion of \eqref{eq:scalarRindler} is different from what we derived in Section \ref{sec:asymminkope} because the spatial slice $\mathbb{H}^{d-1}$  has non-zero curvature. Given the poles, we can reconstruct the two-sided correlator using the product formula.

The retarded energy-energy two-point function in Rindler space was further obtained in \cite{Haehl:2019eae}, for $d=4$ it is given by
\be
    G^{(d=4)}_{R,(tt,tt)}(\omega,k) = -&\frac{c_T\pi^2}{240}(k^2+4)(k^2+1)\cr
    &\times\left[\psi\left(\frac{1+i(k-\omega)}{2}\right)+\psi\left(\frac{1-i(k+\omega)}{2}\right)\right].
\ee
Using (\ref{g12gr}), the two-sided correlator is therefore found to be 
\be
    G^{(d=4)}_{12,(tt,tt)}(\omega,k) =\frac{c_T\pi^3}{480}\frac{4+5k^2+k^4}{\cosh (\pi  k)+\cosh (\pi  \omega )}\label{eq:4dRindler},
\ee
which has poles at $\omega_n=k+i(2n-1)$ for $n=1,2,\ldots$ and the corresponding images. Moreover, \eqref{eq:4dRindler} has no zeroes and can be written equivalently using the product formula \eqref{eq:ansatzprod}.

\section{Examples in pure GR}
In this section we consider examples that originate from matter minimally coupled to general relativity in AdS. The utility of the product formula \eqref{eq:ansatzprod} comes from the fact that it fixes the two-sided correlator in terms of the QNMs. The latter can be rather effectively found numerically \cite{Jansen:2017oag} (whenever they are not known analytically), thereby paving the way to straightforwardly computing the two-sided correlator $G_{12}(\omega)$. \\\indent 
\subsection{General strategy}
\indent In this section we study QNMs numerically for various wave equations and use the (truncated) product formula to reconstruct the correlator. More precisely, we can truncate the product formula at a fixed number $n_{\text{max}}$ of QNMs,
\be
G_{12}^{n_{\text{max}}}(\omega) =  {G_{12}(0)  \over \prod_{n=1}^{n_{\text{max}}} (1 - {\omega^2 \over \omega_n^2})(1 - {\omega^2 \over (\omega_n^*)^2})},\label{eq:ProdFormulaTrunc}
\ee
and then systematically improve it by increasing $n_{\text{max}}$ and/or attaching to it a tail of the asymptotic QNMs. The error made in this way can be estimated as follows. As discussed in Section \ref{qnmfromope}, one way to reproduce the OPE (which is realized in many examples that we will consider) is if the poles are asymptotically linear. If asymptotically we have a single line of QNMs $\omega_n\approx r e^{i\theta}n+\ldots$ for $0\leq\theta\leq \frac{\pi}{2}$, then for $\omega\ll\omega_{n_{\text{max}}}$ we have\footnote{Here we assume that the first $n_{\text{max}}$ QNMs are known exactly, and do not take into account numerical error.}
\be\label{eq:ErrorTrunc}
\log {G_{12}^{n_{\text{max}}}(\omega) \over G_{12}(\omega)} = -\frac{2 \omega ^2 \cos (2 \theta )}{r^2n_{\text{max}}}+\ldots,
\ee
For the special case $\theta=\pi/4$ (which holds for $AdS_5$/Schwarzschild), the error is order $1/n_{\text{max}}^2$. If there are multiple lines of QNMs one has to sum over all of them.

Now let us introduce an improved truncation scheme which improves the error estimate (\ref{eq:ErrorTrunc}). Let us assume as above that we have $\omega_n\approx \omega_{n_{\text{max}}}+re^{i\theta}(n-n_{\text{max}})$ for $n_{\text{max}}$ large and $n\ge n_{\text{max}}$. Assuming that we choose $n_{\text{max}}$ big enough, we can then improve the truncated solution by 
\be
    \hat{G}_{12}^{n_{\text{max}}}(\omega)=G_{12}^{n _{\text{max}}}(\omega)\frac{\Gamma\left(\frac{\omega_{n_{\text{max}}}\pm \omega}{re^{i\theta}}+1\right)\Gamma\left(\frac{\omega_{n_{\text{max}}}^*\pm \omega}{re^{-i\theta}}+1\right)}{|\Gamma\left(\frac{\omega_{n_{\text{max}}}}{re^{i\theta}}+1\right)|^4}\label{eq:truncImpr},
\ee
where $\Gamma(a\pm b) \equiv \Gamma(a+b)\Gamma(a-b)$. Eq.\ \eqref{eq:truncImpr} can be generalized to purely imaginary modes with asymptotically linear scaling with the mode number $n$. 

The improved correlator \eqref{eq:truncImpr} correctly captures the large $n$ behavior of QNMs, so that for $\omega_n=re^{i\theta}(n-n_{\text{max}})+\omega_{n_{\text{max}}}+\frac{f}{n^{\gamma}}e^{i\rho}+\ldots$ we instead find
\be\label{eq:ErrorTruncImpr}
\log {\hat{G}_{12}^{n_{\text{max}}}(\omega) \over G_{12}(\omega)} = \frac{4 f \omega ^2 \cos (3 \theta -\rho )}{(2+\gamma) r^3 n_\text{max}^{2+\gamma}}+\ldots,
\ee
which is improved compared to \eqref{eq:ErrorTrunc} by a factor of $n_\text{max}^{-(1+\gamma)}$ for generic values of $\theta$ and $\rho$.

We begin this section with a toy model, the two-point function of R-currents in $\mathcal{N}=4$ SYM with zero spatial momentum. The correlator and the QNMs are known analytically and can be compared with the truncated counterpart. We then move on to scalar QNMs in a black brane background, which can be calculated using a variety of methods (see \cite{Horowitz:1999jd,Berti:2009kk} and references therein). In practice we use the $\mathtt{QNMSpectral}$ package \cite{Jansen:2017oag}. The truncated correlator is then  obtained using \eqref{eq:ProdFormulaTrunc}. We further show the applicability to stress tensor correlators by studying metric fluctuations, and also consider the scalar correlator in a charged brane background. In these examples, the QNMs grow asymptotically linearly with the mode number $n$, so the truncated product formula can be improved using \eqref{eq:truncImpr}. For $\omega\in \mathbb{R}$ we further numerically solve the bulk wave equation to extract the correlator and compare against the product formula with perfect agreement.

\subsection{R-currents}

\indent Our first example is the two-point function of R-currents  with zero spatial momentum\footnote{When $k=0$ the correlators are the same for any polarization.} in $\mathcal{N}=4$ SYM \cite{Myers:2007we}, which is the simplest known example of a two-point function in a higher dimensional black hole. The retarded Green's function is
\begin{align}
G_R(\omega,0)=-\frac{N^2 }{32\pi^2}\left[i\omega+\omega^2\left(\psi\left(\frac{(1-i)\omega}{2}\right)+\psi\left(-\frac{(1+i)\omega}{2}\right)\right)\right],
\end{align}
where $\psi$ is the digamma function and $\beta=2\pi$.\\
\indent Using (\ref{g12gr}), we find the two-sided Wightman function
\begin{align}\label{rcurrentswightman}
G_{12}(\omega,0)=\frac{N^2}{32\pi}\frac{ \omega^2}{\cosh(\pi \omega)-\cos(\pi \omega)}.
\end{align}
The poles of this function in the upper right quadrant are at
\begin{align}
\omega_n=(1+i)n.
\end{align}
The spacing between QNMs is $r=\sqrt{2}$ at angle $\theta=\pi/4$, and the subleading constant correction vanishes, so the OPE predictions (\ref{spacingangle}) and (\ref{deltas}) are indeed satisfied with $\Delta=3$. \\
\indent The residues  are
\begin{align}
\lambda_n=\frac{N^2}{32\pi}\frac{(1+i) n^2(-1)^n}{\sinh(\pi n)}.
\end{align}
These satisfy the singularity sum rules (\ref{sumrules}), since 
\begin{align}
\text{Re}\left(\sum_{n=1}^{\infty} ((1+i)n)^{m}\frac{(1+i)n^{2}(-1)^n}{\sinh(\pi n)}\right)=0,\hspace{10 mm}\text{for odd }m>0.
\end{align}
Let us notice that the sum converges exponentially fast due to the ${1 \over \sinh (\pi n)}$ factor.
\begin{figure}[t!]
  \centering
  \includegraphics{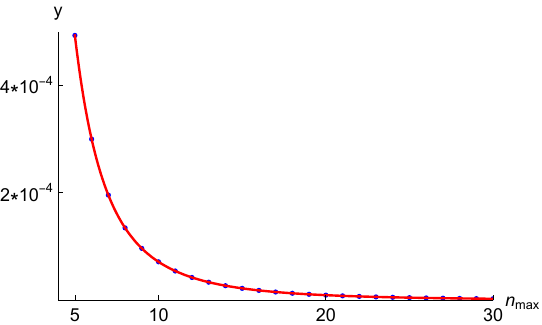}
\caption{We plot $y=\log (G_{12}^{n_{\text{max}}}(\omega,0)/G_{12}(\omega,0))$ at $\omega=1$ for R-currents as a function of $n_{\text{max}}$ and compare against the red line $y=\frac{0.083}{n_{\text{max}}^3}-\frac{0.12}{n_{\text{max}}^4}+\frac{0.078}{n_{\text{max}}^5}$. The coefficient of $n_{\text{max}}^{-3}$ agrees with \eqref{errorrcurrents} which predicts $\frac{1}{12}\approx 0.083$.} 
\label{fig:ToyFunc1}
\end{figure}
\indent We now assume that the QNMs $\omega_n=(1+i)n$ are given as input. Using the product formula \eqref{eq:ansatzprod}, we reproduce the expected result
\begin{equation}
   G_{12}(0,0) \prod_{n=1}^{\infty }\frac{1}{(1-\frac{\omega^2}{\omega^2_n})(1-\frac{\omega^2}{(\omega_n^*)^2})} = G_{12}(0,0)\frac{\pi^2 \omega^2}{\cosh  \pi \omega-\cos\pi\omega}
\end{equation}
On the other hand, truncating to include the first $n_{\text{max}}$ QNMs, we plot in Figure \ref{fig:ToyFunc1} $\log (G_{12}^{n_{\text{max}}}(\omega,0)/G_{12}(\omega,0))$ at $\omega=1$ as a function of $n_{\text{max}}$. In this case, the modes are given by $\omega_n=\sqrt{2} e^{i\frac{\pi}{4}}n$, with no subleading terms in $1/n$. One then finds along the lines of \eqref{eq:ErrorTrunc} that 
\be \label{errorrcurrents}
\log {G_{12}^{n_{\text{max}}}(\omega,0) \over G_{12}(\omega,0)} = \frac{\omega ^4}{12 n_{\text{max}}^3}+\ldots.
\ee
For $\omega=1$ this decays as $\frac{1}{12}n^{-3}_{\text{max}}  $, in agreement with the observed values in \Fig{fig:ToyFunc1}.

\subsection{Scalar QNMs in black branes}\label{sec:scalarEinstein}
\begin{figure}
\centering
\begin{subfigure}{.5\textwidth}
  \centering
  \includegraphics[width=1\linewidth]{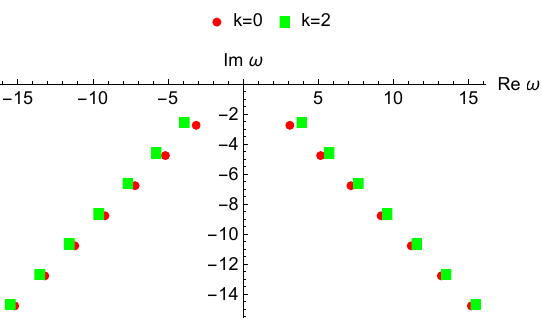}
  \caption{}
  \label{fig:scalarQNMs}
\end{subfigure}%
\begin{subfigure}{.5\textwidth}
  \centering
  \includegraphics[width=1\linewidth]{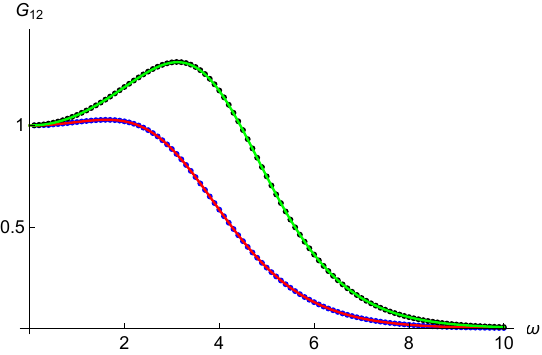}
  \caption{}
  \label{fig:scalarCorr}
\end{subfigure}
\caption{(a) QNMs for a massless scalar field with $k=0,2$, with temperature set to $T=\pi^{-1}$.  (b) $G_{12}$ for $k=0$ and $k=2$. The lines correspond to the results obtained using the first $27$ modes and a tail of modes with $r=2\sqrt{2}$ and $\theta=\frac{\pi}{4}$, while the points correspond to the result obtained by numerically solving the bulk wave equation and extracting the correlator.}
\end{figure}
We now consider a more nontrivial example for which we resort to numerics: the scalar two-point function with $\Delta=4$ on $S^1_\beta\times\mathbb{R}^3$ dual to a massless scalar field in an AdS black brane background\footnote{It is straightforward to do the same computations for generic values of $\Delta$.}. This is equivalent to the scalar perturbation of the metric in pure GR and the correlator will therefore be the same as for $\langle T_{xy}T_{xy}\rangle$\footnote{Up to overall normalization.}. The metric is given by\footnote{We follow the conventions in \cite{Kovtun:2005ev}.}
\be 
ds^2 = \frac{1}{u}(-f(u)\,dt^2+d\vec{x}^2)+\frac{du^2}{4u^2f(u)},\label{eq:blackBraneMetric}
\ee
where $f(u)=1-u^2$ and $u=\frac{r_0^2}{r^2}$, $r_0$ being the horizon radius. In \eqref{eq:blackBraneMetric} without loss of generality we set the AdS radius and $r_0$ to 1, which fixes the temperature $\beta=\pi$. The wave equation for a scalar field $\phi(u,t,z)=\phi(u)e^{-i\omega t+ik z}$ dual to an operator with $\Delta=4$ is then given by 
\be 
\left(\p_u^2-\frac{1+u^2}{uf}\p_u+\frac{\omega^2-k^2 f}{4uf^2}\right)\phi(u)=0.\label{eq:EOMscalar}
\ee
Imposing infalling boundary conditions $\phi(u)\sim (1-u)^{-i\omega/4}$ at the horizon $u\to 1$, we can extract the retarded correlator and hence the two-sided correlator using (\ref{g12gr}). For further details see e.g.\ \cite{Son:2002sd,Kovtun:2005ev,Kovtun:2006pf}.

In Figure \ref{fig:scalarQNMs} we display the first $7$ QNMs obtained from \eqref{eq:scalarQNMeq} for a scalar field with $k=0$ and $k=2$. The truncated product formula with $n_{\text{max}}=27$ modes and improved with $r=2\sqrt{2}$ and $\theta=\frac{\pi}{4}$ is shown in Figure \ref{fig:scalarCorr}, where we also compared against the result obtained from numerically solving the ODE \eqref{eq:EOMscalar} in Mathematica. 

As shown analytically in Section \ref{qnmfromope}, the OPE predictions for the linear and constant term in the QNM asymptotics are satisfied for this model. We can see this numerically as well. A numerical fit in the two cases gives
\begin{equation}
\begin{aligned}
&\omega_n \simeq c(k) + 2.00 (1+i) n \,, \\
&c(0) \simeq 1.219 + 0.7789 i \,, \,\,\,\, c(2) \simeq 1.28 + 0.76 i \,. 
\end{aligned}
\end{equation}
Comparing with \eqref{spacingangle} we find
\begin{equation}
\frac{4 \pi \sin \theta}{r} \simeq 3.14 \simeq \pi \simeq \beta ,
\end{equation}
consistently. Likewise, for the subleading term in the OPE expansion we find for $k=0$
\be 
\frac{4s\cos(\theta-\phi)}{r}+2\simeq 4.00,
\ee 
in agreement with \eqref{deltas} for $\Delta=d=4$. For $k=2$ with $n_{\text{max}}\sim80$ modes we get $\simeq 4.05$.

The picture discussed here does not change qualitatively as we change $\Delta$ or $d$. As we vary $\Delta$ away from $4$ we get more contributions in the large $\omega$ OPE expansion. We considered $\Delta$s for which both the stress tensor and the double-trace stress tensor operators appear, and checked that the product formula correctly reproduces the numerical solution. One difference as we change $d$ is that the asymptotic angle at which the QNMs approach infinity changes to $e^{- i {\pi \over d}}$, see \eqref{eq:asymqnms}. As a noteworthy example, it has been recently shown \cite{Biggs:2023sqw} that QNMs of the gravity dual to the BFSS matrix model \cite{Banks:1996vh} can be computed using the black brane geometry in $d={14 \over 5}$.

\subsection{Metric fluctuations}
\begin{figure}
\centering
\begin{subfigure}{.5\textwidth}
  \centering
  \includegraphics[width=1\linewidth]{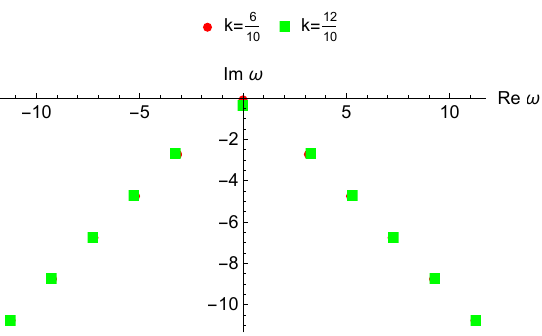}
  \caption{}
  \label{fig:txtxQNMs}
\end{subfigure}%
\begin{subfigure}{.5\textwidth}
  \centering
  \includegraphics[width=1\linewidth]{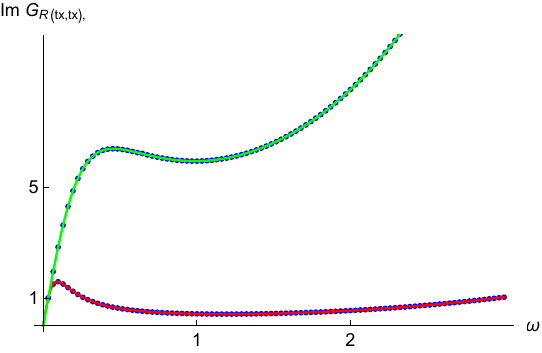}
  \caption{}
  \label{fig:txtxCorr}
\end{subfigure}
\caption{(a) QNMs in the shear channel for $k=\frac{6}{10}$ and $k=\frac{12}{10}$. (b) ${\text{Im }} G_{R,(tx,tx)}$ for $k=\frac{6}{10}$ and $k=\frac{12}{10}$, where the lines were obtained from the improved truncated product formula with $n_{\text{max}}=13$, $r=2\sqrt{2}$ and $\theta=\frac{\pi}{4}$. The points were obtained from $\mathtt{NDSolve}$. For small $k$ the peak is located at $\omega=\frac{k^2}{4}$, due to the hydrodynamic shear mode \cite{Kovtun:2006pf}. As above, we have chosen the normalization for the product formula to agree with the result from $\mathtt{NDSolve}$ at the lowest value of $\omega$ used in the latter, as well as chosen the absolute normalization arbitrarily.}
\end{figure}
In this section, we will observe numerically that the product formula (\ref{eq:ansatzprod}) correctly reproduces the correlator for metric fluctuations. The metric perturbations $h_{\mu\nu}$ are classified according to their symmetry properties with respect to rotations along the transverse directions of the propagation. This leads to three gauge invariant combinations $Z_i=Z_i(h_{\mu\nu})$, corresponding to three independent scalar functions determining the stress tensor two-point functions at the boundary. However, note that the analysis of the zeroes of the Wightman function in Appendix \ref{nozeroesappendix} is only applicable for the scalar wave equation, since the potential for metric fluctuations contains singularities at positions depending on the spatial momentum and frequency (see \cite{Loganayagam:2022teq} for a recent discussion).   It would be interesting to generalize the analysis to metric fluctuations as well. We leave this problem for future work.\\
\indent The scalar perturbation is equivalent to a $\Delta=4$ scalar, which was studied in Section \ref{sec:scalarEinstein}. For the shear channel $Z_1$ and sound channel $Z_2$, the wave equations are given by \cite{Kovtun:2005ev}
\be
    \left(\p_u^2+p_i \p_u+q_i\right)Z_i(u)=0,
\ee 
where
\be
    p_1 &= \frac{(\omega^2-k^2f)f+2u^2\omega^2}{uf(k^2f-\omega^2)}\\
    q_1 &= \frac{\omega^2-k^2f}{4uf^2}\\
    p_2 &= -\frac{3\omega^2(1+u^2)+k^2(2u^2-3u^4-3)}{uf(3\omega^2+k^2(u^2-3))}\\
    q_2 &= \frac{3\omega^4+k^4(3-4u^2+u^4)+k^2(4u^2\omega^2-6\omega^2-16u^3f)}{4uf^2(3\omega^2+k^2(u^2-3))}.
\ee 
Here $f(u)=1-u^2$ and we have set $\beta=\pi$. Given the gauge-invariant observables $Z_i$, the stress tensor two-point functions for various polarisations can be extracted, see \cite{Kovtun:2005ev,Kovtun:2006pf} for details. 

\begin{figure}[t!]
\centering
\begin{subfigure}{.5\textwidth}
  \centering
  \includegraphics[width=1\linewidth]{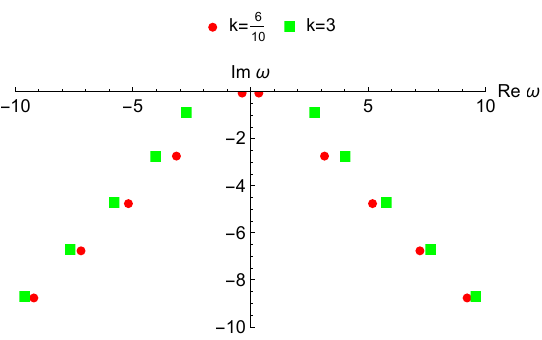}
  \caption{}
  \label{fig:ttttQNMs}
\end{subfigure}%
\begin{subfigure}{.5\textwidth}
  \centering
  \includegraphics[width=1\linewidth]{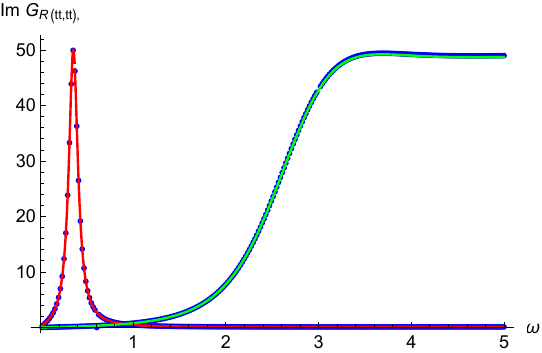}
  \caption{}
  \label{fig:ttttCorr}
\end{subfigure}
\caption{(a) QNMs in the sound channel for $k=\frac{6}{10},3$. (b) $\text{Im } G_{R,(tt,tt)}$ for $k=\frac{6}{10},3$ with $n_{\text{max}}=13$, $r=2\sqrt{2}$ and $\theta=\frac{\pi}{4}$. For small $k$ there is a sharp peak located at $\omega=\frac{k}{\sqrt{3}}$ due to the hydrodynamic sound mode \cite{Kovtun:2006pf}. Here we have chosen the normalization for the product formula to agree with the result from $\mathtt{NDSolve}$ at the lowest value of $\omega$ used in the latter, as well as chosen the absolute normalization arbitrarily.}
\end{figure}

In Figure \ref{fig:txtxQNMs} we have plotted the QNMs in the shear channel for $k=\frac{6}{10}$ and $k=\frac{12}{10}$. In Figure \ref{fig:txtxCorr} we plotted the imaginary part of the retarded correlator $G_{tx,tx}$ for $\omega\in\mathbb{R}$ using the improved truncated product formula and compared against the numerical solution. Similarly, in Figure \ref{fig:ttttQNMs} we plotted the QNMs in the sound channel for $k=0.6$ and $k=3$, and the comparison between the improved truncated product formula and the numerical solution of the bulk wave equation in Figure \ref{fig:ttttCorr}.
The asymptotic behavior of the QNMs for metric fluctuations was obtained in \cite{Cardoso:2004up,Natario:2004jd} in general dimensions. For both the shear and the sound channel these are given by
\be\label{eq:shearSoundOPE}
\omega^{\text{asy}}_n = \frac{4\pi}{\beta}e^{i\frac{\pi}{d}}\sin \left(\frac{\pi }{d}\right)\left(n+\frac{d-2}{4}-\frac{i \log 2}{2 \pi }\right),
\ee 
which is the same as a scalar with $\Delta=d$. Note that in both cases the hydrodynamic modes are not included. It is then seen that the OPE of $T_{tx}T_{tx}$ is correctly reproduced, since 
\be
\partial_\omega \log G_{12,(tx,tx)}(\omega) = -\frac{\beta}{2}+\frac{d-2}{\omega}+\ldots,
\ee
where we used \eqref{logsum}, \eqref{spacingangle} and \eqref{deltas} together with \eqref{eq:shearSoundOPE}, $\Delta=d$ and the fact that there is a purely imaginary hydrodynamic mode and its image contributing as $-\frac{2}{\omega}$. Likewise, we find that
\be
\partial_\omega \log G_{12,(tt,tt)}(\omega) = -\frac{\beta}{2}+\frac{d-4}{\omega}+\ldots,
\ee
in agreement with expectations for the $T_{tt}T_{tt}$ OPE. The  term $-\frac{4}{\omega}$ is again due to the hydrodynamic mode and its images.

\subsection{Charged black brane}

\begin{figure}[t!]
\centering
\begin{subfigure}{.5\textwidth}
  \centering
  \includegraphics[width=1\linewidth]{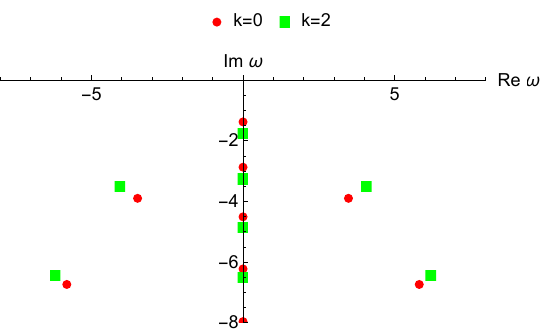}
  \caption{}
  \label{fig:ChargedQNMS}
\end{subfigure}%
\begin{subfigure}{.5\textwidth}
  \centering
  \includegraphics[width=1\linewidth]{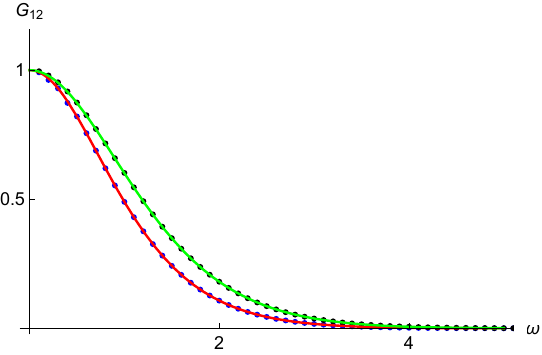}
  \caption{}
  \label{fig:ChargedCorr}
\end{subfigure}
\caption{(a) QNMs in a charged black brane with $Q=1$ and $k=0,2$. (b) $G_{12}$ for $Q=1$ and $k=0,2$. Here we have used $n_{\text{max},2}=19$ and $n_{\text{max},1}=24$, and we have extended with the asymptotics $r_{2}=3.60$, $\phi_{2}=0.89$, $r_{1}=1.76$ and $\phi_{1}=\frac{\pi}{2}$.}
\end{figure} 
In this section we consider a scalar correlator in a charged state, dual to a charged black brane. The QNMs in a charged AdS black hole were analyzed in \cite{Wang:2000dt,Wang:2000gsa,Natario:2004jd}. The main new feature in this case compared to the uncharged case is the presence of an infinite line with purely imaginary modes. By modifying the truncated product ansatz \eqref{eq:ProdFormulaTrunc} to include such imaginary modes we will again verify good agreement with the numerical solution on the real line. 

The metric for a charged black brane is given by, see e.g.\ \cite{Carmi:2017jqz},
\be
    ds^2=-r^2f(r)\,dt^2+\frac{dr^2}{r^2f(r)}+r^2\,d\vec{x}^2,
\ee
where $f(r)=1-\mu r^{-4}+Q^2r^{-6}$. We fix the temperature  $T=\frac{1}{\pi}(1-\frac{Q^2}{2})$ by setting the horizon at $r=1$, which further implies that $\mu=1+Q^2$.

In Figure \ref{fig:ChargedQNMS} we plot the QNMs with $Q=1$ and $k=0,2$, which contain a new infinite line of imaginary modes that were absent in the uncharged case. In Figure \ref{fig:ChargedCorr} we compare the numerical solution to the improved truncated product formula, modified to include pure imaginary modes.

To compare with \eqref{betamultilinlines} we fit the two lines of QNMs. For the $k=0$ case, using about $50$ complex modes and $80$ imaginary modes we find
\begin{equation}
\begin{aligned}
&\omega_{1,n} \simeq i(-0.867+1.76 n) \,, \\
&\omega_{2,n} \simeq (1.33+1.21 i)+(2.27+2.79 i) n \,.
\end{aligned}
\end{equation}
The OPE predictions are then satisfied,
\begin{equation}
\begin{aligned}
&\frac{4\pi\sin \theta_2}{r_2} +  \frac{2\pi}{r_1}  \simeq 6.28  \simeq \beta,  \\
&\frac{4 s_2 \cos\left(\theta_2-\phi_2\right)+2r_2}{r_2}+\frac{2s_1+r_1}{r_1} \simeq 4.00 \sim 2 \Delta-d \,.
\end{aligned}
\end{equation}
For $k=2$ and a similar number of modes we get $2 \Delta-d \simeq 4.13$, consistent with the fact that we expect slower convergence for higher $k$.

\section{Higher derivative corrections}\label{higherderivative}
The examples considered so far have all been at infinite $\lambda$, but the product formula (\ref{eq:ansatzprod}) is equally applicable when higher derivative corrections are taken into account. In this section we analyze several instructive examples of higher derivative terms in the Lagrangian, and confirm the product formula and the predictions from the OPE.
\subsection{Gauss-Bonnet black holes}
Let us first calculate the QNMs in Gauss-Bonnet gravity, from which we compute the two-point function and compare with the numerical solution to the wave equation. The Gauss-Bonnet QNMs were previously discussed in \cite{Grozdanov:2016fkt,Grozdanov:2016vgg}.

Consider a black brane in Gauss-Bonnet gravity, with the metric \cite{Boulware:1985wk,Cai:2001dz}
\begin{equation}\label{eq:GBmetric}
ds^2 = -\frac{r^2f(r)}{L^2 f_\infty}\,dt^2+\frac{L^2 \,dr^2}{r^2 f(r)}+\frac{r^2}{L^2}\,(dx^i)^2,
\end{equation}
where $f_\infty=\frac{1-\sqrt{1-4\lambda_{\text{GB}}}}{2\lambda_{\text{GB}}}$, $\lambda_{\text{GB}}$ is the Gauss-Bonnet coupling and
\begin{equation}
    f(r)= \frac{1}{2\lambda_{\text{GB}}}\left(1-\sqrt{1-4\lambda_{\text{GB}}\left(1-\frac{r^4_+}{r^4}\right)}\right),\hspace{10 mm}\lambda_{\text{GB}}\le \frac{1}{4}.
\end{equation}
The AdS radius is given by $\tilde{L}^2=L^2/f_\infty$. The Hawking temperature is 
\begin{equation}
T = \frac{r_+}{\pi L^2\sqrt{f_\infty}}.
\end{equation}
From now on we set $L=1$ and further fix the temperature in terms of $\lambda_{\text{GB}}$ by setting $r_+=1$.

\begin{figure}
\centering
\begin{subfigure}{.5\textwidth}
  \centering
  \includegraphics[width=1\linewidth]{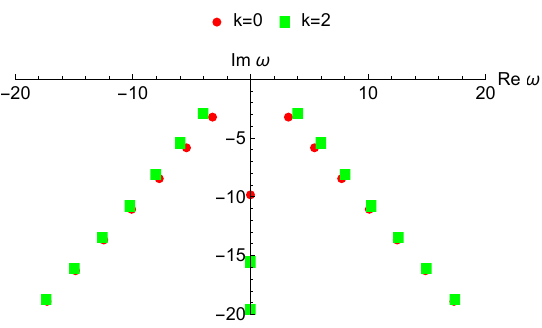}
  \caption{}
  \label{fig:GBQNMs}
\end{subfigure}%
\begin{subfigure}{.5\textwidth}
  \centering
  \includegraphics[width=1\linewidth]{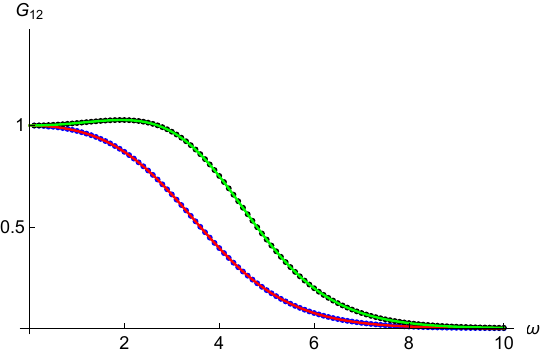}
  \caption{}
  \label{fig:GBCorr}
\end{subfigure}
\caption{(a) Gauss-Bonnet QNMs for $\lambda_{\text{GB}}=\frac{8}{100}$ and $k=0,2$. (b) Scalar correlator with $\Delta=4$ for $\lambda_{\text{GB}}=\frac{8}{100}$ and $k=0,2$. Here we have used $n_{\text{max},1}=17$ purely imaginary modes and $n_{\text{max},2}=44$ complex modes.}
\end{figure}

The QNMs can be calculated as above by passing to Eddington-Finkelstein coordinates\footnote{One needs to take into account the extra factor of $f_\infty^{-1}$ in the $dt^2$ term in the metric \eqref{eq:GBmetric}.}, and are shown for a scalar operator with $\Delta=4$ in Figure \ref{fig:GBQNMs}. Likewise, the numerical solution of the correlator is shown in Figure \ref{fig:GBCorr} and compared against the improved truncated product formula.

A numerical fit gives for the two lines of QNMs (again we restrict to the $k=0$ case)
\begin{equation}
\begin{aligned}
&\omega_{1,n} \simeq 7.77 i n \,, \\
&\omega_{2,n} \simeq (2.53+2.54 i) n \,.
\end{aligned}
\end{equation}
This gives
\begin{equation}
\frac{4\pi\sin \theta_2}{r_2} + \frac{2\pi}{r_1} \simeq 3.30,
\end{equation}
while the inverse temperature reads
\begin{equation}
\beta = \pi \sqrt{\frac{1-\sqrt{1-4\lambda_{\text{GB}}}}{2 \lambda_{\text{GB}}}} \simeq 3.29,
\end{equation}
consistently with (\ref{betamultilinlines}). To check the subleading sum rule more QNMs are needed, and we leave this problem to future work.  It would be also curious to generalize this analysis to metric perturbations \cite{Buchel:2009sk,Huang:2022vet}.

\subsection{$\phi^2 W^2$ coupling}
We now introduce a higher derivative term that displays novel behavior for QNMs. In particular, we will find a line of imaginary modes which asymptotically behave as $\omega_n\propto in^3$.

Let us consider the Lagrangian
\begin{equation}
\mathcal{L} = -\frac{1}{2} \int d^5 x \, \sqrt{-g}\left[ \left( \partial \phi\right)^2 +  m^2 \phi^2 +\alpha W^2 \phi^2 \right], 
\end{equation}
where $W$ is the Weyl tensor. A similar interaction $W^2\phi$ was analyzed in \cite{Myers:2016wsu,Grinberg:2020fdj} as a model for higher derivative corrections to the one-point function. We consider again the black brane metric \eqref{eq:blackBraneMetric} with $\beta=\pi$. The squared Weyl tensor in these coordinates is
\be
W_{\mu \nu \rho \sigma} W^{\mu \nu \rho \sigma} = 72 u^4 \,.
\ee
The equation of motion reads
\be 
\left(\p_u^2-\frac{1+u^2}{uf}\p_u+\frac{\omega^2-k^2 f}{4uf^2}- \frac{\Delta(\Delta-4)}{4u^2 f(u)} - \frac{18u^2}{f} \alpha\right)\phi(u)=0.\label{eq:EOMscalarAlpha}
\ee
The coupling $\alpha$ is measured in units of the AdS radius $L = 1$.
\begin{figure}
\centering
\begin{subfigure}[b]{0.45\textwidth}
\centering
\includegraphics[width=\textwidth]{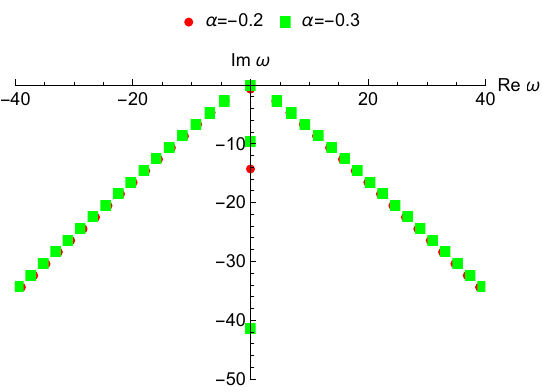}
\caption{}
\label{fig:phi2w2}
\end{subfigure}
\hfill
\begin{subfigure}[b]{0.45\textwidth}
\centering
\includegraphics[width=\textwidth]{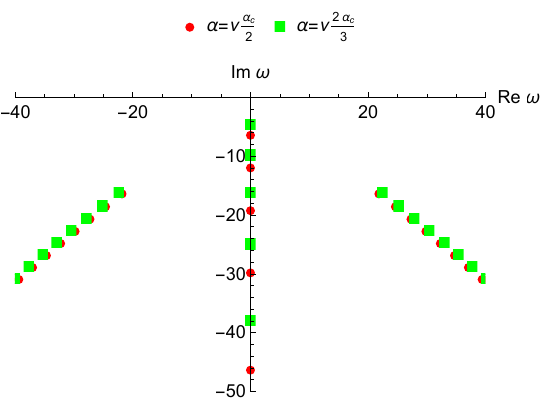}
\caption{}
\label{fig:phi2w2wkb}
\end{subfigure}
\caption{(a) Massless QNMs for $\alpha \lesssim 0$. Here we have a line of imaginary modes with $n^3$ scaling. (b) Large $\nu$ scalar black brane QNMs with an $\alpha W^2\phi^2$ coupling, with $\alpha_c < \bar{\alpha} < 0$ and $\nu = 18$. The spacing between the imaginary modes starts linearly, and gradually increases when going towards the asymptotic $\omega \gg \nu$ region. In both plots $k=0$.}
\label{fig:phi2w2bt}
\end{figure}
We plot the QNMs for $\alpha =-0.2$ and $\alpha =-0.3$ in Figure \ref{fig:phi2w2}. For these values of the coupling we find a line of imaginary QNMs with nonlinear spacing. A numerical fit involving 5 imaginary modes for $\alpha=-0.2$ gives for this line
\begin{equation}
\omega_{1,n} \simeq i (a n^3 + b n^2) \,, \,\,\,\, a \simeq 2.1 \,, \,\,\,\, b \simeq 0.13 \,.
\end{equation}
The OPE expansion predicts that the line of complex QNMs takes the form
\begin{equation}
\omega_{2,n} \simeq r e^{i \theta} n + s e^{i \phi} n^{1/3}+c \,.
\end{equation}
A numerical fit involving about 70 complex modes confirms the predictions and sets
\begin{equation}
c \simeq 0.80 e^{i 0.96} \,, \,\,\,\, r \simeq 2.8 \,, \,\,\,\, \theta \simeq 0.78 \simeq \frac{\pi}{4} \,, \,\,\,\, s \simeq 1.8 \,, \,\,\,\,  \phi \simeq 0 \,.
\end{equation}
From \eqref{spacingangle}
\begin{equation}
\beta = \frac{4 \pi \sin \theta}{r} \approx 3.1 \approx \pi ,
\end{equation}
consistently. Finally, we also confirm the prediction made in \eqref{predictiontwolines}, that is
\begin{equation}
\frac{r^{4/3} a^{-1/3}}{2 \cos \left(\frac{\pi}{6}-\frac{4}{3} \theta \right)} \approx 1.8 \approx s \,. 
\end{equation}
Note that $b, c$ will contribute at higher orders in the $\omega^{-1}$ expansion.

Here we only considered QNMs for $\alpha \lesssim 0$. More generally, the structure of the QNMs as a function of $\alpha$ displays a rich structure. To gain a qualitative understanding, let us consider the WKB limit $\omega, \nu, \alpha \to \infty$ with $\alpha / \nu = \bar{\alpha}$ and $\omega/\nu$ fixed. The WKB potential reads
\be
V(z) = f(r)\left(1+\frac{72 \bar{\alpha}}{r^8}\right) = f(r)\left(1-\frac{\bar{\alpha}}{\alpha_c}\frac{1}{r^8}\right) \,,
\ee
where $r=1/u^2$ is the usual radial coordinate. We start by considering $\bar{\alpha} <0$. In this case, since the $\overline{\alpha}$ term dominates close to the singularity, the potential goes to $+ \infty$ as $r \to 0$. We can distinguish 3 cases: 
\begin{itemize}
\item $\bar{\alpha} = -1/72 = \overline{\alpha}_c$. The WKB potential is positive definite and has a global minimum at the horizon $r=1$ ($z \to \infty$). Here
\be
\lim_{z\to\infty} \partial^n_z V(z) = 0 \,, \quad n \ge 0 \,,
\ee
but $\lim_{z\to \infty} V(z)^{-1}\partial_z^2V(z) = - 16$. Therefore we can still apply \eqref{BSapprox}, and we find a new line of poles at $\omega_n = - (2+4n)i$ for $n\ge0$.
\item $\bar{\alpha} < \overline{\alpha}_c$. The global minimum moves outside the horizon to a finite value of $z$ where $V(z)$ is negative. This minimum leads to linearly spaced ($\partial^2_z V(z_{\text{min}}) \ne 0$) purely imaginary modes in the upper half plane. These modes correspond to bound states, since for $\omega_n = i |\omega_n|$ the ingoing solution behaves as $e^{- |\omega_n| z}$ as $z \to \infty$. As mentioned in Section \ref{qnmbh}, they are associated to an instability of the black hole.
\item $\overline{\alpha}_c<\bar{\alpha}<0$. The minimum moves inside the horizon, and the potential hosts virtual bound states corresponding to linearly spaced QNMs on the lower imaginary axis.
\end{itemize}
On the other hand when $\bar{\alpha}>\overline{\alpha}^*>0$, with $\overline{\alpha}^* \simeq 0.12$, the potential develops a metastable minimum outside the horizon, corresponding to weakly damped QNMs. If $\overline{\alpha}^*>\bar{\alpha}>0$ the metastable minimum becomes complex and the QNM spectrum is qualitatively unchanged compared to the $\overline{\alpha} = 0$ case. 

Note that the WKB analysis holds for $\omega \sim \nu$, while to compare with the OPE predictions we need to analyze the asymptotic behavior of QNMs in the limit $\omega \gg 1, \Delta, k$. The qualitative features of the QNMs are the same as discussed above. The main difference is in the spacing between modes in the case $\overline{\alpha}_c < \bar{\alpha} < 0$. As shown in Figure \ref{fig:phi2w2wkb}, in the large $\nu$ regime the separation between these modes indeed starts linearly, but increases gradually as $\omega$ becomes larger than $\nu$, until we find $\omega_n \propto i n^3$ as noted previously.

\section{Beyond the strong coupling limit}\label{beyondstrong}
In the bulk of this paper we have discussed the analytic properties of the two-point function in the holographic regime of infinite $N$ and infinite $\lambda$. In order to understand the range of applicability of our results, it is important to analyze which of these properties survives at finite coupling and finite $N$.\\
\indent Let us first discuss the effect of $1/N$ corrections. At one loop, the two-point function develops a branch cut, corresponding to late-time hydrodynamic tails \cite{Caron-Huot:2009kyg}. As a result, the Wightman function is not an analytic function of frequency, so its poles and zeroes are no longer enough to determine the full function. It follows that the results presented here have limited applicability at finite $N$ (at least without some modification). \\
\indent We now turn to the case where the coupling is finite but $N$ is infinite. In the rest of this section, we will present several instructive examples of lower-dimensional models where the holographic properties (meromorphy and no zeroes) extend to arbitrary coupling. We first consider the SYK model \cite{Maldacena:2016hyu,kitaev,Sachdev_1993,Polchinski:2016xgd}, a chaotic 0+1 dimensional theory with a built-in disorder average. We then discuss a 1+1 dimensional generalization of the SYK model known as the SYK chain \cite{Choi:2020tdj,Gu:2016oyy}. In both cases, $G_{12}$ is meromorphic and has no zeroes, which shows that the analytic properties of a holographic CFT have a chance of extending beyond the holographic regime. We conclude with several examples of non-chaotic theories, in which both properties are violated.    
\subsection{SYK} 
The SYK model is a quantum mechanical system of $N$ fermions, with a random coupling involving an even number $q$ fermions at a time. The Hamiltonian is given by \cite{Maldacena:2016hyu,kitaev,Sachdev_1993,Polchinski:2016xgd}
\begin{align}
H&=i^{q/2}\sum_{1< i_1<\ldots i_1< N}j_{i_1\ldots i_q}\psi_{i_1}\cdots\psi_{i_q},\hspace{10 mm}\langle j_{i_1\ldots i_q}^2\rangle=\frac{J^2(q-1)^2}{N^{q-1}}.
\end{align}
We consider the limit $N\to \infty$ with $q$ fixed, followed by the limit $q\to \infty$. In this regime, the two-sided Wightman function is given by \cite{Maldacena:2016hyu,Tarnopolsky:2018env}
\begin{align}
G_{12}(t)\propto\frac{1}{\cosh\left(\frac{\pi t v}{\beta}\right)^{2/q}},\hspace{10 mm}\beta J=\frac{\pi v}{\cos\left(\frac{\pi v}{2}\right)}.
\end{align}
\indent Fourier transforming, we find 
\begin{align}
  G_{12}(\omega)\propto \Gamma \left(\frac{1}{q}-\frac{i \beta  \omega }{2 \pi  v}\right)
   \Gamma \left(\frac{1}{q}+\frac{i \beta  \omega }{2 \pi  v}\right).
   \end{align}
   For any value of the coupling $v$, this is a meromorphic function of $\omega$. The poles all lie on the imaginary axis, 
   \begin{align}
\omega_n=\pm \frac{2\pi i (1/q+n-1)v}{\beta},\hspace{10 mm}n=1,2,\ldots
   \end{align}
   Moreover, there are no zeroes, echoing the property discussed in the holographic context. In this model, the Wightman function is structureless as a function of the coupling, since $G_{12}(\omega)$ is only a function of $\omega/v$. In order to get a richer structure, we next turn to the SYK chain.
\subsection{The SYK chain}
The SYK chain is a generalization of the SYK model to a lattice model in 1+1 dimensions. This lattice model contains both on-site interactions between $q$ fermions and nearest neighbor interactions between $q/2$ fermions from each site.  We refer the reader to \cite{Choi:2020tdj,Gu:2016oyy} for the full definition of the model.\\
\indent As above, we take the limit $N\to \infty$, followed by $q\to \infty$. We consider the two-point function of energy density operators in this limit. The two-point function was computed in \cite{Choi:2020tdj} to be (here we have used (\ref{g12gr}) and set $\beta=2\pi$)
\begin{align}
G_{12}(\omega)\propto\frac{i}{\sinh(\pi \omega)}\partial_\theta\log\left(\frac{f(-\omega,\theta)}{f(\omega,\theta)}\right)\big|_{\theta=\theta_v},\label{g12chain}
\end{align}
where
\begin{align}
f(\omega,\theta)&=\frac{i\Gamma\left(1-\frac{h}{2}-\frac{i\omega}{2v}\right)\sin\left(\frac{\pi h}{2}+\frac{\pi i \omega}{2v}\right)\sinh\left(\frac{\pi \omega}{2}\right)}{\Gamma\left(\frac{1}{2}-\frac{h}{2}+\frac{i\omega}{2v}\right)}\cos\theta(\sin\theta)^h{_2F_1}\left(\frac{1+h-\frac{i\omega}{v}}{2},\frac{1+h+\frac{i\omega}{v}}{2},\frac{3}{2},\cos^2\theta\right)\notag\\
&+\frac{\Gamma\left(\frac{1-h}{2}-\frac{i\omega}{2v}\right)\cos\left(\frac{\pi h}{2}+\frac{\pi i\omega}{2v}\right)\cosh\left(\frac{\pi \omega}{2}\right)}{2\Gamma\left(1-\frac{h}{2}+\frac{i\omega}{2v}\right)}(\sin\theta)^h{_2F_1}\left(\frac{h-\frac{i\omega}{v}}{2},\frac{h+\frac{i\omega}{v}}{2},\frac{1}{2},\cos^2\theta\right).
\end{align}
The parameter $h$ is given by
\begin{align}
h=\frac{1}{2}\left(1+\sqrt{9+4\gamma(\cos k-1)}\right),
\end{align}
where $k$ is the spatial momentum and $\gamma$ controls the relative strength of the on-site and intersite couplings, with $0\le \gamma\le 1$. Finally, we have defined 
\begin{align}
\theta_v=\frac{\pi}{2}(1-v),
\end{align}
where $0<v< 1$ is the overall coupling strength.\\
\begin{figure}
    \centering
    \begin{subfigure}[b]{0.48\textwidth}
         \centering
         \includegraphics[width=\textwidth]{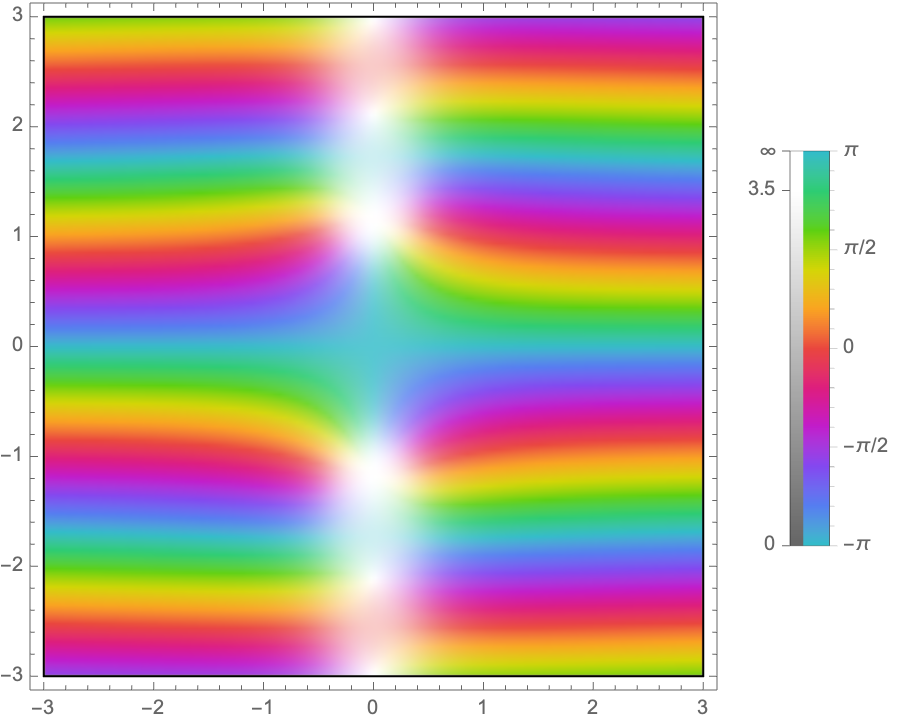}
         \caption{}
     \end{subfigure}
     \hfill
   \begin{subfigure}[b]{0.48\textwidth}
         \centering
         \includegraphics[width=\textwidth]{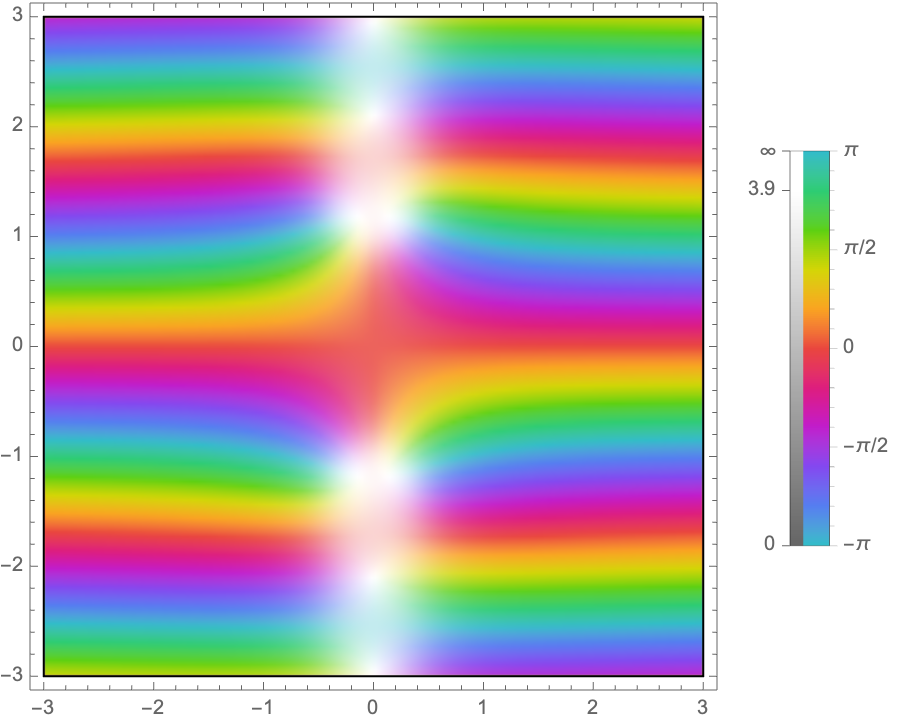}
         \caption{}
     \end{subfigure}
      \caption{(a) A plot of $G_{12}(\omega)$ in the complex $\omega$ plane for $k=1.297$, $\gamma=1$, and $v=.8$. As shown in the legend, the coloring captures the phase and the brightness signifies the magnitude. The poles, shown in white, are all on the imaginary axis. Putative zeroes would be shown in black, and are absent. (b) Here the parameters are taken to be $k=1.35$, $\gamma=1$, and $v=.8$. The first two poles have moved off the imaginary axis, as pointed out in \cite{Choi:2020tdj}. There are still no zeroes.}
      \label{g12chainplot}
\end{figure}
\indent The two-point function (\ref{g12chain}) defines a meromorphic function of frequency for any coupling. In Figure \ref{g12chainplot} we plot $G_{12}(\omega)$  for several values of the parameters. The poles of this function are all close to, but not necessarily on, the imaginary axis. Moreover, there are no zeroes, so this model provides a nontrivial example where the properties of the holographic Green's function extend beyond the holographic regime. We have checked numerically that (\ref{g12chain}) has no zeroes for various parameter values. It would be interesting to prove this analytically.

\subsection{Non-chaotic theories}

In theories which are not chaotic, we do not expect the no-zeroes or meromorphy properties to hold. For example, consider $\mathcal{N}=4$ SYM at zero 't Hooft coupling, as analyzed in \cite{Hartnoll:2005ju}. Setting $\beta=2\pi$, the Wightman function for ${\cal O}={\text{Tr }} F^2$ at zero spatial momentum is given by 
\begin{align}
G_{12}(\omega,0)= \frac{N^2}{8\pi}\frac{\omega^4}{\sinh^2(\pi \omega/2)}.
\end{align}
This is an analytic function of $\omega$. However, we see that there is a zero at $\omega=0$, so the no zeroes property no longer holds. \\
\indent When a nonzero spatial momentum is introduced, the correlator develops branch cuts in the complex $\omega$ plane \cite{Hartnoll:2005ju},
\be 
G_{12}(\omega,k)&=-\frac{N^2 \left(k^2-\omega ^2\right)^2 }{8\pi^2 ik\sinh(\pi \omega)}\Big[(i-\omega)\log\left(\frac{\omega+k}{\omega-k} - i 0 \right) + 2i\log
   \frac{\Gamma \left(-\frac{i (k+\omega)}{2}\right)}{\Gamma \left(\frac{i (k-\omega)}{2}\right)}-(\omega\to-\omega)\Big].
\ee
It follows that the analyticity property is broken at finite momentum. Let us briefly discuss the properties of this function as we increase $k$. First of all we see that it has a zero at $\omega=k$ (the same that we observed for $k=0$). At this point we have $\omega-k=0$ and therefore our local operator carries a null momentum, and thus can be interpreted as a light-ray operator that annihilates the vacuum \cite{Kravchuk:2018htv}. Resolving the identity as in \cite{Kologlu:2019bco}, we can write the thermal expectation value as a sum of double commutators which have zeros at certain integer-spaced values of scaling dimensions. Due to the integrability of the free theory the whole spectrum consists of integer-valued scaling dimension operators and we get zero. For the same reason we do not expect $G_{12}(\omega,k)$ to have zeros for real $\omega$ in interacting theories. For $k>\omega$ the correlator decays exponentially fast with the rate $e^{- \pi k}$ as expected on general grounds \cite{Son:2002sd,Banerjee:2019kjh}.

It is important to understand whether analyticity and no-zeroes are restored at small but finite 't Hooft coupling. A resummation is likely necessary to address this problem, since perturbation theory in $\lambda$ breaks down at late times \cite{Festuccia:2006sa}. An alternative possibility is that there is a transition between the holographic and free field theory behaviors at some intermediate coupling (see also \cite{Grozdanov:2016vgg,Grozdanov:2018gfx,Romatschke:2015gic} for related discussions).

Another simple example of a nonchaotic system is generalized free field theory. The generalized free field result takes a very simple form for the case $S^1 \times \mathbb{R}^{d-1}$, where the two-sided correlator is given by the vacuum block in \eqref{eq:OPEmink}, see \cite{Manenti:2019wxs}.\footnote{Recall that for actual CFTs, we are always in the black hole phase on $S^1 \times \mathbb{R}^{d-1}$.} Ignoring the 
$\theta(\omega^2 - k^2)$, we see that the two-sided correlator exhibits branch cuts at $\omega = \pm k$ and a power-law behavior along the imaginary axis $\omega \to \pm i \infty$ in contrast to the result in the black hole phase. Let us also mention that below the Hawking-Page transition the correlator to leading order is given by the generalized free field result on $S^1\times S^{d-1}$. 

It is an interesting question what happens for interacting QFTs in AdS at finite temperature. Perturbatively, we do not expect to see any change \cite{Alday:2020eua}. However it could be that perturbation theory breaks down at large (but not too large) times, which happens in other related cases \cite{Festuccia:2006sa}.

Finally, vector models in $d=3$ at large $N$ are characterized by weakly broken higher spin symmetry \cite{Maldacena:2012sf}, and in this sense are not chaotic. This manifests itself for example through the fact that higher spin currents have anomalous dimension $O(1/N)$. Another way to see that these theories are not chaotic is that the Lyapunov exponent is $O(1/N)$ in this case \cite{Chowdhury:2017jzb}. Thermal correlators in the $O(N)$ models have been studied for example in \cite{Petkou:1998fb,Petkou:1998fc,Witczak-Krempa:2012qgh}. The basic result is that they do not exhibit quasi-normal modes as expected in chaotic theories, and they have branch cuts in the $\omega$-plane. Similarly, we expect Chern-Simons matter theories at finite temperature \cite{Gur-Ari:2016xff} not to exhibit meromorphy and no-zero structure, but we have not explicitly checked this.

\section{Conclusions and further discussion}

The emergence of semi-classical gravity in holography is associated with strongly coupled quantum dynamics. Thermal correlators are particularly interesting observables in this regard since the dual geometry contains a black hole. However, understanding thermal correlators in strongly coupled systems is a challenging task. In particular, developing efficient bootstrap methods for probing them is an open problem.  

In this paper we have considered the holographic thermal two-point function and we noted that it exhibits an intriguing and non-obvious property: thermal two-sided correlators do not have zeros in the complex energy plane. From the dual geometry point of view, this property is very closely associated with the presence of the black hole horizon. Together with meromorphy, the no-zero property leads to the product representation of the thermal correlator \eqref{eq:ansatzprod}. We have derived this representation for holographic systems. However, we believe it may hold more generally.

\vspace{0.3cm}
\noindent {\bf Thermal product hypothesis (TPH):} The two-sided thermal two-point correlator in a chaotic large $N$ system is a meromorphic function with no zeros in the $\omega$-plane. As such it is given by a product over its poles.
\vspace{0.3cm}

From holography we expect that large $N$ chaotic theories at finite temperature are described by some kind of stringy black holes. Then the content of TPH is that at the level of the thermal two-point function, stringy black holes behave like ordinary black holes. We checked this property analytically in the SYK model and numerically in the SYK chain in Section \ref{beyondstrong}.  In particular, in the latter case it looks quite nontrivial; it would be instructive to derive the no-zero property of correlators in the SYK chain using an analytic argument. We also checked in Section \ref{higherderivative} that TPH holds if we include certain higher derivative corrections to GR.

 A few comments are in order.  In the context of CFTs, we expect that TPH can only be valid for theories where the anomalous dimensions of higher spin currents are $O(1)$, as opposed to $O(1/N)$. In particular, TPH does not apply to theories with slightly broken higher spin symmetry \cite{Maldacena:2012sf}. The requirement that the theory is chaotic is necessary since there are branch cuts for the correlator in free field theory or in the planar $O(N)$ model \cite{Witczak-Krempa:2012qgh}.\\
 \indent We expect that TPH may hold for CFTs on $S^1 \times \mathbb{R}^{d-1}$, or $S^1 \times S^{d-1}$ above the Hawking-Page phase transition. On $S^1 \times S^{d-1}$, the assumption of large $N$ is required since otherwise the two-sided correlator is given by a sum of $\delta$-functions. On $S^1 \times \mathbb{R}^{d-1}$, large $N$ is required since otherwise we expect to have branch cuts due to long-time tails, see e.g. \cite{Witczak-Krempa:2012qgh}. We require the theory to be above the Hawking-Page transition so that it is described by the black hole geometry. Finally, cuts appear in a simple kinetic theory description of thermal correlators in weakly coupled gauge theories, see e.g. \cite{Romatschke:2015gic}. However, to the best of our knowledge their existence has not been rigorously established, see the discussion in \cite{Hartnoll:2005ju}.

We have also reviewed the observation of \cite{Festuccia:2005pi}, related to the previous work \cite{Fidkowski:2003nf,Kraus:2002iv}, that holographic two-sided correlators decay exponentially fast as $\omega \to \pm i \infty$ in $d>2$. This behavior is associated with a light-like geodesic that bounces off the black hole singularity. We noticed that when combined with the prediction from the OPE there is a dispersive way to express this behavior through the black hole singularity sum rules \eqref{eq:BHSS}. Assuming the product representation of the thermal correlator, we see that the sum rules are trivially satisfied by having a family of QNMs that go to infinity in the complex plane at some angle $\omega \sim e^{i \theta}$ with $\theta \neq {\pi \over 2}$. At finite string coupling $\lambda$ we expect the sum rules to still hold in some range of the $\omega$-plane for which the gravity approximation applies. In this case we can consider finite energy singularity sum rules: $\oint_{C_\Lambda} d\omega' \,(\omega')^m G_{12}(\omega') \simeq 0 $, where $\Lambda$ is a finite energy scale. We can imagine that a similar version of the sum rules might hold at large but finite $N$.

Another open problem is to develop computational techniques to probe the region of large imaginary frequencies in string theory (or gauge theories, see e.g. \cite{Festuccia:2006sa}). In particular, tidal effects should become important near the singularity. These effects are responsible for resolving singularities in the two-point function on the bulk light-cone in one-sided correlators \cite{Hubeny:2006yu,Dodelson:2020lal}, and it seems plausible that tidal effects change the behavior of the two-sided correlator at large imaginary frequencies as well. However, the black hole geometry receives large corrections at a string length away from the singularity, so new insights are likely needed to address this problem.

In Section \ref{qnmfromope} and Section \ref{opesumrules}, we explored how the structure of the QNMs is constrained by the OPE. In Section \ref{sec:qnmhydro} we studied how the known structure of the QNMs puts constraints on the low-energy or hydrodynamic expansion of the correlator. The basic observation here is that the product representation of the correlator and its no-zero property lead to a convenient dispersive representation of the two-point function, see \eqref{logsum}. Moreover, the no-zero property translates into the fact that the discontinuity of the correlator is non-negative. The hydrodynamic expansion then turns into a moment problem for QNMs, and depending on the structure of QNMs bounds on the hydrodynamic expansion can be derived. It would be very interesting to understand such constraints more systematically. Conversely, the known data on the hydrodynamic coefficients can be used to put bounds on the geometry of QNMs (again assuming the no-zero property).  

 Another challenge that could provide important insights into the finite coupling structure of the correlators and the dual geometry is to find models in which the structure of the QNMs can be computed in a controllable setting and the emergence of gravitational features can be understood, see e.g. \cite{Maldacena:2023acv}. It would also be interesting to see if the product representation of thermal correlators generalizes beyond the two-point function \cite{Pantelidou:2022ftm,Loganayagam:2022zmq}.

Finally, the results of \cite{Rey:2005cn,Rey:2006bz} suggest that instanton corrections to weakly coupled thermal correlation functions can be interpreted in terms of AdS black holes. This is reminiscent of the fact that instanton corrections to vacuum correlators at weak coupling can be written as an integral over AdS, where the AdS space is simply the moduli space of instantons \cite{Maldacena:2015iua,Bianchi:1998nk,Dorey:1999pd,Bianchi:2013xsa}. It would be interesting to explicitly compute the first instanton correction to the thermal two-point function at weak coupling in $\mathcal{N}=4$ SYM, and to analyze its properties.

\section*{Acknowledgements}
We thank Luca Delacr\'etaz, Alba Grassi, Luca Iliesiu, Shota Komatsu, Petr Kravchuk, Raghu Mahajan, Kyriakos Papadodimas, Wilke van der Schee, Steven Shenker, Douglas Stanford, Andy Stergiou, Marija Toma\v{s}evi\'c, and Urs Wiedemann for helpful discussions and correspondence. This project has received funding from the European Research Council (ERC) under the European Union’s Horizon 2020 research and innovation programme (grant agreement number 949077). This research was supported in part by the National Science Foundation under Grant No. NSF PHY-1748958.

\appendix
\section{Two-point functions at finite temperature}
\label{app:2pointconventions}
Here we recall some basic definitions of thermal correlators, see e.g. \cite{Meyer:2011gj} for a detailed discussion. The Wightman, retarded, and advanced two-point functions are defined as 
\begin{align}
G_W(t)&=\frac{1}{Z}\text{Tr}\left(e^{-\beta H}\mathcal{O}(t)\mathcal{O}(0)\right)\\
G_R(t)&=i\theta(t)\frac{1}{Z}\text{Tr}\left(e^{-\beta H}[\mathcal{O}(t),\mathcal{O}(0)]\right)\\
G_A(t)&=-i\theta(-t)\frac{1}{Z}\text{Tr}\left(e^{-\beta H}[\mathcal{O}(t),\mathcal{O}(0)]\right).
\end{align}
The two-sided Wightman correlator is 
\begin{align}
G_{12}(t)=G_W\left(t-i\frac{\beta}{2}\right).
\end{align}
In frequency space, the two-point functions are related as 
\be
G_{12}(\omega)&=e^{-\beta\omega/2}G_W (\omega)= \frac{G_R(\omega)-G_R^*(\omega)}{2 i \sinh(\beta\omega/2)}\label{g12grdistri} \\
G_A(\omega)&=G_R(-\omega), ~~~
G_R^*(\omega)=G_R(-\omega^*).
\ee
where in \eqref{g12grdistri} $\omega$ approaches the real axis from above. Note that for real $\omega$, $G_{12}(\omega)=\rho(\omega)/(2\sinh(\beta \omega/2))$, where the spectral density $\rho$ is defined by $\rho(\omega)=2\text{Im }G_R(\omega)$.\\
\indent In general, \eqref{g12grdistri} defines a distribution and not a function, but for holographic correlators $G_R(\omega)$ is a meromorphic function with no poles on the real axis, and we can rewrite \eqref{g12grdistri} as follows
\be
G_{12}(\omega) &=\frac{G_R(\omega)- G_R(-\omega)}{2 i \sinh(\beta\omega/2)}.\label{g12gr}
\ee
This relation can now be analytically continued to $\omega \in \mathbb{C}$ and defines the meromorphic $G_{12}(\omega)$ studied in this paper. The ``no-zero'' property discussed in the main text translates to the statement that the equation
\be
G_R(\omega) = G_R(-\omega)
\ee
has solutions only for Matsubara frequencies
\be
\omega = {2 \pi i \over \beta} n .
\ee

\indent   The Euclidean correlator defined at Matsubara frequencies $\omega_n = {2 \pi i \over \beta} n$ is related to the retarded correlator as follows
\be
\label{eq:EuclRrelation}
G_E(\omega_n) = G_R(i \omega_n) , ~~~ n > 0 .
\ee
\section{Probing the $\omega$-plane in real time}
\label{app:probingomega}

From the point of view of an experimentalist, the complex $\omega$-plane discussed in this paper is not easily accessible. On the other hand, QNMs are known to control the late time behavior of the retarded two-point function which can be accessed more easily. In fact, for the correlators we consider QNMs control correlators at \emph{all times}. By this we mean the following.

Let us consider the retarded correlator for some positive non-zero time $t>0$,
\be
G_R(t) = {1 \over 2 \pi} \int_{- \infty}^{\infty} d \omega\, e^{- i \omega t} G_R(\omega) . 
\ee
We can close the contour into the lower half-plane to get 
\be
G_R(t) = - i \sum_{k} e^{- i \omega_k t} {\rm Res}_{\omega_k} G_R(\omega) , ~~~ t>0 ,
\ee
where the condition $t>0$ is necessary to drop the arc at infinity. In particular, there could be distributional terms localized at $t=0$ which are not captured by QNMs.
These correspond to subtractions in dispersion relations for $G_R(\omega)$.

\begin{figure}[t!]
\centering
\begin{subfigure}[b]{0.5\textwidth}
\centering
\includegraphics[width=\textwidth]{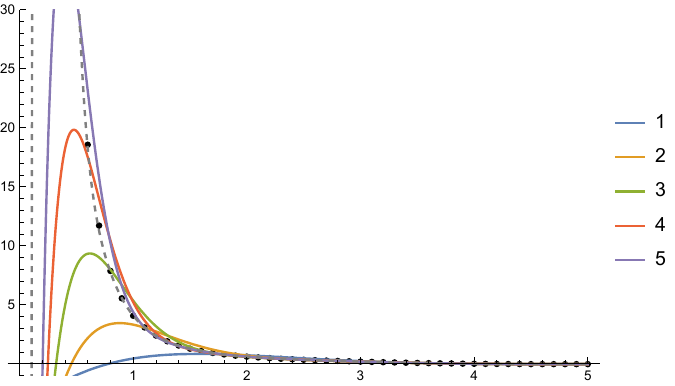}
\caption{}
\label{fig:qnmsalphag0}
\end{subfigure}%
\hfill
\begin{subfigure}[b]{0.5\textwidth}
\centering
\includegraphics[width=\textwidth]{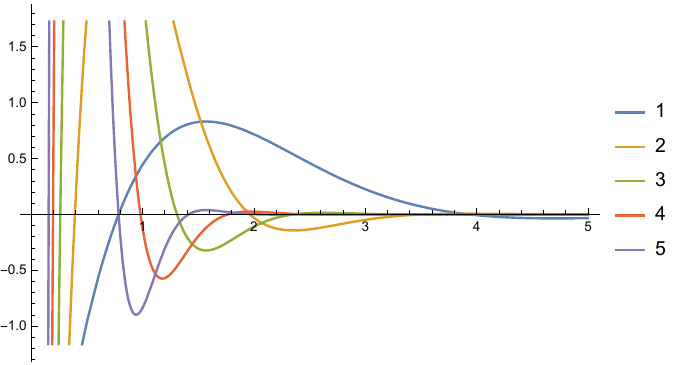}
\caption{}
\label{fig:qnmsalphal0}
\end{subfigure}
\caption{(a) We set $\beta = 2 \pi$ and plot the retarded function $G_R(t)$ as a function of time $t$ for R-currents in ${\cal N}=4$ SYM. The black dots correspond to the exact result. Various colors correspond to the result coming from summing over the first $k$ QNMs. We see that at late times the leading QNM dominates, but at shorter times higher QNMs become important. (b) The contribution of a single QNM at $\omega_n = (1-i)n$. We see that they form a nice hierarchical structure when we decrease $t$ from infinity. First only the $n=1$ QNM matters, then the second QNM becomes important, etc. }
\label{fig:grprod}
\end{figure}

Let us next express the residues in terms of QNMs. For this purpose we write
\be
G_R(\omega) =G_R(-\omega) +  2 i \sinh(\beta \omega/2) G_{12}(\omega) , ~~~ \omega \in \mathbb{C} \ . 
\ee
In writing the formula above we used meromorphy of $G_R(\omega)$, see Appendix \ref{app:2pointconventions} and \eqref{g12gr}.

Considering next $\omega$ in the lower half-plane we get
\be
{\rm Res}_{\omega_k} G_R(\omega) &= 2 i \sinh(\beta \omega_k/2) {\rm Res}_{\omega_k} G_{12}(\omega) \nn \\
&=  - {i \sinh(\beta \omega_k/2) G_{12}(0) \over \prod_{n \neq k}^\infty\left(1 - {\omega_k^2 \over \omega_n^2}\right)\left(1 - {\omega_k^2 \over (\omega_n^*)^2}\right)} {\omega_k \over 1 - {\omega_k^2 \over (\omega_k^*)^2}} . 
\ee

Restricting thus to QNMs with $\text{Re } \omega_k>0$ we get the following expression for the retarded two-point function in terms of the QNMs,
\be
\label{eq:GRQNM}
G_R(t) &=  - i \sum_{k} \Big( e^{- i \omega_k t} {\rm Res}_{\omega_k} G_R(\omega) + e^{i \omega_k^* t} {\rm Res}_{-\omega_k^*} G_R(\omega) \Big) \nn \\
&= - 2 G_{12}(0) \sum_{k} {\rm Re} \left[ {e^{- i \omega_k t} \sinh(\beta \omega_k/2)  \omega_k \over \left(1 - {\omega_k^2 \over (\omega_k^*)^2}\right) \prod_{n \neq k}^\infty\left(1 - {\omega_k^2 \over \omega_n^2}\right)\left(1 - {\omega_k^2 \over (\omega_n^*)^2}\right)}  \right] .
\ee
The sum converges for any $t>0$, with more and more QNMs becoming important at earlier times.

As an example we consider the thermal two-point function of R-currents, see \eqref{rcurrentswightman}. In Figure \ref{fig:grprod} we plot $G_R(t)$ and its approximation by a few QNMs. We see that as we decrease $t$ from infinity, higher QNMs become important in an ordered fashion. Note also that by measuring the residue of the first two QNMs we get access to some information about the high-energy tail through \eqref{eq:GRQNM}. Indeed, in the ratio of the residues $G_{12}(0)$ cancels and we get some prediction for the infinite products.

\section{Analytic properties of AdS wave equations}\label{nozeroesappendix}
Here we review some analytic properties of holographic correlators based on the bulk equations of motion \cite{Festuccia:2005pi,festucciathesis}. In particular, we will recall the argument that Wightman correlators have no zeroes, which is a consequence of scattering theory in quantum mechanics \cite{Newton1966ScatteringTO}.\\
\indent Consider the scalar wave equation (\ref{waveequation}) in an AdS black hole background. The normalizable mode $g$ and non-normalizable mode $\tilde{g}$ are specified by the boundary conditions at $z\to 0$\footnote{Below we drop dependence on the spatial momentum since we are mainly concerned with the analytic behavior as a function of $\omega$.}
\begin{equation}\begin{aligned}
    &g(\omega,z)\sim z^{\frac{1}{2}+\nu}\\
    &\tilde{g}(\omega,z)\sim z^{\frac{1}{2}-\nu}, 
\end{aligned}
\end{equation}
while the ingoing solution $h_R$ and outgoing solution $h_A$ are specified by the boundary conditions at the horizon $z\to\infty$
\begin{equation}\begin{aligned}
    &h_R(\omega,z)\sim e^{i\omega z}\\
    &h_A(\omega,z)\sim e^{-i\omega z}.
\end{aligned}
\end{equation}
\indent Various properties under conjugation and $\omega\to-\omega$ can readily be found from these boundary conditions. These solutions can further be expressed in terms of each other, in particular
\begin{equation}
    g(\omega,z) = \frac{1}{2i\omega}(f(-\omega)h_R(\omega,z)-f(\omega)h_A(\omega,z)),
\end{equation}
where $f(\omega)$ is the so-called Jost function given by the Wronskian 
\begin{align}
f(\omega)= h_R\partial_z g-g\partial_z h_R.
\end{align}
Since the Wronskian is independent of $z$, we can evaluate it at $z=0$. This gives 
\begin{align}\label{josthr}
f(\omega)=2\nu \lim_{z\to 0}\left(z^{\nu-1/2}h_R(\omega,z)\right).
\end{align}
Therefore the analytic properties of the Jost function are the same as those of the ingoing mode $h_R(\omega,z)$.

We further consider the physical solution proportional to the normalizable mode 
\begin{equation}
    \psi(\omega,z) = C(\omega)z^{\frac{1}{2}+\nu},\qquad z\to0,
\end{equation}
where $C(\omega)$ is fixed by the normalization
\begin{equation}
    \psi(\omega,z) = e^{i\omega z+i\delta}+e^{-i\omega z-i\delta},\qquad z\to\infty,
\end{equation}
and $\delta$ is the phase shift. It was shown in \cite{Festuccia:2005pi} that
\begin{equation}
    C^2(\omega)=\frac{4\omega^2}{f(\omega)f(-\omega)}
\end{equation}
and that the Wightman correlator is given by
\begin{equation}
    G_W(\omega) = \frac{4\nu^2}{2\omega}\frac{C^2(\omega)}{1-e^{-\beta\omega}}. 
\end{equation}
The analytic properties of the Wightman function can therefore be read off from those of the Jost function $f(\omega)$. In particular, for a regular potential with the asymptotic behavior \eqref{eq:VAsymp}, the Jost function is a meromorphic function with simple poles at the Matsubara frequencies $\omega=-i\frac{2\pi n}{\beta}$ with $n=1,2,\ldots$. It therefore follows that the Wightman function has no zeroes. Moreover, the zeroes of the Jost function correspond to the QNMs.

Let us review the argument that the only poles of the Jost function appear at the Matsubara frequencies. For simplicity, we begin by assuming that $\nu^2=\frac{1}{4}$ so that the potential behaves like $V\sim z^{\epsilon-2}$ as $z\to 0$ with $\epsilon>0$. In this case we can borrow the techniques of scattering theory with zero angular momentum $l=0$. It is straightforward to generalize the results to generic real values of $\nu^2>0$, as we will see later. The ingoing solution can be written as a Volterra equation as follows,
\begin{equation}\label{eq:VolterraIngoing}
    h_R(\omega,z)= e^{i\omega z} -\frac{\gamma}{\omega} \int_z^\infty dz' \,\sin(\omega(z-z'))V(z')h_R(\omega,z'),
\end{equation}
where we have multiplied $V(z)$ by a parameter $\gamma$, which we will eventually set to one. The solution to (\ref{eq:VolterraIngoing}) can be shown to define an absolutely convergent power series in $\gamma$ at finite $z$, see e.g.\ Sec 12.1 in \cite{Newton1966ScatteringTO}, if
\begin{equation}\label{eq:condIngoing}
    \alpha=\int_z^\infty dz'\, z' |V(z')|e^{(|\text{Im }\omega|-\text{Im }\omega)z'}<\infty.
\end{equation}
\indent Assuming the potential is regular for finite $z$, only the large $z$ region in \eqref{eq:condIngoing} could give rise to a divergence. In particular, for any decaying potential $\alpha$ is finite for $\text{Im }\omega>0$. Because of the presence of the horizon, we restrict to potentials which are exponentially decaying as $V=\sum_n a_n e^{-\frac{4\pi n}{\beta}z}$ as $z\to\infty$. For the purposes of studying whether $\alpha$ is finite or not, we can then choose $z>z_0$ for some large enough $z_0$ and replace the potential with the sum of exponentials. Consider therefore
\begin{equation}
    \alpha =\sum_{n=1}^{\infty}a_n \int_z^\infty dz'\, z' e^{(-\frac{4\pi n}{\beta}-2 \text{Im } \omega)z'}, \hspace{5mm} \text{Im } \omega<0.
\end{equation}
The integral converges for $\text{Im }\, \omega>-\frac{2\pi n}{\beta}$, corresponding to the $n$th Matsubara frequency. The region of analyticity can however be extended beyond this point with simple poles at $\omega=-i\frac{2\pi nm}{\beta}$ for $m=1,2,\ldots$. This can be seen explicitly by solving \eqref{eq:VolterraIngoing} order-by-order in $\gamma$, with a new simple pole arising at each order, see Section 12.1.1 in \cite{Newton1966ScatteringTO} for details\footnote{It is also possible to write down the exact solution for the exponential potential and it is meromorphic with poles arising from a factor of $\Gamma(1-i\frac{\beta\omega}{2\pi})$, see Section 14.3 in \cite{Newton1966ScatteringTO}.}. It then follows from (\ref{josthr}) that the Jost function $f(\omega)$ is meromorphic with simple poles at the Matsubara frequencies. \\
\indent We note that at isolated points in parameter space, the residue of the pole of the Jost function at one of the Matsubara frequencies might vanish. For example, for the black brane potential we have
\begin{align}
V(k,z)\sim e^{-\frac{2\pi z}{\beta}}\left(k^2+\nu^2+\frac{d(d-2)}{4}\right)+\ldots,\hspace{10 mm}z\to \infty.
\end{align}
It follows that $a_1=0$ when 
\begin{align}\label{kspecial}
k^2=-\Big( \nu^2 + \frac{d(d-2)}{4} \Big).
\end{align}
This coincides with the first pole-skipping point
\cite{Grozdanov:2017ajz,Blake:2017ris,Blake:2018leo,Grozdanov:2018kkt,Blake:2019otz}. For the special value (\ref{kspecial}) of $k$, $G_{12}(\omega)$ has a pole at the first Matsubara frequency.

Consider now the case $\nu^2\neq \frac{1}{4}$. The solution for $h_R(\omega_,z)$ can be found from the $\nu^2 =\frac{1}{4}$ solution $h_{R,\nu=\frac{1}{4}}(\omega,z)=h_{R}(\omega,z)$ by 
\begin{equation}\label{eq:highnu}
    h_{R,\nu}(\omega,z)= h_{R}(\omega,z)+\left(\nu^2-\frac{1}{4}\right)\int_z^\infty dz'\, {\cal G}(\omega;z,z')\frac{h_{R,\nu}(\omega,z')}{(z')^2},
\end{equation}
where ${\cal G}$ is the Green's function
\begin{equation}
    (-\partial_z^2+V(z)-\omega^2){\cal G}(\omega;z',z) = -\delta(z-z').
\end{equation}
One can show that ${\cal G}$ is an analytic function of $\omega$ and that the solution to \eqref{eq:highnu} has the same analytic structure as $h_{R}(\omega,z)$. The solution $h_{R,\nu}$ then has the same poles as $h_{R}$.  

Consider the Jost function $f(\omega)=W(h_R,g)$ at spectral points $\omega=\omega_n\neq 0$, where $f(\omega_n)=0$ by definition. One can show that  $f'(\omega_n)\neq 0$ for $\omega_n$ in the lower half plane and on the positive imaginary axis, corresponding to resonances and bound states respectively, and the Jost function therefore has simple zeroes at those spectral points \cite{Rakityansky2022Jost}. This follows from the use of the wave equation to relate $f'(\omega_0)$ to an integral of the square of $h_R(\omega_0,z)$, which converges and is non-zero for resonances and bound states. For virtual states with $\omega_n=-i|\omega_n|$ this no longer is true and it is in principle possible that the zeroes can be of higher order. This arose in the study of higher-derivative corrections \cite{Grozdanov:2018gfx}, which found that QNMs on the imaginary axis for certain values of the coupling collided and branched into complex QNMs.

\section{OPE in momentum space}
\label{app:OPEderiv}

Let us consider the OPE for a pair of identical scalar operators at finite temperature,
\be
\langle {\cal O}(\tau, \vec{x}) {\cal O}(0,0) \rangle_{\beta} = \sum_{{\cal O}_{\Delta,J}} a_{\Delta,J} C_J^{({d-2 \over 2})}\Big( {\tau \over \sqrt{\tau^2 + \vec x^2}} \Big) (\tau^2 + |\vec x|^2)^{{\Delta - 2 \Delta_{{\cal O}} \over 2}} ,
\ee
where we set $\beta = 1$.

We would like to perform the Fourier transform of this OPE expansion to derive the asymptotic expansion of the two-sided correlator $G_{12}(\omega, k)$. We start with the spatial Fourier transform. The result takes the following form \cite{Manenti:2019wxs}
\be
\langle {\cal O}(\tau, \vec{k}) {\cal O}(0,0) \rangle_{\beta} &= \sum_{{\cal O}_{\Delta,J}} a_{\Delta,J} \sum_{j=0}^{[J/2]} c_{J,j} {2 \pi^{{d-1 \over 2}} \tau^{J-2 j}  \over \Gamma({2\Delta_{{\cal O}} - \Delta + J-2 j \over 2})} \Big( {2 \tau \over  k} \Big)^{\beta_- - 1/2} K_{\beta_- - 1/2}(\tau k), \\
\beta_- &= {d-\Delta-J + 2j \over 2} , \\
c_{J,j} &= (-1)^j {\Gamma(J-j+{d-2 \over 2}) \over \Gamma({d-2 \over 2}) \Gamma(j+1) \Gamma(J-2j+1)} 2^{J-2j}  \ ,
\ee
where recall that $k = | \vec k | $.

Next we set $\tau = {1 \over 2} + i t$ and perform the Fourier transform
\be
G_{12}(\omega,  k) &= \int d t\, e^{i \omega t} \left\langle {\cal O}\left({1 \over 2} + i t, \vec{k}\right) {\cal O}(0,0) \right\rangle_{\beta} \nn \\
&= \theta(\omega) \theta(\omega^2- k^2) e^{- {\beta \omega \over 2}}\omega^{2 \Delta_{{\cal O}}-d} \sum_{{\cal O}_{\Delta,J}} {a_{\Delta,J} \over (\beta \omega)^{\Delta} } G_{\Delta,J} \Big({ k \over \omega} \Big) \ ,
\ee
where the relevant block takes the form
\be
G_{\Delta,J} (\zeta) &= \sum_{j=0}^{[J/2]} {2 \pi^{{d-1 \over 2}} c_{J,j}  \over \Gamma({\Delta + J-2 j \over 2})} \int_0^{\text{arccosh}{1/\zeta}} d x \,\frac{\pi  2^{\beta_- +\frac{1}{2}} \zeta ^{\frac{1}{2}-\beta_- } \cosh \left(\left(\beta_-
   -\frac{1}{2}\right) x\right) (1-\zeta  \cosh (x))^{-\beta_- +2
   j-J-\frac{1}{2}}}{\Gamma \left(2 j-J-\beta_- +\frac{1}{2}\right)} \nn \\
&=\frac{\pi ^{\frac{d}{2}+1} 2^{d+\Delta -2 \Delta_{{\cal O}} +1} \Gamma (d+J-2) \left(1-\zeta ^2\right)^{\frac{1}{2} (2\Delta_{{\cal O}} -d-\Delta -J )}}{\Gamma
   (d-2) \Gamma (J+1) \Gamma \left(\frac{J}{2}-\frac{\Delta }{2}+\Delta_{{\cal O}} \right) \Gamma \left(-\frac{d}{2}-\frac{J}{2}-\frac{\Delta
   }{2}+\Delta_{{\cal O}} +1\right)} \ _2 F_1\left({1-J \over 2},-{J \over 2}, {d-1 \over 2}, \zeta^2\right) .
\ee
The integral can be computed explicitly in terms of the \texttt{AppellF1} function. We can then compute the small $\zeta$ expansion of the block.

 \section{An exact expression for $G_{12}$}\label{exactexpression}
In \cite{Dodelson:2022yvn} an exact expression for the thermal scalar two point function in $d = 4$ in the holographic regime has been presented. The expression is given in terms of the connection coefficients for the Heun functions computed in \cite{Bonelli:2022ten}  and the Nekrasov-Shatashvili partition function $F_{NS}$, a special function appearing in the context of $\mathcal{N}=2$ 4d supersymmetric gauge theories \cite{Seiberg:1994rs, Seiberg:1994aj, Nekrasov:2002qd, Alday:2009aq, Nekrasov:2009rc}. The idea of applying $\mathcal{N}=2$ technologies to spectral problems started with \cite{Nekrasov:2009rc}, and was applied in the context of black hole perturbations for the first time in \cite{Aminov:2020yma} (for subsequent related works see \cite{Bonelli:2021uvf, Bianchi:2021mft, Bianchi:2021xpr, Consoli:2022eey, Bhatta:2022wga, Fioravanti:2021dce, Fioravanti:2022bqf, Gregori:2022xks}).

\begin{figure}[t]
\centering
\begin{subfigure}[b]{0.48\textwidth}
\centering
\includegraphics[width=\textwidth]{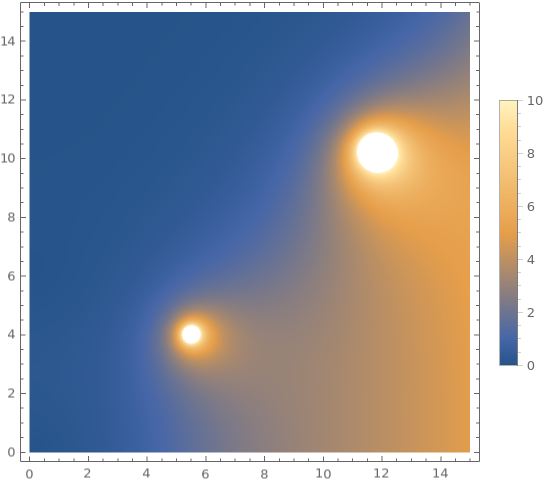}
\caption{}
\end{subfigure}
\hfill
\begin{subfigure}[b]{0.48\textwidth}
\centering
\includegraphics[width=\textwidth]{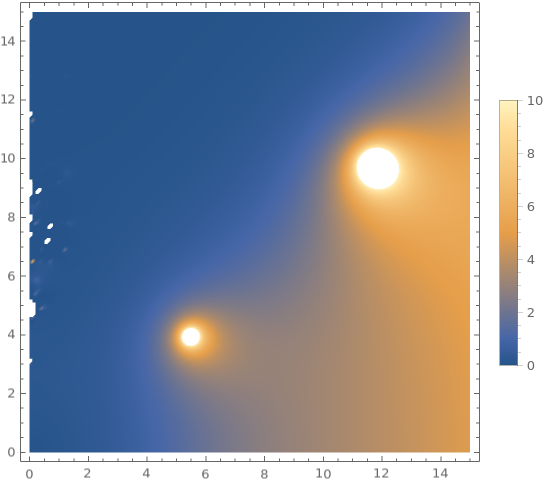}
\caption{}
\end{subfigure}
\caption{(a) A plot of $\text{Im } G_R(\beta \omega)$ in the complex $\beta \omega$ plane for $k=0$ and $\Delta =\frac{5}{2}$ from the product formula. Here the first 3 modes have been used. (b) Here $\text{Im } G_R(\beta \omega)$ is computed from a numerical evaluation of the exact result found in \cite{Dodelson:2022yvn}. The $F_{NS}$ series is truncated at order $t^5$. The slight mismatch in the position of the second pole is expected to be resolved by truncating the series at higher orders. The white spots along the imaginary axis are due to the presence of unphysical poles in the relation \eqref{eq:Matone}.}
\label{fig:imgrinstantons}
\end{figure}

The expression for $G_{12}$ for a theory dual to a black brane geometry in the bulk reads
\be\label{eq:g12frominst}
G_{12}(\omega, \zeta) &= \pi \, 2^{-2a_1} e^{- \partial_{a_1} F_{NS}} \frac{1}{\Gamma\left(2a_1+2\right)\Gamma\left(2a_1+1\right)}  \times\notag \\ &\times \prod_{\theta=\pm1}\left(\sum_{\sigma=\pm1} \frac{e^{-\sigma \partial_a F_{NS}} t^{\sigma a}\Gamma\left(1-2\sigma a\right)\Gamma\left(-2a\right)}{\Gamma\left(\frac{1}{2}-\sigma a + \theta a_t\right)^2\prod_\pm \Gamma\left(\frac{1}{2}-\sigma a+a_1\pm a_\infty\right)}\right)^{-1} \,,
\ee
where
\begin{align}
a_1 = \frac{\Delta_{\mathcal{O}}-2}{2} \,, \quad a_t = i \frac{\beta \omega}{4\pi} \,, \quad a_\infty = \frac{\tilde{\beta}\omega}{4\pi} \,, \quad t = \frac{1}{2}
\end{align}
are the parameters of the wave equation of a scalar in the black brane background in the variables $x = r^2/(1+r^2)$, $\chi(x) = \left(f(r) \frac{dx}{dr}\right)^{\frac{1}{2}} \psi(x)$, that is
\be\label{eq:Heun}
\left(\partial_x^2 + \frac{\frac{1}{4}-a_1^2}{(x-1)^2} - \frac{\frac{1}{2}-a_1^2-a_t^2+a_\infty^2+u}{x(x-1)} + \frac{\frac{1}{4}-a_t^2}{(x-t)^2}+\frac{1}{4 x^2} + \frac{u}{x(x-t)}\right) \chi(x) = 0 \,.
\ee
Here $u = \frac{\beta^2\omega^2}{8 \pi^2}\left(1-2\zeta^2\right)-a_1^2$. The parameters $u, a$ are related to each other through $F_{NS}\left(a_i, a, t\right)$ as follows:
\be
u = - \frac{1}{4}-a^2+a_t^2+t \partial_t F_{NS} \,.
\label{eq:Matone}
\ee
$F_{NS}$ is given as a convergent power series in $t$. Since for the black brane $t = 1/2$, \eqref{eq:g12frominst} cannot be analytically expanded perturbatively for $t \sim 0$. However since $1/2$ is a small number and the series is convergent, one can still truncate the series to a given order and compute \eqref{eq:g12frominst} numerically. In Figure \ref{fig:imgrinstantons} we compare $\text{Im} \, G_R$ obtained from the product formula with the numerical evaluation of \eqref{eq:g12frominst}.

Equation \eqref{eq:g12frominst} dramatically simplifies as $|\omega| \to \infty$. The key observation is that for large $\omega$ \eqref{eq:Heun} reduces to an hypergeometric equation in the variable $\tilde{x} = (x-t)/(1-t)$,
\be\label{eq:hyp}
\left( \partial_{\tilde{x}}^2 + \frac{\left(\frac{1}{4}-a_t^2\right)(1-\tilde{x})+\left(\frac{1}{4}-a_1^2-\left(\frac{1}{4}-a_\infty^2\right)(1-\tilde{x})\right)\tilde{x}}{\tilde{x}^2(1-\tilde{x})^2} \right) \chi(\tilde{x}) = 0 \,.
\ee
In order for this to be consistent with \eqref{eq:Heun} and \eqref{eq:Matone}, at leading order in $\omega$ we must have
\be\label{eq:largeomdict}
F_{NS} \simeq \left(\frac{1}{4}-a_1^2-a_t^2+a_\infty^2\right)\log(1-t) \,, \quad a \simeq a_t \,.
\ee
Note that the previous expression gives $F_{NS}$ at a specific value of $a = a_t$, so it cannot be used to evaluate $\partial_a F_{NS}$ in \eqref{eq:g12frominst}. By solving \eqref{eq:Matone} order by order in $t$ and then expanding for large $\omega$ we find
\begin{equation}
\partial_a F_{NS} = i c \beta \omega + \mathcal{O}(\omega^0) \,, \quad c \in \mathbb{R} \,.
\label{eq:largeomegadaF}
\end{equation}
Substituting \eqref{eq:largeomdict} and \eqref{eq:largeomegadaF} in \eqref{eq:g12frominst} and expanding for large positive $\omega$ we get the leading order with its nonperturbative corrections.
On the other hand, to compute power law corrections we would need corrections to \eqref{eq:largeomdict}. 

\begin{figure}[t]
\centering
\includegraphics[scale=0.5]{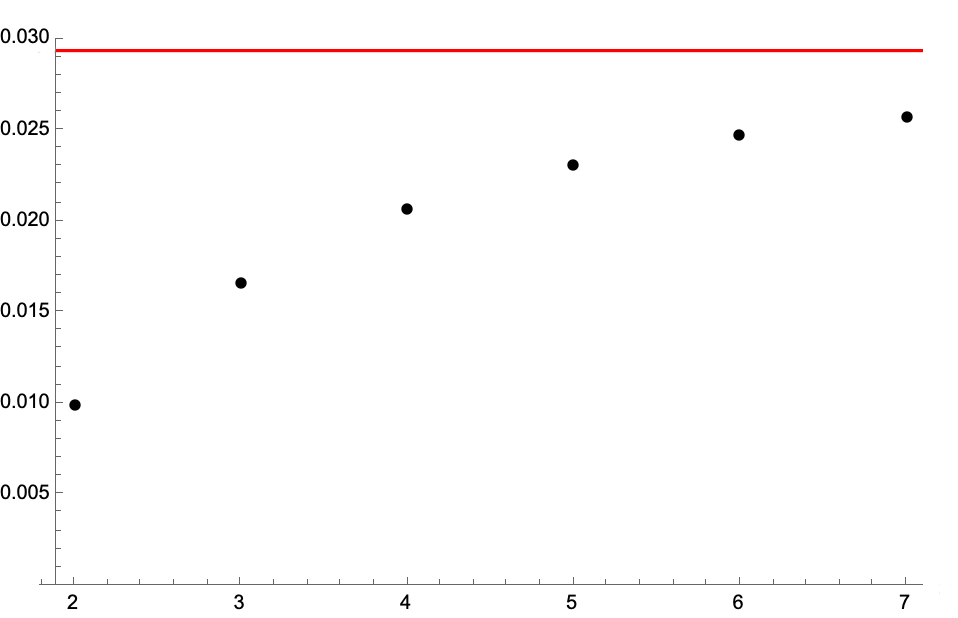}
\caption{The red line is the prediction for $c$ in \eqref{eq:Fpredic}. The black dots are the values of $c$ obtained by truncating the $t$ series in $F_{NS}$ at order $t^{n_{\text{max}}}$ as a function of $n_{\text{max}}$.}
\label{fig:cinst}
\end{figure}

Up to the first nonperturbative correction we get
\begin{equation}
G_{12}\left(\omega, 0\right) \simeq e^{-\frac{\beta \omega}{2}}\omega^{2\Delta_{\mathcal{O}}-4} \frac{4^{\frac{5}{2}-\Delta_{\mathcal{O}}} \pi^3 }{\Gamma\left(\Delta\right)\Gamma\left(\Delta_{\mathcal{O}}-1\right)}\left(1 - 4 e^{-\frac{\beta \omega}{2}} \cos \left(\pi (\Delta_{\mathcal{O}}-2)-\frac{\beta \omega}{2}\left(2 c+{1 \over 2}+{2 \log 2 \over \pi} \right)\right)\right) \,.
\label{eq:nonpertfrominst}
\end{equation}
Comparing with \eqref{eq:nonpertfestuccia} with $d = 4$ and $\beta = \tilde{\beta}$ we find the following prediction for $c$,
\be\label{eq:Fpredic}
c = {1 \over 4} - {\log 2 \over \pi} \simeq 0.0294 \,.
\ee
The prediction is confirmed numerically, see Figure \ref{fig:cinst}.

\section{Subleading corrections to black brane QNMs}\label{subleadingcorr}
In this appendix we compute the large $n$ asymptotics of QNMs in the AdS black brane to the next subleading order beyond the constant term. We start with the black brane potential at $k=0$, 
\begin{align}\label{potentialblackbrane}
V(z)=\left(r^2-\frac{1}{r^{d-2}}\right)\left(\nu^2-\frac{1}{4}+\frac{(d-1)^2}{4r^d}\right).
\end{align}
Following \cite{Natario:2004jd,Cardoso:2004up,Musiri:2005ev}, we need to compute the series expansion of the potential near the singularity. We work in the conventions of \cite{Musiri:2005ev}, so that the tortoise coordinate is defined by $dz=dr/f(r)$, with $z=0$ corresponding to $r=0$. The expansion of the tortoise coordinate around $r=0$ is
\begin{align}
z=-\frac{r^{d-1}}{d-1}-\frac{r^{2d-1}}{2d-1}+\ldots
\end{align}
Plugging into (\ref{potentialblackbrane}), we find
\begin{align}
V(z)=-\frac{1}{4z^2}+\frac{d^2+4\nu^2-8d\nu^2}{4(2d-1)(-(d-1)z)^{\frac{d-2}{d-1}}}+\ldots.
\end{align}
\indent The asymptotic expansion of the QNMs can now be computed to first order in perturbation theory, by matching the solution at the singularity to the normalizable solution at the boundary and imposing ingoing boundary conditions. This computation was done in \cite{Musiri:2005ev}, and we can read off the answer from Equation 100 in that paper. The asymptotic expansion takes the form (\ref{eq:asymqnms}), with 
\begin{align}
c={\pi \over \beta}\frac{(8d\nu^2-d^2-4\nu^2)\cos\left(\frac{\pi}{2(d-1)}\right)\Gamma\left(\frac{d-2}{d-1}\right)\Gamma\left(\frac{d}{2(d-1)}\right)}{2d(2d-1)\left(\frac{d}{2(d-1)}\sin\left(\frac{\pi}{d}\right)\right)^{\frac{1}{d-1}}\Gamma\left(\frac{d-2}{2(d-1)}\right)^3},
\end{align}
where recall that $\nu=\Delta - {d \over 2}$.

Using this expression it is straightforward to compute the coefficient for the first sum rule in \eqref{eq:firstsumruleOPE},
\be
\label{eq:deltac3explicit}
 \delta c_3 &=-\sin
   ^2\left(\frac{\pi }{d}\right) \frac{\log (4) \sin \left(\frac{2 \pi }{d}\right)
   \left(\pi ^2 \left(4-3 (d-2 \Delta )^2\right)+\log ^2(4)\right)}{3 \pi  \beta ^2} \nn \\
   &-\sin
   ^2\left(\frac{\pi }{d}\right) \frac{\pi  (d-2 \Delta )
   \cos \left(\frac{2 \pi }{d}\right) \left(\pi ^2 (d-2 \Delta -2) (d-2 \Delta +2)-3
   \log ^2(4)\right)}{3 \pi  \beta ^2} \nn \\
&+ c \frac{8 \sin \left(\frac{\pi }{d}\right) \left(\pi  \sin
   \left(\frac{\pi  (d-2)^2}{2 (d-1) d}\right) \left(4 \zeta
   \left(\frac{1}{d-1}\right)-(d-2 \Delta +2) \zeta
   \left(\frac{d}{d-1}\right)\right)+\log (4) \zeta \left(\frac{d}{d-1}\right) \sin
   \left(\frac{\pi  (3 d-4)}{2 (d-1) d}\right)\right)}{\beta } \nn \\
   &-4 c^2 \zeta \left(\frac{2 d}{d-1}\right) \cos \left(\frac{\pi  (d-2)}{(d-1)
   d}\right) ,
\ee
where $\zeta(x)$ is the Riemann zeta function.

\section{Shear viscosity and the scalar two-point function}
\label{app:shearvis}

Let us consider the Wightman two-point function of stress-energy tensors in position space. It takes the form
\be
\la T_{\mu \nu}(x) T_{\rho \sigma} (0) \ra &= {c_T \over x^{2 d}} \left( {1 \over 2!} (I_{\mu \rho}(x)I_{\nu \sigma}(x) + I_{\mu \sigma}(x) I_{\nu \rho}(x) ) - {1 \over d} \eta_{\mu \nu} \eta_{\rho \sigma} \right) , \\ 
I_{\mu \nu}(x) &= \eta_{\mu \nu} - 2 {x_\mu x_\nu \over x^2} \ ,
\ee
where we work with mostly plus signature. Consider next
\be
\int d^d x\, e^{- i q \cdot x} \la T_{\mu \nu}(x) T_{\rho \sigma} (0)  \ra .
\ee
This problem was solved for example in \cite{Gillioz:2018mto}. The result takes the form
\be
\la T_{\mu_1 \mu_2}(q) T_{\nu_1 \nu_2} (0) \ra &= c_T {\pi^{d/2+1} \theta(q^0)\theta(-q^2) (-q^2)^{d/2} \over 2^{d-1}(d+1)\Gamma(d-1)\Gamma({d+2 \over 2})} \sum_{n=0}^2 {2^{n+1} \over n! (2-n)!} {(-{d \over 2})_n \over (-d)_n} \nn \\
&\hspace{-2 mm}\times \left( {1 \over 2} {q_{\mu_1} q_{\nu_1} ... q_{\mu_n} q_{\nu_n} \over (-q^2)^n} \eta_{\mu_{n+1} \nu_{n+1}} ... \eta_{\mu_{2} \nu_{2}}  + \text{permutation} - \text{trace} \right) . 
\ee
For the shear viscosity computation we choose $\la T_{x y}(\omega,\vec{0},q^z) T_{x y}(0) \ra $. In this case only the $n=0$ term of the $I_{ac}(x)I_{bd}(x)$ part contributes, with the following result
\be
\la T_{x y}(\omega,\vec{0},q^z) T_{x y}(0) \ra = c_T {\pi^{d/2+1} \theta(\omega)\theta(\omega^2 - (q^z)^2) (\omega^2 - (q^z)^2 )^{d/2} \over 2^{d}(d+1)\Gamma(d-1)\Gamma({d+2 \over 2})} .
\ee
Let us compare this with the Wightman function of the normalized scalar primaries $\la \cO(x) \cO(0) \ra = {c_\cO \over x^{2d}}$ of scaling dimension $\Delta = d$,
\be
\la  \cO(\omega,\vec{0},q^z)  \cO(0) \ra =c_\cO {\pi^{d/2+1} \theta(\omega)\theta(\omega^2 - (q^z)^2) (\omega^2 - (q^z)^2 )^{d/2} \over 2^{d-1} \Gamma(d)\Gamma({d+2 \over 2})} .
\ee
From this we find that
\be
c_\cO = {d-1 \over 2(d+1)} c_T .
\ee
This is the normalization for the scalars that we need to use in our computations for $\Delta = d$ in order to compare to the stress-energy computation.

The shear viscosity $\eta$ is defined as \cite{Policastro:2001yc}
\be
\eta = \lim_{\omega \to 0} {1 \over \omega} \text{Im } G_R(\omega, 0). 
\ee
In a theory with a gravity dual it is given by
\be
\eta = {\sigma_{abs}(0) \over 16 \pi G_N}, 
\ee
where $\sigma_{abs}(0)$ is the black hole cross-section at zero frequencies, which is given by the area of the horizon (per unit CFT volume)
\be
\sigma_{abs}(0) = {\rm Area} . 
\ee
Recall that the black hole entropy density is in the same way $s = {{\rm Area} \over 4 G_N}$. This produces the famous relation $\eta/s = {1 \over 4 \pi}$ \cite{Kovtun:2004de}.
Therefore the prediction is
\be
 \lim_{\omega \to 0} {1 \over \omega}  {d-1 \over 2(d+1)} c_T \text{Im } G_R(\omega, 0)|_{\Delta=d} = {{\rm Area}  \over 16 \pi G_N},
\ee
where $G_R(\omega, 0)$ is the scalar two-point function with the normalization used in this paper, namely $\la \cO(x) \cO(0) \ra = {1 \over x^{2d}}$. 

Recall that we have \cite{Kovtun:2008kw}
\be
c_T &= {\pi^{-d/2 - 1} \Gamma(d+2) \over 8 (d-1) \Gamma(d/2)} {1 \over G_N}, \nn \\
{\rm Area} &= \left({4 \pi \over d}\right)^{d-1} \beta^{1-d} ,
\ee
where we set $R_{{\rm AdS}}=1$.

In this way we get the following prediction for the holographic correlator for $\Delta=d$
\be\label{g120deltad}
G_{12}(0) = \frac{\left(\frac{2}{d}\right)^d \pi ^{\frac{3 d-1}{2}}}{\Gamma \left(\frac{d+1}{2}\right)} \beta ^{-d}.
\ee

\section{Computation of QNMs}
In the main body of this work we have numerically calculated the QNMs in various examples using the publicly available $\mathtt{QNMSpectral}$ package for Mathematica \cite{Jansen:2017oag}. The method used is described in detail in \cite{Jansen:2017oag}, which we briefly review here. In holographic theories, the calculation of QNMs boils down to solving wave equations on top of a black hole background in AdS with specified boundary conditions, namely ingoing boundary conditions at the horizon and normalizable boundary conditions at the AdS boundary. As discussed in detail in Appendix \ref{nozeroesappendix}, the QNMs are solutions for discrete values of $\omega\in\mathbb{C}$ such that the solution with ingoing boundary conditions at the horizon is proportional to the normalizable mode and the Wronskian therefore vanishes. 

The method of \cite{Jansen:2017oag} finds a solution to the wave equation numerically by discretizing the radial direction on a grid of $n+1$ points, reducing the wave equation to a generalized eigenvalue problem of $(n+1)\times(n+1)$-matrices. The boundary conditions are imposed by expanding in a set of functions which manifestly obey the specified boundary conditions. The output is $n+1$ eigenvalues, some of which correspond to physical QNMs while some are unphysical. To filter out unphysical solutions, following \cite{Jansen:2017oag} we compute the QNMs twice with different grid sizes\footnote{Typically we choose $n=200$ and $n=400$ with the precision set to $n/2$.} and keep only those that appear in both cases. 

The computational time and the number of physical QNMs that are obtained depend on the grid size $n$ and the precision used when solving the generalized eigenvalue equation. These aspects were studied for AdS-Schwarzschild in Appendix A of \cite{Jansen:2017oag}. It was found that when the precision is set to $n/2$, the computational time grows roughly like $t\sim t_0n^{3.3}$ for large enough $n$, and the number of physical QNMs grows linearly with $n$.

Let us review in a bit more detail how this works for the $\Delta=4$ scalar $\phi(U,t,x)=\phi(U)e^{-i\omega t+ik z}$ in a black brane background. A convenient way to impose the correct the boundary conditions for QNMs is to pass to Eddington-Finkelstein coordinates
\be 
    ds^2 = -F(U)\,dt^2+2G(U)\,dt\,dU+\frac{1}{U^2}\,d\vec{x}^2,
\ee
where $F(U)=\frac{1}{U^2}-U^2$ and $G(U)=-\frac{1}{U^2}$. We further consider the wave equation in terms of the field $\psi(U)=U^{-3}\phi(U)$, which is given by
\be \label{eq:scalarQNMeq}
    \Big[\left(-k^2 U^2+3 i U \omega-9 U^4-3\right)+U \left(2 i U \omega-7 U^4+3\right)\p_U -\left(U^4-1\right) U^2 \p_U^2\Big]\psi(U)=0.
\ee
The main points are the following: 1) after going to EF coordinates, at the horizon the ingoing mode approaches a constant $\psi\propto 1+\ldots$, while the outgoing mode oscillates rapidly $\psi\propto(1-U)^{\frac{i\omega}{2}}+\ldots$ and 2) after rescaling by $U^{-3}$, at the boundary the non-normalizable mode diverges as $\psi\propto U^{-3}+\ldots$, while the normalizable mode goes to zero linearly, $\psi\propto U+\ldots$. 

In order to discretize and solve \eqref{eq:scalarQNMeq} numerically, \cite{Jansen:2017oag} uses a pseudo-spectral method where a function $f(x)$ is approximated on a grid $x_i$ with $i=0,1,2\ldots n$ as 
\be\label{eq:interp}
    f(x)=\sum_{j=0}^n f(x_j)C_j(x),
\ee
where $C_j(x_i)=\delta_{ij}$. In particular, the so-called cardinal function $C_j(x)$ is given by 
\be\label{eq:cardinal}
    C_i(x)= \prod_{j=0,j\neq i}^n \frac{x-x_j}{x_i-x_j},
\ee
and the interpolation \eqref{eq:interp} is exact on the grid points $x_i$. The grid points in \cite{Jansen:2017oag} are chosen to be the Chebyshev grid\footnote{These live in $[-1,1]$ but can shifted and rescaled to $[0,1]$ to apply to \eqref{eq:scalarQNMeq}.}
\be\label{eq:grid}
x_i = \cos\left(i\pi/n\right),\qquad i=0,1,2\ldots n.
\ee
Inserting the choice of grid points \eqref{eq:grid} into \eqref{eq:cardinal}, the functions $C_i$ can be written as a linear combination of Chebyshev polynomials $T_i(x)$, see \cite{Jansen:2017oag} for further details. \\
\indent In particular, with this choice the interpolation \eqref{eq:interp} can never approximate a function that diverges close to the boundary and oscillates rapidly as we approach the horizon. The boundary conditions corresponding to QNMs are therefore automatically implemented. By separating the $\omega^0$ and the $\omega^1$ terms in \eqref{eq:scalarQNMeq}, the wave equation evaluated at the grid points can be put into the form of an $(n+1)\times(n+1)$ generalized eigenvalue equation. The $n+1$ eigenvalues then corresponds to candidate QNMs $\omega_n$. However, for a fixed number of grid points $n$, only a subset of the $(n+1)$ eigenvalues correspond to actual QNMs. To select the physical solutions, one can do the computation with different grid sizes $n_1$ and $n_2$, and keep only the solutions which agree between the two \cite{Jansen:2017oag}.\footnote{This is not guaranteed to always correctly select the QNMs but in practice seems to work well.}

\bibliographystyle{JHEP}
\bibliography{mybib}
\end{document}